\documentstyle[prd,aps,floats,graphicx] {revtex}

\begin{document} 
\draft

%\twocolumn[\hsize\textwidth\columnwidth\hsize\csname
%@twocolumnfalse\endcsname

\title{Tadpole and Anomaly Cancellation Conditions in D-brane Orbifold
models}

\author{Maria E. Angulo$^{\ast}$, David Bailin$^{\ast}$ and Huan-Xiong
Yang$^{\ast,  \star}$ \\  {\sl $\ast$ Center  for Theoretical Physics,
University of  Sussex,}\\{\sl Brighton  BN1  9QJ, UK} \\ {\sl  $\star$
Zhejiang   Institute of Modern  Physics,  Zhejiang University,}\\ {\sl
Hangzhou, 310027, China }}

\date{\today}

\maketitle

\begin{abstract}

We derive  and    generalize   the  $\mathbf{RR}$     twisted  tadpole
cancellation     conditions necessary to     obtain  consistent $D=4$,
${\mathbf Z}_N$ orbifold  compactifications of Type IIB string theory.
At least  two different types of branes  (or  antibranes with opposite
$\mathbf{RR}$ charges)   are introduced  into  the  construction.  The
matter  spectra   and their contribution    to  the non-abelian  gauge
anomalies are computed.  Their relation  with the tadpole cancellation
conditions is  also reviewed.  The presence  of  tachyons is  a common
feature for some of the non-supersymmetric systems of branes.

\end{abstract}

%\vskip2pc]

\section{Introduction}

A crucial aspect to consider  in the construction of consistent string
theories and their compactified versions is whether the theory is free
of  ultraviolet (UV)   divergences.    In the  perturbative  heterotic
superstring theory, the absence of UV divergences is guaranteed by the
modular  invariance of the  torus   amplitude (the one-loop   oriented
closed string vacuum amplitude) \cite{Pol1}.  This is not the case for
orbifold  or  orientifold compactifications  of Type   I and  Type  II
superstring theories, where open and  unoriented closed string sectors
(corresponding to  the cylinder,  M\"{o}bius  strip and   Klein-bottle
amplitudes  respectively) are also  present  in the theory.  For these
amplitudes there  is no  modular group and  UV divergences  may remain
present in  the  theory.   This  problem  arises  in the  orbifold and
orientifold constructions of Type I and Type II string theories in the
presence of D-branes, where open and  unoriented closed strings appear
besides oriented closed strings.  Fortunately,  we can still construct
consistent string theories of this  type if the  the theory is free of
Ramond-Ramond  ($\mathbf{RR}$)  tadpole divergences.  The  absence  of
tadpoles is a  neccessary constraint to  guarantee the  consistency of
the equations of  motion of the  ${\mathbf{RR}}$ form potentials.   In
superstring theory, we  encounter   two different sources of   tadpole
divergences:  Neveu-Schwarz-Neveu-Schwarz  ($\mathbf{NSNS}$)  tadpoles
and $\mathbf{RR}$ tadpoles.   The presence of $\mathbf{NSNS}$ tadpoles
represents a shift of the vacuum state in perturbation theory and they
can be removed by expanding the background fields (dilaton and metric)
around the  corrected   solutions to the   equations  of  motion.  The
existence of  $\mathbf{RR}$ tadpole divergences indicates the presence
of unbalanced $\mathbf{RR}$ charges  in the theory which  would couple
to the $\mathbf{RR}$ form  potentials producing inconsistent equations
of motion for  those potentials \cite{Pol1}.   To obtain  a consistent
superstring theory we need, at  least, to impose  the vanishing of the
$\mathbf{RR}$ tadpoles even if in some (non-supersymmetric) cases, the
theory   is not  free   of  $\mathbf{NSNS}$  tadpole divergences.   In
supersymmetric models, both contributions  to the tadpoles cancel each
other.  Even  then,  a  separate  cancellation  of  the  $\mathbf{RR}$
tadpoles is required to ensure the consistency of the theory.

Before the ``second  string revolution'', the heterotic  string theory
was thought  to  be the  only  candidate in which  the  unification of
General Relativity  and  the Standard Model  (SM)  could be  realized.
With   the discovery of  D-branes \cite{PolRR},  all five perturbative
descriptions of superstring  theory  were found to  be  related to one
another and each of them is conjectured to be a different perturbative
limit of the same  non-perturbative theory (M-  theory).  As a result,
the previous emphasis on the heterotic  string became less pronounced.
Type I and  Type II string theories are  now strong candidates for the
unification of gravity and gauge interactions within the string theory
framework, if suitable types and numbers of D-branes are included.  So
far,  several  types of  orbifold   and  orientifold models based   on
compactifications  of Type  I and  Type  II string theories have  been
successfully constructed  in D=4 and D=6  dimensions, with  or without
space-time  supersymmetry  \cite{Al1} -\cite{Leigh}.   An  alternative
approach to model building based on Type  II superstring theory is the
so-called  ``bottom-up'' approach \cite{Al2}.   It  searches for local
configurations of  D-branes    at  a $R^6/{\mathbf   Z}_N$    orbifold
singularity, in such a way that the world-volume  theory is similar to
that  of the Standard   Model,   before embedding this  local   theory
(embedding of  the  orbifold point group  only)  in a  global orbifold
theory (embedding of the lattice as well).  This approach is appealing
because  some  properties  of  the  model, such   as    the number  of
generations and the gauge group of the theory depend only on the local
configuration of the D-branes  at the orbifold  singularities.  Simple
configurations of  Type  IIB   superstring  models  with  world-volume
theories close to the  Standard Model have been  obtained by placing a
stack of D3 branes on top of orbifold singularities with matter fields
localized   in  the   4-dimensional  world-volume.   The  presence  of
additional  branes beyond  D3 (e.g.   D7) is  required by  the tadpole
cancellation conditions which at the same time  determine how and when
the open strings should be added.

Dp-branes are p-dimensional hyperplanes describing the dynamics of the
endpoints of   open   strings.  These endpoints  carry   gauge quantum
numbers and allow gauge interactions and chiral matter to exist within
the D-brane worldvolume  while  gravity remains  present in  the bulk.
D-branes  are carriers of   $\mathbf{RR}$ charges \cite{PolRR}.  These
charges can be calculated by looking at the $\mathbf{RR}$ tadpole.  If
the Dp-brane has a non-zero  $\mathbf{RR}$ charge, then  it acts as  a
source term in the  equations of motion for  the $A_{p+1}$ form field.
According to the  Gauss law, all the  field lines in the compact space
must end   on an  opposite  charge.   The way   to  cancel the  charge
contribution from  one  brane in a   compact space is by  adding other
Dp-branes   in such a way  that  all the  tadpoles  are cancelled.  In
orientifold  constructions, the planes that are  left invariant by the
worldsheet symmetry element    of the orientifold   group (orientifold
planes), also carry $\mathbf{RR}$ charges.  The one-loop closed string
amplitude     (Klein-bottle) contains a    $\mathbf  {RR}$ tadpole and
D-branes carrying  opposite    $\mathbf{RR}$ charges  are   needed  to
neutralize those of    the  orientifold planes.   Therefore,   in  the
orientifold  framework  of Type IIB  string theories,  the presence of
Dp-branes  and  with them  the open  string  sector  of  the theory is
required  for consistency reasons.   However,  it is also possible  to
consider the  presence of   open strings from  the start   in orbifold
compactifications of Type IIB superstring  theories as long as we keep
the theory free  of $\mathbf{RR}$ tadpoles and  thus consistent.  This
is the idea behind the ``bottom-up'' approach.

In the orientifold theory, the one-loop  vacuum amplitudes include the
torus, the Klein-bottle, the M\"{o}bius  strip and the cylinder but as
mentioned before, only  the  last three  act as  sources  for massless
tadpoles.   The Klein-bottle    corresponds  to the contribution    of
unoriented   closed   strings    to the  one-loop   vacuum  amplitude.
Alternatively, this amplitude can  be pictured as a tree-level  closed
string amplitude where    the closed  strings propagate  between   two
orientifold planes.  The  cylinder (or  annulus) amplitude corresponds
to the contribution of open strings to  the one-loop vacuum amplitude,
or equivalently, to the  closed string tree-amplitude where the closed
strings  propagate between   two   D-branes.   The  M\"{o}bius   strip
corresponds to the tree-level closed string amplitude where the closed
srings propagate between one brane and  one orientifold plane.  In the
bottom-up  approach,   the absence of   orientifold  planes leaves the
cylinder worldsheets as the only source for  tadpole divergences.  The
different tadpole contributions  can be classified according to  their
volume dependence.  According to whether we are computing the tadpoles
from the untwisted  or the twisted  sectors of the orbifold, different
volume  dependences  arise.   In   the   ``bottom-up'' approach,  only
cylinder amplitudes contribute to the computation of the tadpoles.  In
orientifold models, there are additional contributions coming from the
Klein-bottle and M\"{o}bius strip amplitudes.

In this paper we present a detailed derivation  of the twisted tadpole
cancellation  conditions necessary for   obtaining a consistent  $D=4$
Type    IIB  superstring theory  within   the  ``bottom-up'' approach,
compactified on a 6-dimensional orbifold in  the presence of different
sets of D-branes.  We  also compute the  contribution from the  chiral
matter  to the anomalies and  analyze whether anomally cancellation is
guaranteed  when the theory is free  of tadpole divergences \cite{Al1}
-\cite{Bada}.  The plan of the paper is as follows.   In section 2, we
give a generic introduction   to   the open string mode    expansions,
Hilbert space and partition functions for $D=4$  Type IIB orbifolds in
the  presence  of  branes.  Section  3  is   devoted to  the  study of
orbifolds of Type IIB superstring theory in the presence of sets of D9
and D5  branes.  We evaluate  the chiral  fermion  spectra of the open
string sector  and derive its  contribution  to  the gauge  anomalies.
Section 4  lists   equivalent results  when  in  the theory  there are
present sets  of    D3-D7,  D3-D9,  D3-D5, D9-D7    and  D5-D7  branes
respectively, some of which  break supersymmetry.  Section 5 discusses
the conclusions   and   some final remarks.    An  appendix   with the
properties of   the   Jacobi  theta functions  that   appear   in  the
computation of  the partition functions  is given  at  the end to help
understanding the calculations.

\section{D=4 Orbifolds with D-branes}
\renewcommand{\theequation}{2.\arabic{equation}}\setcounter{equation}{0}

In the construction of the four dimensional  theory ($\mu = 0,1,2,3$),
six of the spatial  dimensions of  the original 10-dimensional  theory
($i = 4,5,6,7,8,9$) are     compactified  on a $T^6/{\mathbf     Z}_N$
orbifold.    The  six-dimensional  orbifold     is   obtained from   a
6-dimensional torus  $T^6$ with  ${\mathbf Z}_N$ discrete  symmetry by
dividing out this discrete symmetry.   This ${\mathbf Z}_N$  invariant
$T^6$ can be realized  as a root lattice  of a rank-6 Lie algebra,  on
which  the elements of  the  orbifold group  act crystallographically.
For ADE semisimple   Lie  algebras, if we   choose   the simple  roots
$\alpha_i$  ($i=4,5, \ldots, 9$) as  the basis  vectors of the compact
subspace,  $X \equiv \sum^9_{i=4}  X_{i} \alpha_{i}$, the torus can be
defined as:
\begin{equation} \label{2.1}
X_{i} \cong X_{i} + 2\pi R_{i}\; ,
\end{equation}
where $i=4,5,\dots,9$ and $R_i$ is the $i$-th  component of the vector
$\vec{R} \equiv \sum^{9}_{i=4}  R_{i} \vec{\alpha}_{i}$, which belongs
to the 6-dimensional lattice $\Lambda = \{ \sum_{i=4}^{9} n_i \alpha_i
|n_i \in Z \}$. We denote the  elements of the ${\mathbf Z}_N$ abelian
point group as:
\begin{equation} \label{2.2} {\mathbf Z}_{N} = \{ 1,
\theta, {\theta}^2, \dots, {\theta}^{N-1}\}. \end{equation}
The orbifold fixed points are determined by the following condition:
\begin{equation} \label{2.3} X_{f} \equiv
\theta  X_{f} + \sum^{9}_{i=4}n_{i}     \alpha_{i}\; ,  \nonumber   \\
\;\;\;(n_{i} \in Z). \end{equation}
Generally, the point group elements act non-diagonally with respect to
the simple roots $\alpha_{i}$ $(i=4,5,  \dots, 9)$.  Alternatively, we
can choose  a   more  convenient basis  $e_{j}$  $(j=3,4,5)$   for the
orbifold basis vectors,
\begin{equation} \label{2.4}
 e_{\pm i} \cdot e_{\pm  j}  =0, \, \,  \,  e_{\pm i} \cdot e_{\mp  j}
=\delta_{ij} \,\, \,\, (i,j = 1, 2, \cdots , 5) \end{equation}
where  now the basis  vectors  are eigenvectors  of the  abelian point
group elements and the orbifold action is diagonal with respect to the
new basis.  The simple roots $\alpha_{i}\; (i=4,5,\dots,9)$ are linear
combinations of the complex basis vectors.   With respect to these new
basis vectors, the  lattice $\Lambda$  and  its dual $\Lambda^*$   are
identical.   The  orbifold  action on an   arbitrary   coordinate $X =
\sum_{j=1}^5 (X_j e_{-j} + X_{-j} e_j)$, is now given by:
\begin{equation} \label{2.5} 
\theta^k   X_{\pm j} =  \left  \{   \begin{array}{l} X_{\pm j}  \hfill
(j=1,2) \\ \exp(\pm 2 \pi i k v_{j}) X_{\pm j} \hfill \,\,\, (j=3,4,5)
\end{array} \right. 
\end{equation}
where $v_{j}$ $(j=3,4,5)$ defines the orbifold twist vector.  Type IIB
superstring theory without the presence of D-branes is a closed string
theory.   Under the  orbifold  compactification, modular invariance of
the one-loop scattering  amplitudes   requires the theory to   contain
twisted  closed string states in addition  to the toroidally untwisted
closed string states.  In the $k$-th twisted closed string sector, the
worldsheet obeys the following monodromies:
\begin{equation} \label{2.6}
X_{\pm j} (\sigma + 2\pi, \tau) = \left \{ \begin{array}{l} X_{\pm j}
(\sigma,  \tau) \hfill (j=1,2)  \\  \exp(\pm  2  \pi ik  v_{j}) X_{\pm
j}(\sigma, \tau) \hfill \,\,\,\, (j=3,4,5)
\end{array} \right.  \end{equation} 
where  $k=0,1,2,\dots, N-1$ and  $k=0$  refers to the untwisted closed
string sector.  Superconformal  symmetry of the worldsheet implies the
following transformation rules  under   the orbifold action  for   the
fermionic partners:
\begin{equation} \label{2.7}
{\psi}_{\pm j} (\sigma + 2\pi + \tau) = \left \{
\begin{array}{l} {{\psi}_{\pm j}} (\sigma + \tau) \hfill (j=1,2) \\ 
\exp(\pm i 2 \pi k v_{j}) {\psi}_{\pm j}(\sigma + \tau) \hfill
\,\,\,\,\, (j=3,4,5).
\end{array} \right. 
\end{equation} 
with   similar  monodromies  for  the  right-moving $\tilde{\psi}_{j}$
fermions.  Based upon  the above monodromy  properties, we  can easily
obtain the  mode expansions, Hamiltonian  (in  SCFT),  as well as  the
modular  invariant partition functions  of  the closed string  sector.
For a more detailed discussion about  modular invariance of the closed
string sector we  refer   the reader  to \cite{Bai2}   and  references
therein.

Orbifold models of Type  IIB superstring theory cannot describe  gauge
interactions unless open strings are also included. As we have already
mentioned in Section I, this is equivalent to adding D-branes into the
theory.  The type and  number of Dp-branes  that can be added into the
theory is strongly  constrained  by the twisted  $\mathbf{RR}$ tadpole
cancellation conditions.  To our knowledge,  there are two fundamental
ways of  adding  Dp-branes into  orbifold  constructions  of Type  IIB
string theory.  The  first type of  orbifold construction involves the
introduction of  at  least two  different types of  D-branes.   Take a
D9-D5  brane   system  as    example.    D9-branes  embed   the   full
10-dimensional  spacetime and   their configuration is   automatically
symmetric   under the ${\mathbf    Z}_N$ orbifold action.  However  D5
branes, which  wrap the three  non-compact spatial real dimensions and
only $one$ of the compactified complex dimensions,  must be located at
the orbifold singularities  so  that they  exactly  embed one  of  the
compactified  complex planes  ($i=3,4,5$).    Different D5-branes  are
$either$ parallel $or$ perpendicular to each other in the compactified
sub-target-space.    In the  second   type  of  orbifold  construction
\cite{Blum1}  - \cite{Cvetic}, only one  type  of Dp-branes is allowed
(i.e.   D5-branes).   These D5-branes, which   generally are  extended
along some ``root vectors''  of  the Coxeter lattice, intersect   each
other  at nontrivial  angles  so  that  the total $T^6$  configuration
preserves  the desired    ${\mathbf  Z}_{N}$ symmetry.   Our work   is
restricted to orbifold theories of the first type.

Once  D-branes  are introduced, there   will   be open  string  states
stretched between them, describing the  matter fields and their  gauge
interactions.  The open strings obey  various boudary conditions which
determine the matter fields and gauge group. In the remaining parts of
this section,  we will write the  mode expansions of  the open strings
under the different boundary conditions.  For concreteness, we work in
the light-cone   gauge and  focus in   each  of the   possible complex
dimensions.

\subsection{Open string mode expansions}

\subsubsection {$j$-th noncompact dimension $(j=1,2)$}

Open strings satisfy Neumann (N) boundary conditions ($\partial_\sigma
X^{\pm j}(\tau , 0) =  \partial_\sigma X^{\pm j}(\tau ,  \pi) = 0$) in
the  directions parallel  to  the brane   but Dirichlet   (D) boundary
conditions ($X^{\pm  j}(\tau , 0)  = 0$ and  $ X^{\pm j}(\tau , \pi) =
Y^{\pm j}$) in the perpendicular directions.  Therefore, the endpoints
of  the  open strings  are  only  free   to  move along  the  parallel
directions to the brane.   We assume  that in  each of  the noncompact
dimensions,  open strings  satisfy Neumann   boundary conditions.  Its
momentum in these directions is  also continuous.  The mode expansions
for the $X^{\pm j}$ worldsheet coordinates are given by:
\begin{equation}
\label{ncb}
X^{\pm j}(\sigma, \tau)   = x^{\pm j}  +  2\alpha ' p^{\pm   j} \tau +
i\sqrt{2\alpha         '}{\sum_{n    \in    Z}}^{\prime}   \frac{1}{n}
{\alpha_{n}}^{\pm j} \exp(-i n \tau) \cos (n \sigma)\; ,
\end{equation}
where $\sum ^{\prime}$ excludes  the  $n =  0$ contribution.  For  the
open string $\alpha_0^{\pm  j}  =   \sqrt{2\alpha '}p^{\pm  j}$    and
$(\alpha_n^{\mu})^*     =    \alpha_{-n}^{\mu}$.   For the  worldsheet
fermions, the mode expansions are:
\begin{equation}
\label{ncr} 
\left\{ \begin{array}{l} {\psi}^{\pm  j}(\sigma,  \tau) = \sqrt{\alpha
'}\sum_{n   \in  Z} {  d_{n}}^{\pm j}  \exp[-i  n  (\tau + \sigma)] \\
{\tilde{\psi}}^{\pm j}(\sigma, \tau) = \sqrt{\alpha '}\sum_{n \in Z} {
d_{n}}^{\pm j} \exp[-i n (\tau - \sigma )]\; , \end{array} \right.
\end{equation}
in the ${\mathbf R}$ sector and
\begin{equation}
\label{ncns} 
\left\{  \begin{array}{l} {\psi}^{\pm  j}(\sigma, \tau) = \sqrt{\alpha
'}\sum_{r \in Z+1/2} {  b_{r}}^{\pm j} \exp[-i  r (\tau + \sigma) ] \\
{\tilde{\psi}}^{\pm j}(\sigma, \tau) =  - \sqrt{\alpha '}\sum_{r \in Z
+1/2} { b_{r}}^{\pm  j} \exp[-i r (\tau  -  \sigma )]\;  , \end{array}
\right.
\end{equation} 
in the ${\mathbf NS}$ sector.  After  quantization, the mode expansion
coefficients   are  interpreted as    the  creation and   annihilation
operators actiong on the string Hilbert space, and satisfy:
\begin{eqnarray}
\label{crnc} 
&[ x_{\pm j}, p_{\pm l} ] &=  - i \delta_{j1} \delta_{l1} \nonumber \\
& [  x_{\pm j}, p_{\mp l} ]  &= i \delta_{j2} \delta_{l2} \nonumber \\
&[{\alpha_{n}}^{\pm  j},  {\alpha_{m}}^{\mp l}    ]& =  n  \delta^{jl}
\delta_{n+m,0}  \nonumber \\ &\{  {b_r}^{\pm j}, {b_s}^{\mp  l} \} & =
\delta^{jl} \delta_{r+s,0}  \nonumber \\ &\{ {d_n}^{\pm j}, {d_m}^{\mp
l}\} & = \delta^{jl} \delta_{n+m,0}.
\end{eqnarray}
All   other commutators are   zero.   The  contributions to the  total
Hamiltonian,
\begin{equation}
\label{hamnc1} H = L_0 = H_0 + H_B + H_{NS-R}
\end{equation}
(where $H_{NS-R}=H_{NS}-H_R$) from the  dimensions obeying NN boundary
conditions, take the following expressions:
\begin{equation}
\label{hamnc0} H_0 (NN) = \alpha ' \sum_{j=1,2} p^jp_{-j}
\end{equation}
\begin{equation}
\label{hamncb} H_B (NN) = N_B (NN) - \frac{1}{12} = \sum_{n=1}^{\infty}
({\alpha_{-n}}^{-j}{\alpha_{n}}_{j}                                  +
{\alpha_{-n}}^{j}{\alpha_{n}}_{-j}) - \frac{1}{12}\;\;\;\;\;\;\; (j  =
2).
\end{equation}
In the above  expression, we have  considered that each integer modded
pair of complex bosons   contributes with $-\frac{1}{12}$ towards  the
zero-point energy.  In the  light-cone gauge, the physical  vibrations
are  those that are transverse  to the worldsheet, in $p-2$ dimensions
($j \not=  1$). By  $N_B(NN)$ we  mean  the contribution to  the total
bosonic  number operator from   the dimensions  obeying  $NN$ boundary
conditions.  The Ramond sector contribute:
\begin{equation}
\label{hamncr} H_{R} (NN) = N_F^{R}(NN) + \frac{1}{12} =
\sum_{n=1}^{\infty}n   ({d_{-n}}^{-j}{   d_{n}}_{j}      +{d_{-n}}^{j}
{ d_{n}}_{-j}) + \frac{1}{12}\;\;\;\;\;\;\; (j = 2)
\end{equation}
where each integer  modded pair of  complex fermions  contributes with
$+\frac{1}{12}$ towards the zero-point energy. In the NS sector:
\begin{equation}
\label{hamncns} H_{NS} (NN) = N_F^{NS}(NN) - \frac{1}{24} = 
\sum_{r=1/2}^{\infty}r     ({b_{-r}}^{-j}{     b_{r}}_{j}+{b_{-r}}^{j}
{ b_{r}}_{-j}) - \frac{1}{24}\;\;\;\; (j = 2)
\end{equation}
where  each  half-integer    complex  pair   of  worldsheet   fermions
contributes with $-\frac{1}{24}$ towards  the total zero-point energy.
The  $\mathbf{NS}$   sector  Hilbert  space   is  constructed   from a
non-degenerate ground state ${\vert 0 \rangle}_{NS}$, satisfying:
\begin{equation}
\label{hnsnc} \alpha_{r}^{\pm j} {\vert 0 \rangle}_{NS} =
b_{r}^{\pm j} {\vert 0 \rangle}_{NS} =0 \,\,\,\,\,\,  ( r \in Z + 1/2,
\; r>0)
\end{equation}
and contributes with $-\frac{1}{8}$  to the zero-point energy.  On the
other  hand, the $\mathbf{R}$    ground states  ${\vert  \vec{  s_{j}}
\rangle}_{R}$ are massless but degenerate, forming a spacetime spinor:
\begin{equation}
\label{hrnc} {\vert \vec{s_{j}} \rangle}_{R} \equiv {\vert \pm 
\frac{1}{2} \rangle}_{R}.
\end{equation} 
The   contribution from the ${\mathbf  NS}$  sector  to the total mass
spectrum \cite{Pol1} $\alpha^{\prime}M^2$ is given by:
\begin{equation}
\label{mncns}
-\frac{1}{8}+N_B(NN)+N_F^{NS}(NN)
\end{equation}
and from the ${\mathbf R}$ sector:
\begin{equation}
\label{mncr}
N_B(NN)+N_F^{R}(NN).
\end{equation} 

\subsubsection{$j$-th compact complex dimension with ${\mathbf{NN}}$
boundaries $(j=3,4,5)$}

If the  string has NN  boundary  conditions along some  of the compact
directions, both ends of the open strings are free to move along these
directions. Therefore, there are no winding  modes associated to these
dimensions because the open string  can  continuously wrap and  unwrap
the dimensions.  On the other  hand,  the momentum is quantized  along
those directions because the open  string cannot transfer longitudinal
momentum to the D-brane:
\begin{equation} 
p_{\pm j} = \frac{n_{\pm j}}{R_{\mp j}}
\end{equation}
with   $n_{\pm j}$    being   an integer   and    $R_{\pm   j}  \equiv
\sum_{i=4}^{9}R_{i}\vec{e}_{\pm      j}\cdot   \vec{\alpha}_{i}$. When
considering toroidal  compactification  for  open strings,   the  mode
expansions for the worldsheet coordinates are:
\begin{eqnarray}
\label{nnb} X^{\pm j}(\sigma, \tau) &=& x^{\pm j} + 
2\alpha  '\tau p_{\pm   j}     +  i\sqrt{2\alpha  '}    {\sum_{n   \in
Z}}^{\prime}\frac{1}{n}
 {\alpha_{n}}^{\pm j} \exp(-i n \tau) \cos (n
\sigma).
\end{eqnarray} 
The worldsheet fermions have exactly the same mode expansions as those
in  the   noncompact  complex  dimensions   (\ref{ncr}), (\ref{ncns}).
Moreover,  both cases share  the  same  commutation relations  for the
oscillator ladder  operators (\ref{crnc}).  The ${\mathbf NS}$ ${\vert
0 \rangle}_{NS}$ and  ${\mathbf R}$ ${\vert \vec{ s_{j}} \rangle}_{R}$
ground  states are  also defined  by  (\ref{hnsnc}) and  (\ref{hrnc}).
Therefore, both  cases  contribute in the  same way  to the total mass
spectra  (\ref{mncns})\,(\ref{mncr}),  bearing in  mind  that  now the
momentum  is quantized.  However, in  the  current case, the operators
$x_{\pm j}$ commute   with everything,   including themselves.    This
property is  used  to construct the correct   Hilbert space.  Once  we
include the orbifold projection operator into our theory, the twisting
$\theta ^k$ changes   the  boundary conditions  with  respect to   the
toroidal  compactification.  Now,  the  momentum  modes vanish  in the
presence of the orbifold action unless it acts trivially ($\theta ^k =
1$).  The twisted  sector is trapped at  the fixed point and does  not
feel the shape of the compact dimension.

\subsubsection{$j$-th compact complex dimension with ${\mathbf{DD}}$
boundaries $(j=3,4,5)$}

The brane configuration consists   of two Dp-branes (or  Dp-Dq branes)
with  $p<9$ (or $p$, $q<9$),   both being perpendicular  to the $j$-th
complex plane. Strings with DD boundary conditions stretch between the
D-branes and in the toroidal compactification they are compatible with
the existence of a winding number $\omega$, since they are attached to
the D-brane and cannot unwrap from these compact dimensions.  Denoting
the distance between the  branes in this  complex direction as $Y^{\pm
j}$, we have $X^{\pm j}(\pi, \tau) - X^{\pm j}(0, \tau)  = Y^{\pm j} +
2\pi \omega^{\pm j}R^{\pm j} $  (where $\omega^{\pm j}  \in Z$ are the
string winding numbers along these directions) and the mode expansions
for  an open string winding $\omega^{\pm  j}$ times around each of the
$X^{\pm j}$ compact   dimensions (assuming without loss  of generality
that one of the branes contains the origin) read:
\begin{equation}
\label{ddb} 
X^{\pm j}(\sigma, \tau)   = (Y^{\pm j} +  2\pi\omega^{\pm  j}R^{\pm
j})\frac{\sigma}{\pi}  -  i\sqrt{2\alpha  '}{\sum_{n  \in Z}}^{\prime}
\frac{1}{n} {\alpha_{n}}^{\pm j} \exp(-i n \tau) \sin (n \sigma).
\end{equation}
As before,  once the orbifold projection  operator is  introduced into
our theory, the boundary conditions differ  from those of the toroidal
compactification.  Only for  the  planes  in  which $\theta ^k$   acts
trivially are winding  modes allowed in  the compact directions.   The
mode expansions for the worlsheet fermions are:
\begin{equation} 
\label{cddr}
\left\{  \begin{array}{l} {\psi}^{\pm  j}(\sigma, \tau) = \sqrt{\alpha
'}\sum_{n \in Z}   { d_{n}}^{\pm j} \exp[-i   n  (\tau +  \sigma)]  \\
{\tilde{\psi}}^{\pm j}(\sigma, \tau) =  -\sqrt{\alpha '}\sum_{n \in Z}
{ d_{n}}^{\pm j} \exp[-i n (\tau - \sigma )], \end{array} \right.
\end{equation}
in the ${\mathbf R}$ sector and
\begin{equation}
\label{cddns} 
\left\{ \begin{array}{l} {\psi}^{\pm  j}(\sigma, \tau) =  \sqrt{\alpha
'} \sum_{r \in Z+1/2} { b_{r}}^{\pm j} \exp[-i r  (\tau + \sigma) ] \\
{\tilde{\psi}}^{\pm j}(\sigma, \tau)  = -\sqrt{\alpha '}\sum_{r  \in Z
+1/2} { b_{r}}^{\pm j} \exp[-i r (\tau - \sigma )],
\end{array} \right.  
\end{equation} 
in the ${\mathbf NS}$ sector.   The commutation relations (\ref{crnc})
and    the worldsheet  vacuum   states (\ref{hnsnc})\,(\ref{hrnc}) are
defined in  the  same way as before.   Following  the same analysis as
before, the different contributions to the Hamiltonian read:
\begin{equation}
\label{hamdd0} H_0 (DD) = \alpha '
\sum_{j}\left(\frac{2\pi\omega^{+j}R^{+j}    +      Y^{+j}}{2\pi\alpha
'}\right)\left(\frac{2\pi\omega_{-j}R_{-j}                           +
Y_{-j}}{2\pi\alpha'}\right),
\end{equation}
where the term proportional to the square of  the distance between the
branes is due to the stretching energy of the string.
\begin{equation}
\label{hamddb} H_B (DD) = \sum_j \left(N_B(DD)-\frac{1}{12}\right)=
\sum_j \left(\sum_{n=1}^{\infty} ({\alpha_{-n}}^{-j}{\alpha_{n}}_{j} +
{\alpha_{-n}}^{j}{\alpha_{n}}_{-j}) - \frac{1}{12}\right).
\end{equation}
In the light-cone  gauge, the physical vibrations  are  those that are
transverse to the world sheet, in $p-2$ dimensions ($j \not= 1$).
\begin{equation}
\label{hamddr} H_{R} (DD) = \sum_j \left(N_F^{R}(DD)+\frac{1}{12}\right)=
\sum_j   \left(\sum_{n=1}^{\infty}n     ({d_{-n}}^{-j}{     d_{n}}_{j}
+{d_{-n}}^{j}{ d_{n}}_{-j}) + \frac{1}{12}\right)
\end{equation}
\begin{equation}
\label{hamddns} H_{NS} (DD) = \sum_j \left(N_F^{NS}(DD)-\frac{1}{24}\right)=
\sum_j          \left(\sum_{r=1/2}^{\infty}r            ({b_{-r}}^{-j}
{ b_{r}}_{j}+{b_{-r}}^{j}{ b_{r}}_{-j}) - \frac{1}{24}\right)
\end{equation}
The  contributions to the total mass  spectrum $\alpha 'M^2$ from each
of the complex dimensions are:
\begin{equation}
\label{mddns}
\left(\frac{2\pi\omega^{+j}R^{+j}         +         Y^{+j}}{2\pi\alpha
'}\right)\left(\frac{2\pi\omega_{-j}               R_{-j}            +
Y_{-j}}{2\pi\alpha'}\right)-\frac{1}{8} + +N_B(DD)+N_F^{NS}(DD),
\end{equation}
in the ${\mathbf NS}$ sector and:
\begin{equation}
\label{mddr}
\left(\frac{2\pi\omega^{+j}     R^{+j}       +      Y^{+j}}{2\pi\alpha
'}\right)\left(\frac{2\pi\omega_{-j}            R_{-j}               +
Y_{-j}}{2\pi\alpha'}\right) + N_B(DD)+N_F^{R}(DD),
\end{equation}
in the ${\mathbf R}$ sector.

\subsubsection{ $j$-th compact complex dimension with ${\mathbf{DN
(ND)}}$ boundaries $(j=3,4,5)$}

If the string has  mixed boundary conditions,  the mode expansions for
the $X^{\pm j}$  worldsheet bosonic degrees  of freedom are compatible
with neither the presence of quantized momenta nor winding numbers:
\begin{equation}
\label{dnb} 
X^{\pm j}(\sigma, \tau) = Y^{\pm j}  - i\sqrt{2\alpha '}{\sum_{r \in Z
+  \frac{1}{2}}}\frac{1}{r} {\alpha_{r}}^{\pm j}  \exp(-i r \tau) \sin
(r \sigma).
\end{equation}
The worldsheet fermionic partners obey the following mode expansions:
\begin{equation}
\label{cndr}
\left\{  \begin{array}{l} {\psi}^{\pm j}(\sigma,  \tau) = \sqrt{\alpha
'}\sum_{r \in  Z+\frac{1}{2}}  {  d_{r}}^{\pm  j}  \exp[-i r   (\tau +
\sigma)] \\  {\tilde{\psi}}^{\pm  j}(\sigma,   \tau)  =  -\sqrt{\alpha
'}\sum_{r \in Z+\frac{1}{2}} { d_{r}}^{\pm j} \exp[-i r (\tau - \sigma
)],
\end{array} \right. 
\end{equation} 
in the ${\mathbf R}$ sector and
\begin{equation}
\label{cndns} 
\left\{ \begin{array}{l} {\psi}^{\pm  j}(\sigma, \tau)  = \sqrt{\alpha
'}\sum_{n  \in Z} {  b_{n}}^{\pm j}  \exp[-i n (\tau   + \sigma) ]  \\
{\tilde{\psi}}^{\pm j}(\sigma, \tau) = - \sqrt{\alpha '}\sum_{n \in Z}
{ b_{n}}^{\pm j} \exp[-i n (\tau - \sigma )], \end{array} \right.
\end{equation} 
in the   ${\mathbf NS}$ sector of  the  theory.   The ladder operators
satisfy the following non-trivial commutation relations:
\begin{eqnarray} &
[   {\alpha_r}^{\pm   j},  {\alpha_s}^{\mp   l}  ]    &=   \delta^{jl}
\delta_{r+s,0} \nonumber   \\  & \{  d_r^{\pm  j}, d_s^{\mp  l}  \} &=
\delta^{jl} \delta_{r+s} \nonumber \\ &\{ {b_n}^{\pm j}, {b_m}^{\mp l}
\} & = \delta^{jl} \delta_{n+m,0}. \nonumber
\end{eqnarray}
The various contributions to the total Hamiltonian read:
\begin{equation}
\label{hamdn0} H_0 (ND) = 0,
\end{equation}
\begin{equation}
\label{hamdnb} H_B (ND) = \sum_j\left(N_B(ND)+\frac{1}{24}\right)=
\sum_j \left(\sum_{r=1/2}^{\infty} ({\alpha_{-r}}^{-j}{\alpha_{r}}_{j}
+ {\alpha_{-r}}^{j}{\alpha_{r}}_{-j}) + \frac{1}{24}\right),
\end{equation}
\begin{equation}
\label{hamdnr} H_{R}(ND) = \sum_j\left(N_F^{R}(ND)-\frac{1}{24}\right)=
\sum_j  \left(\sum_{r=1/2}^{\infty}r    ({d_{-r}}^{-j}{     d_{r}}_{j}
+{d_{-r}}^{j}{ d_{r}}_{-j}) - \frac{1}{24}\right),
\end{equation}
\begin{equation}
\label{hamdnns} H_{NS} (ND) = \sum_j\left(N_F^{NS}(ND)+\frac{1}{12}\right)=
\sum_j              \left(\sum_{n=1}^{\infty}r          ({b_{-n}}^{-j}
{ b_{n}}_{j}+{b_{-n}}^{j}{ b_{n}}_{-j}) + \frac{1}{12}\right).
\end{equation}
The $\mathbf{NS}$ ground state becomes massive (with zero-point energy
$\frac{1}{8}$) and degenerate.  It  is therefore described by a spinor
in this sub-target-space:
\begin{equation}
\label{hnsnd} {\vert \vec{s_{j}} \rangle}_{NS} \equiv {\vert \pm 
\frac{1}{2} \rangle}_{NS}.
\end{equation} 
The $\mathbf{R}$  ground state (in $j$-th complex  plane  ), ${\vert 0
\rangle}_{R}$, remains  massless  but is non-degenerate under  DN (ND)
boundaries. It obeys,
\begin{equation}
\label{hrnd} 
\alpha_{r}^{\pm  j}  {\vert 0  \rangle}_{R} =  b_{r}^{\pm  j} {\vert 0
\rangle}_{R} =0 \,\,\,\,\,\, ( r \in Z + 1/2, \; r>0).
\end{equation}
The total contribution to the mass spectrum $\alpha 'M^2$ from each of
the complex dimensions is:
\begin{equation} 
\label{mndns}
\frac{1}{8} + N_B(ND)+N_F^{NS}(ND),
\end{equation}
in the ${\mathbf NS}$ sector and:
\begin{equation} 
\label{mndr}
 N_B(ND)+N_F^{R}(ND),
\end{equation} 
in the ${\mathbf R}$ sector.

\subsection{Partition Functions}

In the orbifold theory under consideration we only need to include the
cylinder amplitudes $\mathcal C$    in order to compute  the   tadpole
cancellation conditions.   The  general expression   for the  cylinder
amplitudes is given by:
\begin{equation} 
\label{ca1} 
C_{pq}        =      \frac{1}{2N}    \sum_{k=0}^{N              -   1}
\int_{0}^{\infty}\frac{dt}{2t}Tr_{pq}[q^{H}(1 + (-1)^F)\theta^k]
\end{equation}
where $q=e^{-2\pi t}$ and $t$ is the cylinder modulus, the proper time
in the open string channel.  The coefficient $\frac{1}{2N}$ comes from
the $\mathsf{GSO}$ and  ${\mathbf Z}_N$ orbifold projection operators.
The  subscript ``$pq$'' means that the  amplitude  is evaluated in the
1-loop open string picture in  which the open string has  one end on a
Dp-brane and the other on a Dq-brane.  For convenience, we rewrite the
amplitude as
\begin{equation} \label{ca2} C_{pq} = \frac{1}{4N} \sum_{k=0}^{N
- 1} \int_{0}^{\infty}\frac{dt}{t}Z_{pq}
\end{equation} 
where the trace    $Z_{pq}$   is referred to as     the $\it{Partition
\,\,function}$:
\begin{equation} \label{pf}
Z_{pq}=Tr_{pq}[(1 + (-1)^F)\theta^ke^{-2\pi t H}].
\end{equation}
The  $\mathsf{GSO}$ projection   operator  remains the  same  for  the
amplitude   of interaction  between   two  anti-branes but  should  be
modified into $\frac{1 - (-1)^F}{2}$ when the interaction is between a
Dp-brane and an anti-Dq-brane (D$\bar q$).  This results from the fact
that the interaction between two  D-branes and a brane-antibrane  pair
have the same  sign for the $\mathbf{NSNS}$  sector but  opposite sign
for the $\mathbf{RR}$ sector \cite{Asen}\cite{Al4}.  In the light-cone
gauge, if an open string obeys $2n$  ($n \leq 4 $)  DD and NN boundary
conditions, it   will obey $8-2n$  mixed (ND-DN)  boundary conditions.
According to this, we assumed the following formal expressions for the
fermion number operators in the ${\mathbf{NS}}$:
\begin{equation} \label{fons}
(-1)^F  =  (-1)^{1 +  \sum_{j=2}^{n}  \sum_{r>0} (b^{-j}_{-r}b_{r,j} +
b^{j}_{-r}b_{r,-j})  + \sum_{j=n+1}^{5} \sum_{n>0} (b^{-j}_{-n}b_{n,j}
+    b^{j}_{-n}b_{n,-j})}   \cdot     \!\!\!         \prod_{k=n+1}^{5}
(b_{0}^{k}b_{0}^{-k} - b_{0}^{-k}b_{0}^{k} )
\end{equation}
and ${\mathbf R}$ sectors,
\begin{equation} \label{for}
(-1)^F =   (-1)^{  \sum_{j=2}^{n} \sum_{n>0}    (d^{-j}_{-n}d_{n,j}  +
d^{j}_{-n}d_{n,-j}) + \sum_{j=n+1}^{5} \sum_{r>0}  (d^{-j}_{-r}b_{r,j}
+  b^{j}_{-r}b_{r,-j})} \cdot  \prod_{k=2}^{n}  (d_{0}^{k}d_{0}^{-k} -
d_{0}^{-k}d_{0}^{k} )
\end{equation}
respectively.  According to the above definitions, both operators obey
the  right    anticommutation relations  and    satisfy  unitarity and
$(-1)^{2F}=1$.   Furthermore, the ${\bf    NS}$  vacuum is   odd under
$(-1)^{2F}$ when the number of  complex dimensions equals $n=2$ and in
the  ${\mathbf R}$ sector  it allows us   to construct Weyl spinors in
even-dimensional   space-times.   The   corresponding   $\mathsf{GSO}$
projection  operators  $P^{\mathsf GSO}_{\pm}=\frac{1 \pm  (-1)^F}{2}$
are hermitian, as expected.

\subsection{Open string Hilbert space}

The full open  string Hilbert space  can be constructed as  the direct
product of the sub-Hilbert spaces corresponding to each of the complex
degrees  of freedom.  In  the light-cone gauge,   we can eliminate the
oscillator variables for  the $j =  1$ complex plane, reducing to four
$(j = 2, 3, 4, 5)$  the number of complex  degrees of freedom.  Let us
take as examples for  the construction of the  full Hilbert space  the
${\bf 99}$  and ${\bf   95_3}$ sectors, where  by   $5_3$ we denote  a
D5-brane embedding  the $j=3$ complex plane in  addition to  the usual
three dimensional non-compact space.

The ${\bf 99}$ sector consists  of open strings stretched between two,
not  necessarily  different, D9-branes.   All  complex planes  obey NN
boundary conditions and as a consequence  the ${\mathbf NS}$ sector of
each of the complex planes  is half-integer moded (\ref{ncns}) and its
full ground  state   (\ref{hnsnc}) is  tachyonic,  has   a  total mass
(\ref{mncns})  of  $\alpha    'M^2   = -1/2$ and     is non-degenerate
(\ref{hnsnc}). It obeys:
\begin{equation}
\label{99ns} \alpha_{n}^{\pm j} {\vert 0 \rangle}_{NS} =
b_{r}^{\pm j} {\vert 0 \rangle}_{NS} =0 \,\,\,\,\,\, (j=2,\ldots,5; \,
n \in Z, \; n>0 \; {\rm and}\; r \in Z + 1/2, \; r>0),
\end{equation}
with fermion number:
\begin{equation} \label{fn} \exp
( i \pi F ) {\vert 0 \rangle}_{NS} \equiv - {\vert 0 \rangle}_{NS}
\end{equation} 
where the minus sign corresponds  to the  contribution from the  ghost
ground state.  This  tachyonic state can be  removed from the spectrum
via the following $\mathsf{GSO}$ projection:
\begin{equation} 
\label{gso} 
\exp{(i \pi F)}  {\vert \textrm{physical}  \rangle  }_{NS} =  {  \vert
\textrm{physical} \rangle }_{NS}.
\end{equation}
In   the  ${\mathbf  R}$ sector,    each   of the  complex planes   is
integer-moded (\ref{ncr})  and its  ground state massless (\ref{mncr})
and degenerate (\ref{hrnc}):
\begin{equation}
\label{99r} 
{\vert \vec{\bf  s}  \rangle}_{R} \equiv {\vert   s_{2}, s_{3}, s_{4},
s_{5} \rangle}_{R}  \,\,\,\,\,\,\,\,\, (s_{a}=\pm\frac{1}{2}; \, a= 2,
3, 4, 5).
\end{equation} 
The $\mathsf{GSO}$ projection in the Ramond  sector can be implemented
either by
\begin{equation} 
\label{gso1} \sum_{a} s_{a} = 0
\qquad (\,\textrm{mod} \, 2\,)
\end{equation}
or
\begin{equation} 
\label{gso2} \sum_{a}s_{a} = 1
\qquad (\,\textrm{mod} \, 2\,).
\end{equation}
The relevant Hilbert  space  is then constructed  by  acting  with the
bosonic and fermionic oscillator creation  operators on the  ${\mathbf
NS}$ and  ${\mathbf R}$ ground states.   When  considering the ${\bf 9
\bar{9}}$ sector, the sign flip in  the $\mathsf{GSO}$ projection does
not project out the  $\mathbf{NS}$  sector ground state and  tachyonic
excitations with $\alpha ' M^2 = -1/2$ remain in the spectrum, whereas
the  would-be massless states are   projected  out.  The ${\bf  95_3}$
sector consists of open strings stretched between one D9-brane and one
D$5_3$-brane.  The complex planes $j = 2,3$ are subject to NN boundary
conditions.  The rest ($j = 4,5$)  obey mixed boundary conditions.  As
before,  the total Hilbert space  is  constructed  by acting with  the
oscillator creation  operators  on the  ${\mathbf  NS}$ (\ref{cndns}),
(\ref{hnsnd}):
\begin{equation}
\label{95ns} {\vert \vec{\bf s}
\rangle}\equiv  {\vert  s_{4}, s_{5}  \rangle}_{NS} \,\,\,\,\,\,\,\,\,
(s_{a}=\pm\frac{1}{2}; \, a= 4,5)
\end{equation}
and ${\mathbf R}$ (\ref{ncr}), (\ref{hrnc}):
\begin{equation}
\label{95r} {\vert \vec{\bf s} \rangle}_{R} \equiv {\vert
s_{2}, s_{3} \rangle}_{R} \,\,\,\,\,\,\,\,\, (s_{a}=\pm\frac{1}{2}; \,
a= 2, 3)
\end{equation} 
massless ground states.  For the  ${\bf 9 \bar{5}_3}$ or ${\bf \bar{9}
{5}_3}$ sectors, the ${\mathbf{NS}}$ ground  states are also massless.
The  fact that the  endpoints of the  open strings are distinguishable
makes  it natural  for   them to carry   extra  degrees of  freedom in
addition to the fields propagating in the  bulk.  It is allowed by all
the symmetries of the theory to  add at each  endpoint of the string a
new but non-dynamical  quantum degree  of  freedom, known as   {\bf\sl
Chan-Paton} degrees of  freedom.   These new non-dynamical  degrees of
freedom have a major effect  on the space-time physics despite obeying
trivial worldsheet  dynamics.   In consistent string  theories,  these
quantum numbers are actually gauge  quantum numbers.  We may label the
open  string  states by  $(\lambda^{M}_{pq})_{ab}|\Psi  ,  ab>$, where
$\Psi$ refers to the  worldsheet degrees of freedom,  ($p, q$)  to the
type  of brane the   string endpoints  are attached to  ($\mathbf{pq}$
sector) and  ($a,    b$) are  the Chan-Paton   indices   labelling the
particular branes of the  stack of Dp or  Dq branes respectively.  The
superindex  $M$ varies depending  upon   the matter being  considered:
gauge bosons, fermions or matter scalars.

Massive string states  have   masses  of the order    of $M_{String}$,
usually far  heavier  than all  the particles  of  the Standard Model.
Thus,  only   massless   string   states  are  interesting    from the
phenomenological point   of  view,  acquiring  small   masses  through
symmetry  breaking  effects.  Massless open  string  states arise from
open strings with  zero length (coincident D-branes), otherwise, there
would be a  contribution to the mass term  coming from  the tension of
the   stretched  string.   If  our   theory  contains $n$   coincident
Dp-branes, each endpoint of the  string can be in  one of $n$  states.
The set of $n^2 \;  (\lambda_{pp})$ Hermitian matrices form a complete
set of states for the two endpoints.  They are known as the Chan-Paton
matrices (or   wavefuctions) and they are   generators  of {\bf U(n)},
describing the gauge interactions.  The theory  of placing D-branes on
top of   orbifold  singularities  is obtained by   keeping  the states
invariant  under  the  combined  geometrical and  Chan-Paton  orbifold
action.   The geometrical  orbifold  action  acts  on the   worldsheet
degrees of freedom while the  Chan-Paton action acts on the Chan-Paton
degrees of freedom.  In general:
\begin{equation} \label{oa} 
\theta      ^k                     (\lambda^M_{pq})_{ab}|\Psi        ,
ab>=({\gamma}_{k,p})_{aa'}(\lambda^M_{pq})_{a'b'}|\theta  ^k   \Psi  ,
a'b'>   ({\gamma}_{k,q}^{-1})_{b'b}    =    \exp(2   \pi  i     c_{M})
({\gamma}_{k,p})_{aa'}(\lambda^M_{pq})_{a'b'}|    \Psi   ,       a'b'>
({\gamma}_{k,q}^{-1})_{b'b}
\end{equation}
where $c_M$ depends  on the type  of matter we  are  considering.  The
projection for the $\lambda^M$ matrices reads:
\begin{equation} \label{proj} 
\lambda^M  =  \exp(2      \pi i  c_{M})    ({\gamma}_{k,p})  \lambda^M
({\gamma}_{k,q}^{-1}).
\end{equation}
The    gamma   matrices   $\gamma  _{k,p}   =    \gamma_{\theta ^k,p}$
($k=0,1,\ldots,N-1$), represent the  embedding of  the ${\mathbf Z}_N$
orbifold point   group actions on the   Chan-Paton degrees of freedom.
These  matrices should form a   unitary, not necessarily  irreducible,
representation of the orbifold group $\bf{Z}_{N}$. Without any loss of
generality, they can be defined as:
\begin{equation} 
\label{g1} 
\gamma_{1,p} = diag \left( I_{n_{0}^{(p)}}, \, \alpha I_{n_{1}^{(p)}},
\ldots, \alpha^jI_{n_{j}^{(p)}}, \ldots, \alpha^{N-1}I_{n_{N-1}^{(p)}}
\right)
\end{equation}
where $\alpha=e^{\frac{2\pi i}{N}}$, $I_{n_i}$ is the $n_i \times n_i$
identity matrix and $\sum_i n_i = n$, the total number of D$p$-branes.
Similarly, for the antibranes,
\begin{equation} 
\label{g3} 
\gamma_{1, \bar{r}} =   diag \left(  I_{m_{0}^{(\bar{r})}}, \,  \alpha
I_{m_{1}^{(\bar{r})}}, \ldots, \alpha^j I_{m_{j}^{(\bar{r})}}, \ldots,
\alpha^{N-1} I_{m_{N-1}^{(\bar{r})}} \right).
\end{equation}
Gauge bosons in consistent  interacting theories must always transform
in the adjoint representation of the gauge group. For {\bf U(n)} gauge
theories, if the endpoints of the string run over the {\bf n} and {\bf
\= n} representations of {\bf  U(n)}, this is automatically satisfied.
Massless gauge    bosons correspond  to   open string  states   in the
${\mathbf    NS}$    sector,    of   the   form  $\lambda^{G}_{pq}   b
^{\mu}_{-1/2}|0, pq>$, with $\mu $ running along the usual non-compact
space time coordinates.   Their projection (\ref{proj})  is then given
by:
\begin{equation} 
\label{gbp1} 
\lambda^{G} = ({\gamma}_{k,p}) \lambda^{G} ({\gamma}_{k,q}^{-1}).
\end{equation}
Using  the expression for  the  $\gamma_{1,p}$ matrices (\ref{g1}), we
get:
\begin{equation}
\label{gbp2} 
(\lambda^{G}_{pq})_{ab}                                              =
({\gamma}_{k,p})_{aa'}(\lambda^{G}_{pq})_{a'b'}
({\gamma}_{k,q}^{-1})_{b'b}   =  exp(\frac{2\pi ia}{N})exp(\frac{-2\pi
ib}{N})(\lambda^{G}_{pq})_{ab},
\end{equation}
where we have used that the $\gamma$ matrices are diagonal.  The above
projection (\ref{gbp2})   is  only satisfied  if  $a=b$,  breaking the
original $U(n)$ and $U(m)$ gauge groups into:
\begin{equation} 
\label{gg1} 
{\bigotimes_{j=0}^{N-1}}    U(n_{j})  \;\;\;\;  {\rm  and}  \;\;  \;\;
{\bigotimes_{j=0}^{N-1}} U(m_{j})
\end{equation}
respectively.   Fermions are described  by  open string states in  the
${\mathbf    R}$ sector, of    the   form $\lambda  {\vert \vec{s_{j}}
\rangle}_{R}$,   with  $s_j  = \pm    1/2$, the weights   of  a spinor
representation  of SO(8).  Before the  orbifold projection,  we have a
${\mathcal  N}  =  4$ supersymmetric  $U(n)$   gauge theory  with four
adjoint fermions  transforming  in the   ${\bf  4}$ of   $SU(4)$.  The
orbifold projection is then given by:
\begin{equation} 
\label{fp} 
\lambda^{F} =   exp(2\pi  ia_j \cdot  s_j)({\gamma}_{k,p}) \lambda^{F}
({\gamma}_{k,q}^{-1})
\end{equation}
where $a_j = a_2,  a_3, a_4, a_5$ defines the  orbifold action  on the
fermions with $a_2+a_3+a_4+a_5 = 0 \; mod \, N$. Using (\ref{g1}):
\begin{equation}
\label{fp2} 
\lambda^{F} = exp(2\pi i(a_j \cdot s_j + \frac{a-b}{N}))\lambda^{F}
\end{equation}
and chiral  fermions transform in bifundamental representations (${\bf
n_{j}}$, ${\bf  \bar  {u}_{j+Na_j\cdot s_j}})$.   When $a_2 =  0$, the
$Z_N$ orbifold action belongs to $SU(3)$, preserving the supersymmetry
of  the closed string sector.   Massless complex scalars in space-time
belong to the ${\mathbf  NS}$  sector of  the theory and  are obtained
from the states $\lambda^S \Psi ^{j}_{-1/2}|0, pq>$, where $j = 3,4,5$
labels the  three  orbifold complex planes.   Their projection  can be
deduced from the orbifold action on the fermions:
\begin{equation} 
\label{sp} 
\lambda^{S}   = exp(2\pi  iv_j\cdot s_j)({\gamma}_{k,p})   \lambda^{S}
({\gamma}_{k,q}^{-1}),
\end{equation}
where  $v_j   = (v_3,  v_4, v_5)$ is   the  orbifold twist vector with
$v_3=a_4+a_5,  v_4=a_3+a_5, v_5=a_3+a_4$.      Supersymmetry   imposes
$v_3+v_4+v_5=0 \; mod \, N$, so $a_j=-v_j$.  For the massless spectrum
of open strings stretched between two  antibranes or one brane and one
antibrane, we obtain analogous results.  Tachyons are scalar states in
the $\mathbf{NS}$   sector of the  form  $\lambda^{t}_{pq}|0, pq>$ and
obey the following projection:
\begin{equation}
\label{tp} 
\lambda^{t} = ({\gamma}_{k,p}) \lambda^{t}({\gamma}_{k,q}^{-1}).
\end{equation}
In general, the matter content is as shown in the following table:
\vspace{0.1cm}
\begin{center}
\begin{tabular}{|c|c|c|c|c|}
\hline  \hline Sector   & Gauge  bosons  & Tachyonic  scalar fields  &
Massless scalar fields & Massless fermions \\\hline \hline $\bf{pp}$ &
${\bigotimes_{j=0}^{N-1}}   U(n_{j}^{(p)})$   & $\sum_{j=0}^{N-1}(n_j,
\bar{n}_{j})$   &  $\sum_{j=0}^{N-1}(n_j, \bar{n}_{j+Nv\cdot  s}) $  &
$\sum_{j=0}^{N-1}(n_j,        \bar{n}_{j+Na\cdot      s})    $      \\
$\bf{\bar{p}\bar{p}}$           &            ${\bigotimes_{j=0}^{N-1}}
U(m_{j}^{(\bar{p})})$    &$\sum_{j=0}^{N-1}(m_j,    \bar{m}_{j})$    &
$\sum_{j=0}^{N-1}(m_j,      \bar{m}_{j+Nv\cdot      s})       $      &
$\sum_{j=0}^{N-1}(m_j, \bar{m}_{j+Na\cdot s}) $ \\ $\bf{p\bar{p}}$ & &
$\sum_{j=0}^{N-1}(n_j,     \bar{m}_{j})$   &    $\sum_{j=0}^{N-1}(n_j,
\bar{m}_{j+Nv\cdot  s}) $ & $\sum_{j=0}^{N-1}(n_j,  \bar{m}_{j+Na\cdot
s})  $ \\ $\bf{\bar{p}p}$ &  &  $\sum_{j=0}^{N-1}(m_j, \bar{n}_{j})$ &
$\sum_{j=0}^{N-1}(m_j,     \bar{n}_{j+Nv\cdot        s})       $     &
$\sum_{j=0}^{N-1}(m_j, \bar{n}_{j+Na\cdot    s}) $\\    $\bf{qq}$    &
${\bigotimes_{j=0}^{N-1}}     U(u_{j}^{(q)})$ & $\sum_{j=0}^{N-1}(u_j,
\bar{u}_{j})$&$\sum_{j=0}^{N-1}(u_j,   \bar{u}_{j+Nv\cdot     s})    $
&$\sum_{j=0}^{N-1}(u_j,       \bar{u}_{j+Na\cdot        s})       $ \\
$\bf{\bar{q}\bar{q}}$           &            ${\bigotimes_{j=0}^{N-1}}
U(w_{j}^{(\bar{q})})$        &$\sum_{j=0}^{N-1}(w_j,     \bar{w}_{j})$
&$\sum_{j=0}^{N-1}(w_j,                \bar{w}_{j+Nv\cdot          s})
$&$\sum_{j=0}^{N-1}(w_j, \bar{w}_{j+Na\cdot  s}) $  \\ $\bf{q\bar{q}}$
&&  $\sum_{j=0}^{N-1}(u_j,  \bar{w}_{j})$   &   $\sum_{j=0}^{N-1}(u_j,
\bar{w}_{j+Nv\cdot s})  $&   $\sum_{j=0}^{N-1}(u_j, \bar{w}_{j+Na\cdot
s})   $ \\ $\bf{\bar{q}q}$   && $\sum_{j=0}^{N-1}(w_j, \bar{u}_{j})$ &
$\sum_{j=0}^{N-1}(w_j,         \bar{u}_{j+Nv\cdot          s})      $&
$\sum_{j=0}^{N-1}(w_j,   \bar{u}_{j+Na\cdot  s}) $   \\  $\bf{pq}$ & &
$\sum_{j=0}^{N-1}(n_j,  \bar{u}_{j})$     &     $\sum_{j=0}^{N-1}(n_j,
\bar{u}_{j+Nv\cdot   s}) $ & $\sum_{j=0}^{N-1}(n_j, \bar{u}_{j+Na\cdot
s})   $     \\   $\bf{\bar{p}\bar{q}}$    &  &  $\sum_{j=0}^{N-1}(m_j,
\bar{w}_{j})$  &  $\sum_{j=0}^{N-1}(n_j,  \bar{w}_{j+Nv\cdot s})  $  &
$\sum_{j=0}^{N-1}(n_j, \bar{w}_{j+Na\cdot s}) $\\ $\bf{p{\bar q}}$ & &
$\sum_{j=0}^{N-1}(n_j,    \bar{w}_{j})$&        $\sum_{j=0}^{N-1}(n_j,
\bar{w}_{j+Nv\cdot s}) $&  $\sum_{j=0}^{N-1}(n_j,   \bar{w}_{j+Na\cdot
s}) $  \\  $\bf{\bar{p}q}$ &   & $\sum_{j=0}^{N-1}(m_j, \bar{u}_{j})$&
$\sum_{j=0}^{N-1}(m_j,          \bar{u}_{j+Nv\cdot          s})     $&
$\sum_{j=0}^{N-1}(m_j, \bar{u}_{j+Na\cdot s}) $ \\ \hline \hline
\end{tabular}
\end{center}
\vspace{0.1cm}
The  corresponding antiparticles transform  in  the  complex conjugate
representation.   If the $pq$  sector has a  tachyonic ground state, a
suitable $\mathsf{GSO}$ projection will project  this state out of the
$\mathbf{NS}$ sector and we  should only consider the  massless scalar
fields for  this   sector.  On  the  other   hand,  the $\mathsf{GSO}$
projection for the $p\bar{q}$ sector will have a  sign flip which will
project out the  massless scalars but will  leave the tachyons. If the
$pq   \;\mathbf{NS}$ ground   state   is  massless, the  corresponding
massless  scalars  of the $p\bar{q}$  will   also  be  present in  the
spectrum.

\section{Orbifold models with D9 and D5 branes}
\renewcommand{\theequation}{3.\arabic{equation}}\setcounter{equation}{0}

\subsection{Gauge group and fermion content}

In  this section  we  discuss the  massless   matter content and   the
consistency conditions for a system in the presence of a number $n$ of
D9-branes, a number $u^{(i)}$ of D$5_i$-branes, a number $m$ of D$\bar
9$-branes  and a  number   $w^{\bar{(i)}}$ of D${\bar  5}_{i}$ branes,
within    the    bottom-up  approach.   D9-branes      embed the  full
10-dimensional space-time and therefore the boundary conditions are NN
in   all directions.  However, D5$_{k}$-branes  wrap  around the usual
four-dimensional non-compact  space-time and only  one  of the compact
complex planes ($k = 3,4,5$).  In this sense, there are three possible
types  of D5-branes.   Between  any  two  D-branes, there is   an open
string, with boundary conditions  summarized   as follows ($i\not=   l
\not= m \not= i$): \\
\vspace{0.1cm}
\begin{center}
\begin{tabular}{|c|c|c|c|c|}
\hline \hline String sector & j = 2 & j = i & j = l & j  = m \\ \hline
\hline $\bf{99}$ & NN & NN & NN & NN \\$\bf{9{5_i}}$ & NN  & NN & ND &
ND \\ $\bf{{5_i}9}$ & NN & NN & DN & DN \\ $\bf{5_{i}5_{i}}$ & NN & NN
& DD & DD \\ $\bf{5_{i}5_{l}}$ & NN & ND & DN & DD \\ \hline \hline
\end{tabular}
\end{center}
\vspace{0.1cm}
A  necessary  condition for supersymmetry is   that there  is an equal
number  of  bosonic and fermionic  states  transforming under the same
representation at each mass level.  In the presence of antibranes, the
system generally  has  broken  supersymmetry.   We assume  a   general
embedding for the  action of the ${\mathbf  Z}_N$ orbifold point group
on the Chan-Paton degrees of freedom:
\begin{eqnarray} \label{g95}
&  \gamma_{1, 9} & = diag  \left( I_{n_{0}},  \alpha I_{n_{1}}, \dots,
\alpha^jI_{n_{j}}, \dots, \alpha^{N-1}I_{n_{N-1}} \right) \nonumber \\
&   \gamma_{1,    5_r}&   =  diag     \left(   I_{u^{(r)}_{0}}, \alpha
I_{u^{(r)}_{1}}, \dots,\alpha^j  I_{u^{(r)}_{j}},  \dots, \alpha^{N-1}
I_{u^{(r)}_{N-1}}\right)\nonumber \\ &   \gamma_{1, \bar 9}  &  = diag
\left( I_{m_{0}}, \alpha  I_{m_{1}}, \dots, \alpha^j I_{m_{j}}, \dots,
\alpha^{N-1}    I_{m_{N-1}}  \right)   \nonumber    \\ &    \gamma_{1,
{\bar5}_{\bar r}}&    = diag  \left( I_{w^{({\bar    r})}_{0}}, \alpha
I_{w^{({\bar r})}_{1}}, \dots, \alpha^j I_{w^{({\bar r})}_{j}}, \dots,
\alpha^{N-1} I_{w^{({\bar r})}_{N-1}} \right)
\end{eqnarray} 
with $\alpha=  e^{\frac{2\pi   i}{N}}$.   For  the time   being,   the
non-negative    integers     $n_{j}$,   $u^{(r)}_{j}$,  $m_{j}$    and
$w^{(r)}_{j}$ are  kept  arbitrary.   The $\mathbf{99}$  $\mathbf{NS}$
ground  state    is tachyonic.   After  imposing   the  $\mathsf{GSO}$
projection,  the  two massless gauge  bosons   and six complex scalars
survive the  projection.    The ${\mathbf  R}$  sector  contains eight
fermionic states  $|s_{2}s_{3}s_{4}s_{5}  \rangle_{R}$,  four of which
are left handed $(s_{2}=   - \frac{1}{2})$ (\ref{99r}).   In  general,
unbroken supersymmetry requires the number of ND complex dimensions to
be a multiple of two \cite{Pol2}.   Before the orbifold projection and
choosing    (\ref{gso1})  as   the  $\mathsf{GSO}$    projection,  the
left-handed space-time fermions are:
\begin{eqnarray} \label{99rl1}
& |\psi_1  \rangle & = | -  \frac{1}{2},- \frac{1}{2},  - \frac{1}{2},
-\frac{1}{2} \rangle  \nonumber   \\ &  |\psi_2   \rangle  &  =   |  -
\frac{1}{2}, -\frac{1}{2},  \frac{1}{2}, \frac{1}{2} \rangle \nonumber
\\ & |\psi_3 \rangle  & = |  - \frac{1}{2}, \frac{1}{2}, -\frac{1}{2},
\frac{1}{2} \rangle  \nonumber   \\  &  |\psi_4   \rangle  &  =   |  -
\frac{1}{2}, \frac{1}{2}, \frac{1}{2}, -\frac{1}{2} \rangle.
\end{eqnarray}
Under the orbifold projection, it follows that
$$   \theta^{k}  |s_{2}s_{3}s_{4}s_{5}\rangle   =   \exp[2   \pi i   k
(a_{3}s_{3} + a_{4}s_{4} + a_{5}s_{5})] |s_{2}s_{3}s_{4}s_{5}\rangle.
$$
Explicitly,
\begin{eqnarray} \label{99rl2}
& \theta^{k} |\psi_1   \rangle & =   | \psi_1 \rangle  \nonumber  \\ &
\theta^{k} |\psi_2 \rangle &  = e^{-2 \pi  i  k a _3}|  \psi_2 \rangle
\nonumber \\ &  \theta^{k}  |\psi_3 \rangle &  =  e^{-2 \pi i k  a_4}|
\psi_3 \rangle \nonumber \\ & \theta^{k} |\psi_4 \rangle & = e^{-2 \pi
i k a_5}| \psi_4 \rangle
\end{eqnarray}
since $a_3+a_4+a_5=0$.    Thus,  it  follows from    (\ref{99rl2}) and
(\ref{fp})  that   the left-handed chiral   fermion   spectrum in  the
$\bf{99}$ sector is
\begin{equation} \label{99fs}
\sum_{j=0}^{N-1}[(n_j, {\bar   n}_j)+\sum_{r=3}^5( n_{j}, \bar{n}_{j +
Nv_r})],
\end{equation}
where   the subindices are understood  modulo  $N$ and $a_j=-v_j$ when
$a_2=0$.  The corresponding  antiparticles, the right-handed fermions,
transform     in        the     complex     conjugate   representation
$\sum_{j=0}^{N-1}[({\bar    n}_j,    n_j)+\sum_{r=3}^5(          {\bar
n}_{j-Nv_r},n_{j})]$.   The projections  for  the open  strings in the
${\bf \bar 9 \bar 9}$ sector are completely analogous  to those in the
$\bf{99}$. The $\bf{{\bar 9} 9}$ sector has an opposite $\mathsf{GSO}$
projection and tachyonic  states survive in the $\mathbf{NS}$  sector.
With a similar analysis,  the fermion content in the $\bf{5_{i}5_{i}}$
sector reads,
\begin{equation} \label{5i5ifs}
\sum_{j=0}^{N-1}[(u_j^{(i)},     {\bar        u}_j^{(i)})+\sum_{r=3}^5
( u^{(i)}_{j}, {\bar{u}^{(i)}_{j + Nv_r}})].
\end{equation}
For the  $\bf{5_{i}5_{l}}$   sectors ($i  \neq  l$), the  $\mathbf{R}$
fermionic  states are of the  form  $|s_{2} s_{m} \rangle$ before  the
orbifold projection, where  $m=3, 4, 5$ as  long  as $m \neq i  \neq l
\neq m$.   We   choose $s_{m}=  -  s_2$   in order to  implement   the
$\mathsf{GSO}$ projection (\ref{gso1}) in the same way as before.  The
possible left-handed states   are then $|s_{2}   s_{m} \rangle =  |  -
\frac{1}{2}, \frac{1}{2} \rangle$. Under the orbifold projection,
\begin{equation} \label{5i5iop}
\theta^{k}   |s_{2}s_{m} \rangle    =  e^{2\pi i     k \frac{ a_m}{2}}
|s_{2}s_{m} \rangle = e^{2\pi i k \frac{-v_m}{2}} |s_{2}s_{m} \rangle.
\end{equation}
This equation leads to the following $\mathbf{R}$ states,
\begin{equation} \label{5i5lfs}
\sum_{j=0}^{N-1} (u^{(i)}_{j}, {\bar{u}}^{(l)}_{j - \frac{Nv_m}{2}}).
\end{equation}
The general spectrum can be summarized as follows,
\vspace{0.1cm}
\begin{center}\scriptsize{
\begin{tabular}{|c|c|c|c|c|}
\hline  \hline Sector &   Gauge bosons  &  Tachyonic scalar   fields &
Massless scalar fields & Fermion (s = - 1/2) \\\hline \hline $\bf{99}$
&      ${\bigotimes_{j=0}^{N-1}}        U(n_{j})$         &          &
$\sum_{r=3}^5\sum_{j=0}^{N-1}(n_j,      \bar{n}_{j+Nv_r})        $   &
$\sum_{j=0}^{N-1}[(n_j,             \bar{n}_j)+      \sum_{r=3}^5(n_j,
\bar{n}_{j+Nv_r})]      $           \\    $\bf{\bar{9}\bar{9}}$      &
${\bigotimes_{j=0}^{N-1}}             U(m_{j})$     &                &
$\sum_{r=3}^{5}\sum_{j=0}^{N-1}(m_j,   \bar{m}_{j+N      v_r})     $ &
$\sum_{j=0}^{N-1}[(m_j,             \bar{m}_j)+      \sum_{r=3}^5(m_j,
\bar{m}_{j+Nv_r})]  $  \\  $\bf{9\bar{9}}$ &  & $\sum_{j=0}^{N-1}(n_j,
\bar{m}_{j})$ &   &            $\sum_{j=0}^{N-1}[(n_j,   \bar{m}_{j})+
\sum_{r=3}^5(n_j,     \bar{m}_{j-Nv_r})] $ \\     $\bf{\bar{9}9}$ &  &
$\sum_{j=0}^{N-1}(m_j,  \bar{n}_{j})$    &   & $\sum_{j=0}^{N-1}[(m_j,
\bar{n}_{j})+ \sum_{r=3}^5(m_j,  \bar{n}_{j-Nv_r})] $ \\ $\bf{5_i5_i}$
&         ${\bigotimes_{j=0}^{N-1}}         U(u_{j}^{(i)})$          &
&$\sum_{r=3}^{5}\sum_{j=0}^{N-1}(u_j^{(i)}, \bar{u}_{j+Nv_r}^{(i)})  $
&          $\sum_{j=0}^{N-1}[(u_j^{(i)},             \bar{u}_j^{(i)})+
\sum_{r=3}^5(u_j^{(i)},        \bar{u}_{j+Nv_r}^{(i)})]            $\\
$\bf{\bar{5_i}\bar{5_i}}$         &          ${\bigotimes_{j=0}^{N-1}}
U(w_{j}^{\bar{(i)}})$   &   &$\sum_{r=3}^{5}\sum_{j=0}^{N-1}(w_j^{\bar
{       (i)}},    \bar{w}_{j+Nv_r}^{\bar        {     (i)}})    $    &
$\sum_{j=0}^{N-1}[(w_j^{\bar       {(i)}},  \bar{w}_j^{\bar   {(i)}})+
\sum_{r=3}^5(w_j^{\bar { (i)}},  \bar{w}_{j+Nv_r}^{\bar  {(i)}})] $ \\
$\bf{5_i\bar{5_i}}$ & & $\sum_{j=0}^{N-1}(u_j^{(i)}, \bar{w}_{j}^{\bar
{(i)}})$  & & $\sum_{j=0}^{N-1}[(u_j^{(i)},  \bar{w}_{j}^{\bar{(i)}})+
\sum_{r=3}^5 (u_{j}^{(i)},    \bar{w}_{j-Nv_{r}}^{\bar{(i)}})]   $  \\
$\bf{\bar{5_i}5_i}$     &  &  $\sum_{j=0}^{N-1}(w_j^{({\bar      i})},
\bar{u}_{j}^{(i)})$    &  &    $\sum_{j=0}^{N-1}[(w_j^{({\bar    i})},
\bar{u}_{j}^{(i)})+      \sum_{r=3}^5     (w_{j}^{({\bar         i})},
\bar{u}_{j-Nv_{r}}^{(i)})]     $      \\     $\bf{5_i5_l}$   &       &
&$\sum_{j=0}^{N-1}(u_j^{(i)},   \bar{u}_{j-\frac{N}{2}v_m}^{(l)}) $  &
$\sum_{j=0}^{N-1}(u_j^{(i)},  \bar{u}_{j-\frac{N}{2}v_m}^{(l)})  $  \\
$\bf{5_l5_i}$           &        &       &$\sum_{j=0}^{N-1}(u_j^{(l)},
\bar{u}_{j-\frac{N}{2}v_m}^{(i)})    $  & $\sum_{j=0}^{N-1}(u_j^{(l)},
\bar{u}_{j-\frac{N}{2}v_m}^{(i)}) $ \\ $\bf{\bar{5}_i\bar{5_l}}$ & & &
$\sum_{j=0}^{N-1}(w_j^{\bar{(i)}},
\bar{w}_{j-\frac{N}{2}v_m}^{\bar{(l)}})              $               &
$\sum_{j=0}^{N-1}(w_j^{\bar                                    {(i)}},
\bar{w}_{j-\frac{N}{2}v_m}^{\bar{(l)}}) $ \\ $\bf{\bar{5}_l\bar{5_i}}$
&       &             &             $\sum_{j=0}^{N-1}(w_j^{\bar{(l)}},
\bar{w}_{j-\frac{N}{2}v_m}^{\bar{(i)}})               $              &
$\sum_{j=0}^{N-1}(w_j^{\bar                                    {(l)}},
\bar{w}_{j-\frac{N}{2}v_m}^{\bar{(i)}}) $ \\ $\bf{5_i\bar{5_l}}$ & & &
$\sum_{j=0}^{N-1}(u_j^{(i)},        \bar{w}_{j+\frac{N}{2}(-v_i      +
v_l)}^{\bar{(l)}})$          &            $\sum_{j=0}^{N-1}(u_j^{(i)},
\bar{w}_{j+\frac{N}{2}v_m}^{\bar{(l)}})$ \\  $\bf{\bar{5_l}5_i}$ & & &
$\sum_{j=0}^{N-1}(w_j^{({\bar  l})}, \bar{u}_{j+\frac{N}{2}(-v_i     +
v_l)}^{(i)})$     &        $\sum_{j=0}^{N-1}(w_j^{({\bar         l})},
\bar{u}_{j+\frac{N}{2}v_m}^{(i)})$  \\  $\bf{\bar{5_i}5_l}$   & &    &
$\sum_{j=0}^{N-1}(w_j^{({\bar   i})},  \bar{u}_{j+\frac{N}{2}(-v_i   +
v_l)}^{(l)})$     &         $\sum_{j=0}^{N-1}(w_j^{({\bar        i})},
\bar{u}_{j+\frac{N}{2}v_m}^{(l)})$   \\   $\bf{5_l\bar{5_i}}$  &  &  &
$\sum_{j=0}^{N-1}(u_j^{(l)},     \bar{w}_{j+\frac{N}{2}(-v_i         +
v_l)}^{\bar{(i)}})$            &          $\sum_{j=0}^{N-1}(u_j^{(l)},
\bar{w}_{j+\frac{N}{2}v_m}^{\bar{(i)}})$    \\   $\bf{95_i}$ &   &   &
$\sum_{j=0}^{N-1}(n_j,     \bar{u}_{j-\frac{N}{2}v_i}^{(i)})         $
&$\sum_{j=0}^{N-1}(n_j,    \bar{u}_{j-\frac{N}{2}v_i}^{(i)})    $   \\
$\bf{5_i9}$                &                &                        &
$\sum_{j=0}^{N-1}(u_j^{i},\bar{n}_{j-\frac{N}{2}v_i})                $
&$\sum_{j=0}^{N-1}(u_j^{(i)},   \bar{n}_{j-\frac{N}{2}v_i})     $   \\
$\bf{\bar{9}\bar{5}_i}$  &  &            &      $\sum_{j=0}^{N-1}(m_j,
\bar{w}_{j-\frac{N}{2}v_i}^{\bar{(i)}})   $&    $\sum_{j=0}^{N-1}(m_j,
\bar{w}_{j-\frac{N}{2}v_i}^{\bar{(i)}}) $ \\ $\bf{\bar{5}_i\bar{9}}$ &
&  & $\sum_{j=0}^{N-1}(w_j^{({\bar i)}}, \bar{m}_{j-\frac{N}{2}v_i})$&
$\sum_{j=0}^{N-1}(w_j^{({\bar    i)}},  \bar{m}_{j-\frac{N}{2}v_i})  $
\\$\bf{9{\bar      5}_i}$   &        &    &     $\sum_{j=0}^{N-1}(n_j,
\bar{w}_{j+\frac{N}{2}(-v_l     +      v_m)}^{\bar{(i)}})   $        &
$\sum_{j=0}^{N-1}(n_j,   \bar{w}_{j+\frac{N}{2}v_i}^{\bar{(i)}})   $\\
$\bf{{\bar    5}_i9}$  &  &    &    $\sum_{j=0}^{N-1}(w_j^{(\bar  i)},
\bar{n}_{j+\frac{N}{2}(-v_l + v_m)}) $ &  $\sum_{j=0}^{N-1}(w_j^{(\bar
i)},      \bar{m}_{j+\frac{N}{2}v_i})$\\ $\bf{\bar{9}5_i}$   &   &   &
$\sum_{j=0}^{N-1}(m_j, \bar{u}_{j+\frac{N}{2}(-v_l+v_m)}^{(i)}) $    &
$\sum_{j=0}^{N-1}(m_j,   \bar{u}_{j+\frac{N}{2}v_i}^{(i)})     $    \\
$\bf{5_i\bar{9}}$   &   &       &         $\sum_{j=0}^{N-1}(u_j^{(i)},
\bar{m}_{j+\frac{N}{2}(-v_l+v_m)})$     & $\sum_{j=0}^{N-1}(u_j^{(i)},
\bar{m}_{j+\frac{N}{2}v_i}) $\\ \hline \hline
\end{tabular}}
\end{center}
\begin{center}
\bf{Table 1: Spectrum in the $\mathbf{95}$ \bf configuration}
\end{center}
\vspace{0.1cm}
Now we are in a position to calculate  the contribution $A_{n_j}$ from
the massless fermions  in the  world-volume of  the  D9 brane  to  the
$SU(n_{j})$ gauge  anomaly    (for simplicity,  we  will   exclude the
contribution from the brane-antibrane sectors):
\begin{eqnarray} 
\label{95ga1}
A_{n_j}  & = & \sum_{r=3}^{5}  ( n_{j +  N  v_r} - n_{j   - N v_r} ) +
\sum^{5}_{i=3} ( u^{(i)}_{j-{N  v_i}/2 } -  u^{(i)}_{j +  {N v_i}/2} )
\nonumber \\ & = & \sum^{5}_{r=3} [ ( n_{j + N v_r}  - n_{j- N v_r } )
- ( u^{(r)}_{j + {N v_r}/2 } - u^{(r)}_{ j - {N v_r}/2 } ) ].
\end{eqnarray}
Using that \cite{Al2}:
\begin{equation} \label{nu}
\left \{  \begin{array}{l}  n_{j} =  \frac{1}{N} \sum^{N-1}_{k=0} \exp
( -   2 \pi i k  j  /N ) Tr\gamma_{k,9}  \\  u^{(r)}_{j} = \frac{1}{N}
\sum^{N-1}_{k=0} \exp ( - 2 \pi i k j /N ) Tr\gamma_{k,5_{r}}
\end{array} \right. 
\end{equation} 
and the mathematical identity
\begin{equation} \label{id}
\sum_{r=3}^{5} \sin(2 \pi k v_r) = -4 \prod_{r=3}^{5} \sin(\pi k v_r)
\end{equation}
we can rewrite (\ref{95ga1}) as
\begin{equation} 
\label{95ga2}
A_{n_j}=          \frac{2i}{N}\sum_{k=0}^{N-1}e^{-2\pi  i      kj/N}\{
[   \prod_{r=3}^{5} 2  \sin   ( \pi    k  v_r  )  ] Tr\gamma_{k,9}   +
\sum_{r=3}^{5} [ 2 \sin (\pi k v_r ) ] Tr \gamma_{k, 5_r}\}.
\end{equation} 
Similarly,  the   contribution from the chiral    matter present  at a
particular  D$5_i$   brane  to  the non-abelian  $SU(u_j^{(i)})$ gauge
anomaly is:
\begin{equation} 
\label{95ga3}
A_{u_j^{(i)}}=    \frac{2i}{N}\sum_{k=0}^{N-1}e^{-2\pi     i    kj/N}
\{ \prod_{r=3}^{5} 2 \sin ( \pi k v_r) ] Tr\gamma_{k,5_i} + \sum_{l,m
\not= i}  [ 2 \sin  (\pi k v_m )   ] Tr \gamma_{k,  5_l} +  2\sin (\pi
kv_i)Tr \gamma_{k, 9}\}.
\end{equation} 

\subsection{Partition functions}

\subsubsection{ $\mathbf{99}$ sector}

In this section, we will provide  a stringy analysis for obtaining the
tadpole   cancellation conditions    of  the   orbifold  models  under
consideration.  First,    we need to    compute  the one-loop cylinder
amplitudes of   all possible open string   sectors  of the  theory.  A
general string state is the product of  three pieces: a zero mode part
(\ref{hamnc0}),   a   part  constructed  using   worldsheet    bosonic
oscillators (\ref{hamncb})  and  a  part  using  (NS or R)  worldsheet
fermionic  oscillators (\ref{hamncr}),  (\ref{hamncns}).   This allows
allows us to factorize the trace in (\ref{ca1}) so that:
\begin{eqnarray}
\label{ca11} 
C_{99}      &     =     &     \frac{1}{2N} \sum_{k=0}^{N     -      1}
\int_{0}^{\infty}\frac{dt}{2t}\left[Tr_{99}^{(H_0)}(\theta^kq^{H_0})
Tr_{99}^{(B)}(\theta^kq^{   H_B})  \left(Tr^{\mathsf GSO(NS)}(\theta^k
q^{H_{NS}})-Tr^{\mathsf           GSO(R)}(\theta^k          q^{H_{R}})
\right)\right]\nonumber\\&\equiv&   \frac{1}{2N} \sum_{k=0}^{N   -  1}
\int_{0}^{\infty}\frac{dt}{2t}\left(Z^{(H_0)}_{99}
Z^{(B)}_{99}Z^{(F)}_{99}\right),
\end{eqnarray}
where $q= e^{-2\pi t}$. The minus sign in  the ${\bf R}$ sector is due
to the space-time statistics and it confirms that the ${\bf R}$ sector
leads to space-time fermions   while  the ${\bf NS}$  sector  contains
space-time bosons.  Quantum states will be labelled by the eigenvalues
of the corresponding Hamiltonian operator and computing the trace over
all the  possible  quantum states means integrating  over  all that is
continuous and summing over all that is discrete.

\begin{center}
\it{Zero modes}
\end{center}

The trace over  the zero mode  contributions factorises into a product
of    contributions from  each   space-time dimension.    Since in the
$\bf{99}$ sector we have NN boundary conditions in all directions, the
contribution from the  zero modes to  the partition  function is given
by:\\

i) The trace of $p^2$ in each of the longitudinal ($\mu = 0, 1, 2, 3$)
non-compact directions ($j = 1, 2$ complex dimensions):
\begin{equation}
\int_{-\infty}^{\infty}\frac{dp_{\mu}}{2\pi} \; q^{\alpha ' p^2_{\mu}}
= iV_{\mu}(8\pi^2 t \alpha ')^{-1/2},
\end{equation}
for each value of  $\mu$   since $\theta^k$  acts trivially on   these
non-compact coordinates. We used  $ <p|p'> = 2\pi \delta  (p - p')$ as
our state normalization and $V_{\mu}$  is the (infinite) volume of the
coordinate $X_{\mu}$.  The $i$ factor  comes from the Wick rotation in
the integration of the $\mu = 0$ component. \\

ii)     When $\theta^k$ acts   trivially    on a compactified  complex
coordinate $z^j$   obeying   NN  boundary   conditions, there   exists
quantized momenta in the worldsheet mode  expansion (\ref{nnb}).  Thus
when $kv_j$  =   integer, we  should also   consider  a sum  over  the
quantized momenta along the compact complex direction $z^j$ obeying NN
boudary conditions.  For each of the complex dimensions, this is given
by:
\begin{equation}
(\sum_{n=-\infty}^{n=\infty}q^{\alpha   '   (\frac{n}{R_j})^2})^2
\rightarrow \frac{V_j}{8\pi^2 \alpha   '   t}\;\;\; {\rm as}  \;\;   t
\rightarrow 0
\end{equation}
since  it  is  in the  $t   \rightarrow 0$  limit  of  the open string
amplitude where  we find the  contribution to  the  tadpoles.  We have
defined $V_j =  (2\pi R_j)^2$, where $2\pi R_j$  is the periodicity of
the $z^i$ complex  plane.  This limit  can be  easily calculated using
Poisson   resummation   formula $\sum_{n=-\infty}^{n=\infty}e^{-\pi  a
n^2}=a^{-1/2}\sum_{m=-\infty}^{m=\infty}e^{-\pi m^2 / a}$.  \\

iii) In each of the (complex)  compactified directions $z^j$ ($j=3, 4,
5$)   in which $\theta^k$ acts   non-trivially,  the momentum $p_j$ is
zero.  If any of the complex planes of  the compact space  ($j = 3, 4,
5$) satisfy NN boundary conditions, the value of the  field is free to
fluctuate in those directions contributing to the trace as follows:
\begin{equation} 
Tr[\theta^k] = \int   dz^r<z^r|\theta^k|z^r> = \int dz^r  \delta \left
( (1-e^{2\pi ikv_r}) z^r \right)=(2\sin\pi kv_r)^{-2},
\end{equation}
where we  have  used $<z^r|z^{r'}>=\delta(z^r-z^{r'})$ and that  $\int
dz\delta(\alpha z)=\frac{1}{|\alpha^2|}$.

For the  $\mathbf{99}$ sector (obeying NN   boundary conditions in all
directions), this contribution gives:
\begin{equation} 
\label{99ii} 
Tr[\theta^k]  =  \int  dz^3dz^4dz^5<z^3z^4z^5|\theta^k|z^3z^4z^5>    =
\prod_{r=3}^{5}(2\sin\pi kv_r)^{-2}.
\end{equation}

iv)   The  contribution  from the  trace  of  the  orbifold projection
operator on   the  Chan-Paton degrees  of  freedom $(Tr\gamma_{k,9})$,
where k denotes the twisted sector and $9$ labels the D9-brane sector.
In the ${\mathbf 99}$ sector we would get a  contribution from each of
the branes giving $(Tr\gamma_{k,9})(Tr\gamma_{k,9}^{-1})$.

\begin{center}
\it{Bosonic partition function}
\end{center}

We compute the trace  over the bosonic oscillator  states in the basis
of the operators $\alpha_{-n}$ and $\alpha_n$. In the light-cone gauge
we get:
\begin{eqnarray} \label{99vi} 
Z_{99}^{(B)}=[     q^{-1/12}\prod_{n=1}^{\infty}(1-q^n)^{-2}]    \cdot
[    \prod_{r=3}^5   q^{-1/12}      \times        \prod_{n=1}^{\infty}
\frac{1}{(1-q^ne^{2\pi i kv_r}) (1-q^ne^{-2\pi ikv_r})}].
\end{eqnarray} 
The  first  term  in brackets   represents the contribution   from the
non-compact dimensions.  In   particular, the two real ($j=2$  complex
plane) physical dimensions of  the light-cone gauge.   The rest is the
contribution  from  the three  compact  complex   planes in which  the
orbifold action  takes place.  Each   integer-modded  complex pair  of
bosons contributes with $-\frac{1}{12}$ towards the zero point energy.
Making use of  the expressions given in Appendix  A, the total bosonic
contribution to the $\mathbf{99}$ sector  can be rewritten in terms of
the theta functions as follows:
\begin{eqnarray} 
\label{99bpf} 
Z^{(B)}_{99}  =  \eta   (t)   \times  \prod_{r=3}^{5}\frac{(-2\sin \pi
kv_r)}{\vartheta \left [
\begin{array}{c} \frac{1}{2} \\ \frac{1}{2}+kv_r \end{array} \right](t)
}.
\end{eqnarray} 

\begin{center}
\it{Fermionic partition function}
\end{center}

We need to consider the contributions from both, the $\mathbf{NS}$ and
the $\mathbf{R}$ sectors of the theory:
\begin{equation} \label{99fpf}
Z^{(F)}_{99}=Z^{(NS)}_{99} - Z^{(R)}_{99}.
\end{equation} 
In the  $\mathbf{99}$ sector, the  $\mathbf{NS}$ fermionic oscillators
are  half-integer modded (\ref{ncns}).     The contribution from  this
sector is given by:
\begin{eqnarray} \label{99nspf}
Z^{(NS)}_{99} = & & q^{-1/24}\prod^{\infty}_{n=1}(1+q^{n-1/2})^2 \cdot
[\prod_{r=3}^{5}  q^{-1/24}    \prod^{\infty}_{n=1}(1+q^{n-1/2}e^{2\pi
ikv_r}) (1+q^{n-1/2}e^{-2\pi  ikv_r})]  \nonumber \\ &   & - q^{-1/24}
\prod^{\infty}_{n=1}(1-q^{n-1/2})^2 \cdot [  \prod_{r=3}^{5} q^{-1/24}
\prod^{\infty}_{n=1}   (1-q^{n-1/2}e^{2\pi ikv_r})(1-q^{n-1/2}e^{-2\pi
ikv_r})].
\end{eqnarray} 
The two terms correspond to  the trace  computations with and  without
the    insertion   of  the    $(-1)^{F}$   ${\mathsf   GSO}$ operator,
respectively.  Since the oscillators  are  fermionic, their square  is
zero and no more terms can appear. Each half-integer modded worldsheet
fermion complex  pair  contributes  with $-\frac{1}{24}$ towards   the
zero-point energy.   The $\mathbf{R}$  worldsheet fermions are integer
modded (\ref{ncr}),   their contribution  to  the  partition  function
being:
\begin{eqnarray} 
\label{99rpf} 
Z^{(R)}_{99}   &=&    2q^{1/12}\prod^{\infty}_{n=1}(1+q^n)^2     \cdot
[    \prod_{r=3}^{5}       q^{1/12}(2      \cos  \pi         k    v_r)
\prod^{\infty}_{n=1}(1+q^{n}e^{2\pi    i k  v_r})   (1+q^{n-1}e^{-2\pi
ikv_r})].
\end{eqnarray}
Each integer modded  pair contributes with  $\frac{1}{12}$ towards the
zero-point energy.  The  $Tr[(-1)^F]$  vanishes  in  the  $\mathbf{R}$
sector as the expansion is integer-moded.  Using  the formula (A.5) of
the appendix, the total  contribution from the worldsheet fermions can
be rewritten as:
\begin{equation} 
\label{99fpf2} 
Z^{(F)}_{99}     =   \eta^{-4}(t)\!\!\!\!\!\!\sum_{a,   b      =    0,
1/2}\!\!\!\!\!\! \eta_{a b}\; \vartheta \left  [ \begin{array}{c} a \\
b
\end{array} \right ] (t) \prod_{r=3}^{5} \vartheta \left [
\begin{array}{c} a \\ b+kv_r
\end{array} \right ](t),
\end{equation}
where $\eta_{a b} = (-1)^{2(a+b+2ab)}$.

\begin{center}
\it{General expression}
\end{center}

Combining all  contributions, the final  expression  for the partition
function     in     the    $\mathbf{99}$  sector  $Z_{99}=Z^{H_0}\cdot
Z^{B}_{99}\cdot Z^{F}_{99}$ reads:
\begin{eqnarray} 
\label{99fpft}
Z_{99}       &  =   &  iV_4(8\pi^2\alpha      't)^{-2}(Tr\gamma_{k,9})
(Tr\gamma_{k,9}^{-1})\prod_{r=3}^{5}(2\sin     \pi  k v_r)^{-2}  \cdot
\!\!\!\!\!\!    \sum_{a, b   =   0,  1/2}  \!\!\!\!\!\!    \eta_{a  b}
\frac{\vartheta \left [\begin{array}{c}  a  \\ b \end{array} \right  ]
(t)}{\eta^{3}(t)} \cdot\prod_{r  =3}^{5}  \frac{(-2\sin \pi  k v_r   )
\vartheta \left [ \begin{array}{c}  a  \\ b+kv_r \end{array}  \right ]
(t)}{\vartheta \left [ \begin{array}{c} \frac{1}{2} \\ \frac{1}{2} + k
v_r
\end{array} \right ] (t)},
\end{eqnarray} 
where a sum over quantized momenta should also be included if $kv_i$ =
integer.  In the one-loop open string picture in which we computed the
partition function, the $\mathbf{NS}$  open string sector  corresponds
to  taking $a = 0$ ($b  = 0,  1/2$) and  the  $\mathbf{R}$ open string
sector corresponds to taking $a = 1/2$ and $b  = 0$ (since it vanishes
for $b = 1/2$) \cite{Pol1}.  In the dual  picture (see section C), the
contribution to the  $\mathbf{NSNS}$ closed string tadpole divergences
is contained in the $Z \left [\begin{array}{c} 0, 1/2 \\ 0
\end{array} \right ] (t) $ piece.  On the other hand, the
$\mathbf{RR}$  tadpole divergences are   contained within the $Z \left
[\begin{array}{c} 0 \\ 1/2\end{array} \right ]  (t) $ component of the
total partition function.    By   virtue of  the   Riemann  identities
satisfied  by the $\vartheta$ functions,  the total partition function
vanishes  when  the model  is  supersymmetric.    This  result is  not
expected in  other models where supersymmetry  is broken.   Take as an
example a brane-antibrane pair, which  breaks all supersymmetries.  In
this   case,   the contribution  to   the  $\mathbf{RR}$ closed string
amplitude   picks up a   minus  sign and  as   a consequence the total
partition function no longer vanishes.

\subsubsection{$\mathbf{55}$ sectors}

We  denote by $\mathbf{5_{i}}$ a D5-brane   that wraps around the four
dimensional non-compact space-time ($j = 1, 2$) and one of the compact
$j = 3, 4, 5$ complex planes.  Thus, there  are DD boundary conditions
in the $l$-th and $m$-th  directions transverse to the D$\mathbf{5_i}$
branes. Oscillator  mode  expansions with DD boundary  conditions have
integer   modes but include windings instead   of  momenta.  A general
$\mathbf{5_i5_l}$ brane system satisfies DD boundary conditions in the
$m$-th direction perpendicular  to  both  D5-branes and  mixed  DN(ND)
boundary  conditions in the other two  complex  directions if $i \not=
l$. If $i = j$, the system would obey NN boundary conditions in the $i
= l$  direction and  DD  boundary conditions  in the remaining compact
dimensions.  As before, the non-compact dimensions satisfy NN boundary
conditions.

\begin{center}
\it{Zero modes}
\end{center}

The contribution from  the zero modes to  the partition function  of a
system of $\mathbf{5_i5_i}$ branes is given by:
\begin{eqnarray}
\label{5i5izm}
Z^{(H_0)}_{5_i5_i}&=&iV_4(8\pi^2        \alpha       't)^{-2}(2\sin\pi
kv_i)^{-2}(Tr\gamma_{k,5_i})(Tr\gamma_{k,5_i}^{-1}).
\end{eqnarray}
By  comparing this expression with   the bosonic partition function of
the $\mathbf{99}$ system (\ref{99bpf}), we observe that the difference
arises from the computation  of the trace  of the orbifold operator in
the complex planes of the compact space.  Now, only one of the complex
directions, in contrast to the three of  the $\mathbf{99}$ case, obeys
NN boundary conditions.  We should  also consider a sum over quantized
momenta (windings) if $kv_i$ ($kv_l, kv_m$)  are integers. For the sum
over windings:
\begin{equation}
(\sum_{w=-\infty}^{w=\infty}q^{\frac{1}{\alpha'}(w R_j)^2})^2
\rightarrow  \frac{2\pi^2  \alpha  '  }{V_j t}\;\;\;  {\rm as}  \;\;  t
\rightarrow 0.
\end{equation}
The term that depends on the distance  $Y^2$ between the D$5_i$ branes
(\ref{hamdd0}) is not relevant in the tree channel infrared limit. For
the $\mathbf{5_i5_l}$ ($l \not= i$), we get:
\begin{eqnarray}
\label{5i5lzm}
Z^{(H_0)}_{5_i5_l}&=&iV_4(8\pi^2                                \alpha
't)^{-2}(Tr\gamma_{k,5_i})(Tr\gamma_{k,5_l}^{-1})
\end{eqnarray}
where a sum over windings along the $z^m$ complex plane should be also
included if $kv_m = {\rm integer}$. The contribution from the trace of
the orbifold operator $\theta^k$ is absent because none of the compact
dimensions satisfy NN boundary conditions.

\begin{center}
\it{Bosonic partition function}
\end{center}

The  contribution from the  bosonic oscillator states to the partition
function of a system of $\mathbf{5_i5_i}$ branes is the same as in the
$\mathbf{99}$    case because   the    bosonic mode     expansions are
integer-modded  for both  NN or   DD  boundary  conditions.  For   the
$\mathbf{5_i5_l}$ ($l \not= i$), the bosonic contribution is:
\begin{eqnarray} \label{5i5lbpf} 
Z^{(B)}_{5_{i}5_{l}}=  & & [q^{-1/12}\prod_{n=1}^{\infty}(1-q^n)^{-2}]
\cdot [ q^{-1/12} \prod_{n=1}^{\infty} ( 1-q^n e^{- 2 \pi i k (v_{i} +
v_{l})} )^{-1} ( 1 -  q^n e^{ 2 \pi i  k (v_{i} +v_{l})} )^{-1}] \\& &
\!\!\!\!\!\!\!\!\!     \cdot       [     q^{1/24} \prod_{n=1}^{\infty}
(1-q^{n-1/2}e^{2\pi i kv_i})^{-1} (1-q^{n-1/2}e^{-2\pi i  kv_i})^{-1}]
\cdot   [      q^{1/24}    \prod_{n=1}^{\infty} (1-q^{n-1/2}e^{2\pi  i
kv_l})^{-1} (1-q^{n-1/2}e^{-2\pi i kv_l})^{-1}]. \nonumber
\end{eqnarray}
We  can distinguish the contribution  from  the compact complex  plane
with  DD  boundary  conditions $(m)$   with integer-modded  expansions
(\ref{ddb}) and the contribution from  the compact complex planes with
DN(ND)  boundary conditions $(i,  l \not= m)$ with half-integer modded
expansions  (\ref{dnb}).   The  bosonic  contribution coming  from the
non-compact   dimensions remains  the   same.   Each complex  pair  of
half-integer modded bosons contributes with $\frac{1}{24}$ towards the
zero-point energy  and each  complex  pair  of  integer modded  bosons
contributes  with $-\frac{1}{12}$ to  the zero-point energy.  In terms
of the theta functions:
\begin{eqnarray} \label{5i5lbpf2}
Z^{(B)}_{5_{i}5_{l}} = \frac {\eta (t)} {(2\sin \pi k v_i ) (2\sin \pi
k v_l) } \times \prod^{5}_{r=3} \frac{(- 2 \sin \pi k v_r )}{\vartheta
\left [
\begin{array}{c} \frac{1}{2} \\ 
\frac{1}{2}+k v_r \end{array}\right  ](t) }\cdot \frac{\vartheta \left
[ \begin{array}{c} \frac{1}{2}  \\ \frac{1}{2}+k v_i \end{array}\right
](t) } {\vartheta \left [ \begin{array}{c} 0 \\ \frac{1}{2}+k v_i
\end{array}\right ](t) } \cdot 
\frac{\vartheta \left [
\begin{array}{c} \frac{1}{2} \\ \frac{1}{2}+k v_l \end{array}\right
](t) } {\vartheta \left [ \begin{array}{c} 0 \\ \frac{1}{2}+k v_l
\end{array}\right ](t) }. 
\end{eqnarray} 
\begin{center}
\it{Fermionic partition function}
\end{center}
The fermionic contribution in the $\mathbf{5_i5_i}$ sector is the same
as the fermionic contribution  in the $\mathbf{99}$  sector.  Although
the D-brane configurations are  different and obey different  boundary
conditions,  the fact  that  the  fermionic oscillator expansions  are
integer-modded for the $\mathbf{R}$ sector and half-integer modded for
the   $\mathbf{NS}$  sector  irrespective   of whether   the  boundary
conditions are NN or DD, makes both fermionic partition functions look
the same.  For  the $\mathbf{5_i5_l}$ sector, the  fermionic partition
function can be written as:
\begin{eqnarray}\label{5i5lfpf}
&  &   Z^{(F)}_{5_i5_l}=[q^{-1/48}\prod^{\infty}_{n=1}(1+q^{n-1/2})]^2
\cdot                [q^{-1/24}\prod_{n=1}^{\infty}(1+q^{n-1/2}e^{2\pi
ik(v_i+v_l)})(1+q^{n-1/2}e^{-2\pi ik(v_i+v_l)})]\nonumber \\ &\cdot  &
\prod_{i,j}[q^{1/12}e^{\pi   ikv_i}\prod_{n=1}^{\infty}(1+q^{n}e^{2\pi
ikv_i})(1+q^{n}e^{-2\pi                   ikv_i})]                   -
[q^{-1/48}\prod^{\infty}_{n=1}(1+q^{n-1/2})]^2  \nonumber \\ &\cdot  &
[q^{-1/24}\prod_{n=1}^{\infty}(1-q^{n-1/2}e^{2\pi
ik(v_i+v_l)})(1-q^{n-1/2}e^{-2\pi                   ik(v_i+v_j)}]\cdot
\prod_{i,l}[q^{1/12}ie^{\pi  ikv_i}\prod_{n=1}^{\infty}(1-q^{n}e^{2\pi
ikv_i})(1-q^{n-1}e^{-2\pi     ikv_i})]   \nonumber      \\         &-&
[2q^{1/12}\prod^{\infty}_{n=1}(1+q^{n})]^2   \cdot    [q^{1/12}e^{-\pi
ik(v_i+v_l)}\prod_{n=1}^{\infty}(1+q^{n}e^{2\pi
ik(v_i+v_l)})(1+q^{n-1}e^{-2\pi    ikv_m})]\nonumber   \\  &\cdot    &
\prod_{i,l}[q^{-1/24}\prod_{n=1}^{\infty}(1+q^{n-1/2}e^{2\pi
ikv_i})(1+q^{n-1/2}e^{-2\pi ikv_i})].
\end{eqnarray}
The  first two contributions   come  from the  $\mathbf{NS}$ fermionic
sector and the  last  one from the  $\mathbf{R}$  sector respectively.
The     oscillator  expansions  for  the    $\mathbf{NS}$  sector  are
half-integer modded  in  the   $m$-th direction obeying  DD   boundary
conditions (\ref{cddns}) but integer-modded  in the other two  complex
directions     $i,l$   obeying  mixed    DN(ND)   boundary  conditions
(\ref{cndns}).  The  $\mathbf{R}$ sector has integer-modded expansions
in  the $m$-th direction   (\ref{cddr}) but half-integer expansions in
the remaining two (\ref{cndr}).   Each integer-modded  complex fermion
contributes with $\frac{1}{12}$  and each half-integer  modded complex
fermion with  $-\frac{1}{24}$  to the  zero-point energy respectively.
In terms of theta functions:
\begin{eqnarray} \label{5i5lfpf2}
Z^{(F)}_{5_{i}5_{l}}  &=& \eta^{-4}(t)   \!\!\!\!\!  \sum_{a, b  =  0,
1/2}\!\!\!\!\! \eta_{a b}\; \vartheta \left [ \begin{array}{c} a \\ b
\end{array} \right ] (t) \cdot \frac{\vartheta \left [
\begin{array}{c} \frac{1}{2} - a \\ b+kv_i
\end{array} \right ](t) } 
{\vartheta \left [ \begin{array}{c} a \\ b+kv_i
\end{array} \right ](t) } \cdot 
\frac{\vartheta \left [ \begin{array}{c} \frac{1}{2} - a \\ b+kv_l
\end{array} \right ](t) } 
{\vartheta \left [ \begin{array}{c} a \\ b+kv_l
\end{array} \right ](t) } \cdot
\prod_{r=3}^{5} \vartheta \left [ \begin{array}{c} a \\ b+kv_r
\end{array} \right ](t). 
\end{eqnarray} 
\begin{center}
\it{General expression}
\end{center}
Combining all    the  contributions, the  general   expression for the
$\mathbf{5_i5_i}$ brane system is given by:
\begin{eqnarray}
&                &      Z_{5_i5_i}=iV_4(8\pi                  ^2\alpha
't)^{-2}(Tr\gamma_{k,5_i})(Tr\gamma_{k,5_i}^{-1})\cdot    (2\sin   \pi
kv_i)^{-2}\cdot \!\!\!\!\!\!    \sum_{a,  b  = 0,   1/2}  \!\!\!\!\!\!
\eta_{a b}\frac{\vartheta \left [ \begin{array}{c} a \\ b
\end{array} \right ] (t)}{\eta ^3 (t)}
\cdot \prod_{r=3}^{5} \frac{(-2\sin \pi kv_r)\vartheta \left [
\begin{array}{c} a \\ b+kv_r \end{array} \right ] (t)}{\vartheta \left
[ \begin{array}{c} \frac{1}{2}  \\ \frac{1}{2}+kv_r \end{array} \right
] (t)}.
\end{eqnarray}
For the  $\mathbf{5_i5_l}$    brane system,  the  expression   for the
partition function reads:
\begin{eqnarray} \label{5i5jpf} 
&    &     Z_{5_i5_l}      =      iV_4    (8\pi       ^2\alpha't)^{-2}
(Tr\gamma_{k,5_i})(Tr\gamma_{k,5_l}^{-1}) \cdot \!\!\!\!  \sum_{a, b =
0, 1/2} \!\!\!\! \eta_{a b} \frac{\vartheta \left [
\begin{array}{c} a \\ b \end{array} \right ] (t)}{\eta ^3 (t)}
\cdot \prod_{r=3}^{5}\frac{(-2\sin\pi kv_r) \vartheta \left [
\begin{array}{c} a \\ b+kv_r \end{array} \right ] (t)}
{\vartheta \left [ \begin{array}{c} \frac{1}{2} \\ 1/2+kv_r
\end{array} \right] (t)} \nonumber \\ & & \cdot (2\sin \pi k v_i)^{-1}
\frac{\vartheta \left [ \begin{array}{c} \frac{1}{2} - a \\ b+kv_i
\end{array} \right ] (t) \vartheta \left [ \begin{array}{c}
\frac{1}{2} \\ \frac{1}{2}+kv_i \end{array} \right  ] (t) } {\vartheta
\left [  \begin{array}{c} 0  \\ \frac{1}{2}+kv_i \end{array}  \right ]
(t) \vartheta \left [  \begin{array}{c} a \\ b+kv_i \end{array} \right
](t) } \cdot (2\sin \pi k v_l)^{-1} \frac{\vartheta \left [
\begin{array}{c} \frac{1}{2} - a \\ b+kv_l \end{array} \right ](t)
\vartheta \left [ \begin{array}{c} \frac{1}{2} \\ \frac{1}{2}+kv_l
\end{array} \right ] (t) } {\vartheta \left [
\begin{array}{c} 0 \\ \frac{1}{2}+kv_l \end{array} \right ] (t)
\vartheta \left [ \begin{array}{c}  a  \\ b+kv_l \end{array} \right  ]
(t) }.
\end{eqnarray} 

\subsubsection{$\mathbf{95_{i}}$ and $\mathbf{5_{i}9}$ sectors
($k=3,4,5$)}

A $\mathbf{95_i}$ brane system obeys DN(ND) boundary conditions in the
$l$-th  and  $m$-th     complex coordinates  perpendicular    to   the
$\mathbf{5_i}$  D-branes  and  NN boundary  conditions  in  the $i$-th
complex direction.  As before,  $\mathbf{5_i}$ branes wrap around  the
non-compact dimensions and the $i$-th complex plane.
\begin{center}
\it{Zero modes}
\end{center}
The contribution from the zero modes reads:
\begin{equation}
Z^{(H_0)}_{95_i}=                 iV_4(8\pi                   ^2\alpha
't)^{-2}(Tr\gamma_{k,9})(Tr\gamma_{k,5_i}^{-1})(2\sin\pi kv_i)^{-2},
\end{equation}
where a sum over quantized momenta should  be also included if $kv_i =
{\rm integer}$.

\begin{center}
\it{Bosonic partition function}
\end{center}
The contribution from the bosonic states  to the partition function of
a system of $\mathbf{95_i}$ branes is given by:
\begin{eqnarray}
&                          &                           Z^{(B)}_{95_i}=
[q^{-1/12}\prod^{\infty}_{n=1}(1-q^{n})^{-2}]\nonumber   \\  & \cdot &
[q^{-1/12}\prod_{n=1}^{\infty}(1-q^ne^{2\pi ikv_i})^{-1}(1-q^ne^{-2\pi
ikv_k})^{-1}]\cdot
\prod_{l,m}[q^{1/24}\prod_{n=1}^{\infty}(1-q^{n-1/2}e^{2\pi
ikv_l})^{-1}(1-q^{n-1/2}e^{-2\pi ikv_l})^{-1}].
\end{eqnarray}
The first term in brackets is the contribution from the integer-modded
bosonic oscillators in  the non-compact dimensions. The second bracket
comes  from the   integer-modded   bosonic oscillators  in  the $i$-th
complex compact  plane  obeying  NN  boundary  conditions.   The  last
bracket corresponds to  the half-integer modded bosonic oscillators in
the $l$-th and  $m$-th  complex directions.  Each  half-integer modded
complex worldsheet boson  contributes with $\frac{1}{48}$ towards  the
zero-point energy. In terms of the theta functions:
\begin{eqnarray}
& & Z^{(B)}_{95_i} = \eta (t) \times \frac{(-2\sin\pi kv_i)}{\vartheta
\left [
\begin{array}{c} \frac{1}{2} \\ \frac{1}{2}+kv_i \end{array} \right ]
(t)}\cdot\prod_{l,m}\frac{1}{\vartheta \left [ \begin{array}{c} 0   \\
1/2+kv_l \end{array} \right ] (t)}.
\end{eqnarray}
\begin{center}
\it{Fermionic partition function}
\end{center}
The fermionic partition function looks like:
\begin{eqnarray} \label{95fpf}
Z^{(F)}_{95_{i}}  & = &  \eta^{-4}  (t) \!\!\!\!\!\!   \sum_{a, b = 0,
1/2}\!\!\!\!\!\! \eta_{ab}\; \vartheta \left [ \begin{array}{c} a \\ b
\end{array} \right ] (t) \frac{\vartheta \left [ \begin{array}{c} a \\
b +k v_i
\end{array} \right ] (t) }
{\vartheta \left [ \begin{array}{c} \frac{1}{2}-a \\ b + k v_i
\end{array} \right ] (t) }
\prod_{r=3}^{5} \vartheta \left [  \begin{array}{c} \frac{1}{2}-a \\ b
+ k v_r
\end{array} \right ] (t). 
\end{eqnarray}
\begin{center}
\it{General expression}
\end{center}
Combining all the contributions, the full expression for the partition
function is given by the expression:
\begin{eqnarray}
Z_{95_i}(\theta^k)           =         &i&V_4(8\pi            ^2\alpha
't)^{-2}(Tr\gamma_{k,9})(Tr\gamma_{k,5_i}^{-1})(2\sin\pi    kv_i)^{-2}
\nonumber \\ &\cdot &  \!\!\!\!\!\!\sum_{a,  b = 0, 1/2}  \!\!\!\!\!\!
\eta_{a b}\,\frac{\vartheta \left [ \begin{array}{c} a \\ b
\end{array} \right ] (t)}{\eta ^3 (t)}
\cdot \frac{(-2\sin\pi kv_i)\vartheta  \left  [ \begin{array}{c} a  \\
b+kv_i \end{array} \right  ]  (t)}{\vartheta \left [  \begin{array}{c}
\frac{1}{2} \\ \frac{1}{2}+kv_i
\end{array} \right ] (t)} \cdot
\prod_{l,m}   \frac{\vartheta  \left [\begin{array}{c}\frac{1}{2}-a \\
b+kv_l
\end{array} \right ] (t)}{\vartheta \left [ \begin{array}{c} 0 \\
\frac{1}{2}+kv_l \end{array} \right ] (t)}.
\end{eqnarray}
The  partition function in the $\mathbf{5_i9}$   sector is obtained by
exchanging the roles played   by the $\gamma_{k, 9}$ and   $\gamma_{k,
5_i}$ gamma  functions in the above  expression.  For the rest  of the
D-brane systems  that we considered, we give   the expressions for the
partition functions at the end of the paper in appendix B.  The method
followed for their derivation is the same as the one used so far.

\subsection{Twisted RR Tadpole Cancellation Conditions}

A ${\mathbf RR}$ massless tadpole corresponds to the divergent part of
the ${\bf RR}$ closed  string vacuum amplitude.  Conformal  invariance
of string  theory allows a  tree-level closed  string  amplitude to be
pictured alternatively   as  a one-loop  open  string  amplitude.   In
particular, the cylinder  amplitude corresponds to the contribution of
open strings  to the one-loop  vacuum amplitude or equivalently to the
tree-level closed string amplitude where  the closed strings propagate
between two  D-branes.   Each description  reverses  the roles of  the
worldsheet space and time and is more  appropiate in a different limit
of the moduli space.  The abscence of a modular group for the cylinder
worldsheets makes the $t \rightarrow 0$ cylinder limit quite different
from the $t \rightarrow \infty$ limit.  In  the $t \rightarrow \infty$
limit of the open  string amplitude, the radius  of the circle is very
large.  In order for the string mode to  travel that distance, it must
be light and the open string is in the IR limit.  In the closed string
channel, this process can be seen as  a short distance effect.  On the
other hand, in the $t \rightarrow 0$ limit, the  open string is in the
UV limit.  Now, the string modes do not  need to travel long distances
in making  the loop.  However, this is  the long distance IR  limit of
the closed string and the source for the massless tadpole.  This limit
can be easily computed by using  the modular transformation properties
of the theta  functions.  In  the  previous section we calculated  the
cylinder  amplitudes  in the  one-loop  open string  picture.  In this
section   we will    factorize   the divergent   contribution   to the
$\mathbf{RR}$ tadpoles in the  tree-channel approximation.   The final
step requires the   factorization of the  divergences  into  a sum  of
perfect squares.  It  is then  that  we obtain a  very strong  tadpole
cancellation  condition   for each value $k$    of the twisted sector.
Using the  Jacobi identities (\ref{eq:A6}), (\ref{eq:A7}) satisfied by
the $\vartheta$ functions, we can rewrite the partition functions as:

{\footnotesize
\begin{eqnarray}
Z_{99}       =      (1-1)    iV_4(8\pi^2\alpha't)^{-2}(Tr\gamma_{k,9})
(Tr\gamma_{k,9}^{-1}) \prod_{r=3}^{5} [  2\sin (\pi k v_r) ]^{-2}\cdot
\frac {\vartheta \left [ \begin{array}{c} 0 \\ \frac{1}{2} \end{array}
\right ]  (t)} {\eta  ^3  (t)}\cdot \prod_{r =3}^{5}  \frac  {[ -2\sin
( \pi k v_r )] \vartheta \left [
\begin{array}{c} 0 \\ \frac{1}{2}+kv_r
\end{array} \right ](t)} {\vartheta \left
[\begin{array}{c} \frac{1}{2}   \\  \frac{1}{2}  + k   v_r \end{array}
\right ] (t)}
\end{eqnarray}}
{\footnotesize
\begin{eqnarray} 
Z_{5_{i}5_{i}}= (1-1) iV_4(8\pi ^2\alpha 't)^{-2} (Tr\gamma_{k,5_{i}})
(Tr\gamma_{k,5_{i}}^{-1})  [   2\sin    (    \pi  kv_i)]^{-2}    \cdot
\frac{\vartheta \left [ \begin{array}{c} 0 \\ \frac{1}{2}
\end{array} \right ] (t)} {\eta ^3 (t)}
\cdot \prod_{r=3}^{5} \frac{ [ -2\sin ( \pi kv_r)] \vartheta \left [
\begin{array}{c} 0 \\ \frac{1}{2} + k v_r
\end{array} \right ] (t)}
{\vartheta \left [\begin{array}{c} \frac{1}{2} \\ \frac{1}{2} + k v_r
\end{array}\right](t)} 
\end{eqnarray}}
{\scriptsize
\begin{eqnarray} 
Z_{5_i5_l}=         &(&1-1)        iV_4     (8\pi     ^2\alpha't)^{-2}
(Tr\gamma_{k,5_i})(Tr\gamma_{k,5_l}^{-1})\cdot \frac{\vartheta \left [
\begin{array}{c} 0 \\ \frac{1}{2}
\end{array} \right ] (t)} {\eta ^3 (t)}\cdot 
\prod_{r=3}^{5}\frac{[ -2 \sin (\pi kv_r)] \vartheta \left [
\begin{array}{c} 0 \\ \frac{1}{2}+kv_r \end{array} \right ] (t)}
{\vartheta \left [ \begin{array}{c} \frac{1}{2} \\ \frac{1}{2}+ k v_r
\end{array} \right] (t)} 
\prod_{i,  l}[  2 \sin  (\pi   k  v_i)]^{-1} (  \frac{\vartheta  \left
 [ \begin{array}{c} \frac{1}{2} \\ \frac{1}{2}+kv_i \end{array} \right
 ]  (t) } {\vartheta \left  [  \begin{array}{c} 0 \\  \frac{1}{2}+kv_i
 \end{array} \right ] (t) } )^{2}
\end{eqnarray}} 
{\footnotesize
\begin{eqnarray}
Z_{95_i}=&(&1-1)             iV_4        (8\pi        ^2\alpha't)^{-2}
(Tr\gamma_{k,9})(Tr\gamma_{k,5_i}^{-1})  [2   \sin  (\pi  k v_i)]^{-2}
\frac{\vartheta \left [
\begin{array}{c} 0 \\ \frac{1}{2}
\end{array} \right ] (t)} {\eta ^3 (t)} \cdot
\prod_{r=3}^{5}\frac{ \vartheta \left [
\begin{array}{c} \frac{1}{2} \\ \frac{1}{2}+kv_r \end{array} 
\right ] (t)} {\vartheta \left  [ \begin{array}{c} 0 \\ \frac{1}{2}+ k
v_r  \end{array} \right] (t)} [-  2 \sin (\pi k v_i)] (\frac{\vartheta
\left [
\begin{array}{c} 0 \\ \frac{1}{2}+kv_i \end{array} \right ] (t) }
{\vartheta \left [ \begin{array}{c} \frac{1}{2} \\ \frac{1}{2}+kv_i
\end{array} \right ] (t) } )^{2}. 
\end{eqnarray}} 
The second term in the (1-1) prefactor corresponds to taking $a=0$ and
$b=\frac{1}{2}$  in   the  original expressions  for    the  partition
functions and represents the contribution from the cylinder amplitudes
to the  $\mathbf{RR}$   couplings in   the tree-level closed    string
picture.  The tadpole divergences  at $\frac{1}{t} \to \infty$ in  the
closed string  channel  are easily evaluated  by taking  the $t \to 0$
limit in the one-loop cylinder amplitudes  of the open string channel.
Using the formulas (\ref{eq:A8})-(\ref{eq:A11})  given in appendix  A,
we get as $t \to 0$,  the following contributions to the $\mathbf{RR}$
tadpole divergences:
\begin{equation}
Z_{99}^{(RR)} \cong -iV_4 (8\pi^2\alpha')^{-2} \cdot \frac{2}{t} \cdot
\prod^{5}_{r=3}   \vert   2   \sin(  \pi   k   v_r)  \vert^{-1}  \cdot
(Tr\gamma_{k,9}) (Tr\gamma_{k,9}^{-1})
\end{equation}
\begin{equation}
Z_{5_{i}5_{i}}^{(RR)}   \cong   -iV_4     (8\pi^2\alpha')^{-2}   \cdot
\frac{2}{t} \cdot \prod^{5}_{r=3}  \vert 2 \sin( \pi  k v_r) \vert  [2
\sin    (\pi         k        v_i)]^{-2}\cdot     (Tr\gamma_{k,5_{i}})
(Tr\gamma_{k,5_{i}}^{-1})
\end{equation}
\begin{equation}
Z_{5_{i}5_{l}}^{(RR)}\cong        -iV_4 (8\pi^2\alpha')^{-2}     \cdot
\frac{2}{t} \cdot  \prod^{5}_{r=3} \vert 2  \sin( \pi  k v_r) \vert [2
\sin (\pi k  v_i)]^{-1}[2 \sin (\pi  k v_l)]^{-1} (Tr\gamma_{k,5_{i}})
\cdot (Tr\gamma_{k,5_{l}}^{-1})
\end{equation}
\begin{equation}
Z_{95_{i}}^{(RR)}      \cong      -iV_4 (8\pi^2\alpha')^{-2}     \cdot
\frac{2}{t}\cdot \prod^{5}_{r=3} \vert  2 \sin( \pi  k v_r) \vert^{-1}
\frac{\prod_{r=3}^{5}[2 \sin(\pi  k  v_r)] }{ [2 \sin   (\pi k v_i)] }
\cdot (Tr\gamma_{k, 9}) (Tr\gamma_{k,5_{i}}^{-1}).
\end{equation}
The   expression for the      contribution to the   cylinder amplitude
${\mathcal C}= \sum_{pq} {\mathcal C}_{pq}$ in the asymptotic limit of
$t \to 0$ reads:
\begin{eqnarray}\label{95ca}
{\mathcal C}  \approx    & &  (1-1) [-    iV_4   (8\pi^2\alpha')^{-2}]
\frac{1}{2N}     \int^{\infty}_{0}    \frac{dt}{t^2}  \sum_{k=0}^{N-1}
\prod^{5}_{r=3} \vert 2 \sin  (\pi k v_r) \vert^{-1}  \nonumber \\ & &
\cdot   \left( Tr\gamma_{k, 9}   +  \prod_{r=3}^{5}[2 \sin(\pi k v_r)]
\sum_{i=3}^{5}{    [ \frac{Tr\gamma_{k,5_{i}}}{2  \sin(\pi   k  v_i) }
]}\right)     \cdot  \left(Tr\gamma_{k,  9}^{-1}  +  \prod_{r=3}^{5}[2
\sin(\pi  k v_r)] \sum_{i=3}^{5}{  [ \frac{Tr\gamma^{-1}_{k,5_{i}}} {2
\sin(\pi k v_i) } ]}\right).
\end{eqnarray}
Using that the $\gamma_{k,p}$ matrices are both diagonal and unitary,
\begin{equation}
\left. \begin{array}{lll} \gamma_{k, p}^{-1} = \gamma_{k, p}^{\star} &
\, \, \, & Tr \gamma_{k, p}^{-1} = Tr \gamma_{k, p}^{\star}
\end{array}
\right.
\end{equation}
(\ref{95ca}) can alternatively be written as,
\begin{equation}
 {\mathcal C} \approx (1-1) [- iV_4 (8\pi^2\alpha')^{-2}] \frac{1}{2N}
\int^{\infty}_{0} \frac{dt}{t^2}  \sum_{k=0}^{N-1}     \prod^{5}_{r=3}
\vert 2  \sin  (\pi k v_r) \vert^{-1}  \cdot  \vert  Tr\gamma_{k, 9} +
\prod_{r=3}^{5}[2      \sin(\pi      k       v_r)]     \sum_{i=3}^{5}{
[ \frac{Tr\gamma_{k,5_{i}}}{2 \sin(\pi k v_i) } ]} \vert^{2}
\end{equation}
and the $\mathbf{RR}$ twisted tadpole   cancellation conditions for  a
system in the  presence of different sets  of  D9-branes and D5-branes
are then:
\begin{eqnarray} \label{95tcc} 
{\mathbf \frac{Tr  \gamma_{k,9}} {\prod_{j=3}^{5} [ 2\sin (\pi kv_j)]}
+ \sum_{i=3}^{5} { \frac{Tr \gamma_{k,5_i}} { [2\sin ( \pi kv_i)]} } =
0\;\;\;\;\;\;\;\;\;\;\;\;\;\;\;\;\;\;\; (k = 1, 2, \cdots, N-1)}
\end{eqnarray} 
and they are required to  be satisfied at each fixed  point of the six
dimensional compact space.  If  instead of only having D$p$-branes, we
also allowed   the  presence of  D$\bar   p$-branes, the $\mathbf{RR}$
tadpole cancellation conditions should be modified by replacing:
\begin{equation}
Tr\gamma_{k,p} \longrightarrow Tr\gamma_{k,p} - Tr\gamma_{k,\bar p}
\end{equation}
since the brane-antibrane  pairs have opposite  $\mathbf{RR}$ charges.
Let us now  analyze the   relationship  between tadpole  and   anomaly
cancellation conditions.    For simplicity  we assume that   our model
contains only one type of D$5_i$ branes  and that we have compactified
on   a   ${\mathbf Z}_3$  orbifold.     In  this  scenario,   equation
(\ref{95tcc}) reads:
\begin{eqnarray} \label{95tcc2} 
\frac{Tr \gamma_{k,9}}  {\prod_{j=3}^{5}  [   2\sin  (\pi kv_j)]}    +
\frac{Tr \gamma_{k,5_i}}     {    [2\sin     (    \pi kv_i)]}        =
0\;\;\;\;\;\;\;\;\;\;\;\;\;\;\;\;\;\;\; (k = 1, 2, \cdots, N-1)
\end{eqnarray} 
with    twisted    vector  $v=\frac{1}{3}(1,  1,    -2)$.    Comparing
(\ref{95tcc2})  with (\ref{95ga3}), we  see  that tadpole cancellation
guarantees the  absence of $SU(u_j^i)$ gauge  anomalies.  On the other
hand, to calculate the total  contribution to the gauge anomalies from
all the chiral matter  in the D9  brane, we should include in addition
to   the chiral matter from  the  $\mathbf{99}$ sectors, the remaining
chiral matter   arising  from  all   the   possible ${\mathbf   95_i}$
sectors. Since we are considering $D5_i$  branes parallel to the $z_i$
complex plane  and  we have  nine different   fixed  points with space
transverse to the  D$5_i$ branes at which we  can place them (remember
that the ${\mathbf Z}_3$ orbifold has a total  of 27 fixed points), we
get:
\begin{equation} 
A_{n_j}= \frac{2i}{N}\sum_{k=0}^{N-1}e^{-2\pi i  kj/N}2\sin(\pi  kv_i)
\{ [\!\! \prod_{l, m \not= i} 2 \sin ( \pi k  v_l ) ] Tr\gamma_{k,9} +
9 Tr \gamma_{k, 5_i}\}
\end{equation} 
which cancels out for the ${\mathbf Z}_3$  orbifold as a result of the
tadpole cancellation.    Note   that  since  the   tadpole   condition
(\ref{95tcc2}) must be satisfied at  all nine fixed points, it ensures
that $Tr \gamma_{k, 5_r}$ is the same at all of them and therefore the
factor of nine  in the above expression. Thus  for a system of D9  and
D$5_i$ branes,  tadpole cancellation ensures anomaly cancellation (but
not vice-versa).

\section{Orbifold models with various D-branes}
\renewcommand{\theequation}{4.\arabic{equation}}\setcounter{equation}{0}

\subsection{D3-D7 brane orbifold}

We now  consider a system with  a  number $n$ of  D3-branes, $u^{(i)}$
number    of   D$7_i$-branes, $m$  number  of   D$\bar   3$-branes and
$w^{\bar{(i)}}$ number of D${\bar 7}_{i}$ branes.  D3-branes embed the
4-dimensional  non-compact Minkowski space-time and  sit at some fixed
points of the remaining  6-dimensional internal space, while D7-branes
occupy an 8-dimensional subspace of the full 10-dimensional spacetime.
They  wrap  the 4-dimensional non-compact spacetime    plus two of the
complex planes.  By  D7$_{i}$ we denote  a D7-brane  transverse to the
$i$-th complex plane.  As a T-dual version of  the D9D5 brane orbifold
already discussed, the  orbifold containing D3D7 branes also preserves
spacetime supersymmetry.  We assume the same general embedding for the
action of the  ${\mathbf Z}_N$ orbifold  point group on the Chan-Paton
degrees  of  freedom   (\ref{g95}).   The spectrum     that arises  is
completely analogous to that in the $\mathbf{95}$ sector: \\
\vspace{0.1cm}
\begin{center}\footnotesize{
\begin{tabular}{|c|c|c|c|c|}
\hline    \hline Sector & Gauge bosons    & Tachyonic  scalar fields &
Massless scalar fields & Fermion (s = - 1/2) \\\hline \hline $\bf{33}$
&              ${\bigotimes_{j=0}^{N-1}}     U(n_{j})$     &         &
$\sum_{r=3}^5\sum_{j=0}^{N-1}(n_j,   \bar{n}_{j+Nv_r})         $     &
$\sum_{j=0}^{N-1}[(n_j,         \bar{n}_j)+          \sum_{r=3}^5(n_j,
\bar{n}_{j+Nv_r})]   $        \\           $\bf{\bar{3}\bar{3}}$     &
${\bigotimes_{j=0}^{N-1}}             U(m_{j})$        &             &
$\sum_{r=3}^{5}\sum_{j=0}^{N-1}(m_j,   \bar{m}_{j+N  v_r})     $     &
$\sum_{j=0}^{N-1}[(m_j,         \bar{m}_j)+          \sum_{r=3}^5(m_j,
\bar{m}_{j+Nv_r})]  $  \\  $\bf{3\bar{3}}$ & &  $\sum_{j=0}^{N-1}(n_j,
\bar{m}_{j})$      &         &   $\sum_{j=0}^{N-1}[(n_j, \bar{m}_{j})+
\sum_{r=3}^5(n_j,  \bar{m}_{j-Nv_r})]  $   \\  $\bf{\bar{3}3}$    &  &
$\sum_{j=0}^{N-1}(m_j,   \bar{n}_{j})$   &  &  $\sum_{j=0}^{N-1}[(m_j,
\bar{n}_{j})+  \sum_{r=3}^5(m_j, \bar{n}_{j-Nv_r})] $ \\ $\bf{7_i7_i}$
&         ${\bigotimes_{j=0}^{N-1}}           U(u_{j}^{(i)})$        &
&$\sum_{r=3}^{5}\sum_{j=0}^{N-1}(u_j^{(i)}, \bar{u}_{j+Nv_r}^{(i)})  $
&          $\sum_{j=0}^{N-1}[(u_j^{(i)},             \bar{u}_j^{(i)})+
\sum_{r=3}^5(u_j^{(i)},            \bar{u}_{j+Nv_r}^{(i)})]        $\\
$\bf{\bar{7_i}\bar{7_i}}$         &          ${\bigotimes_{j=0}^{N-1}}
U(w_{j}^{\bar{(i)}})$   &   &$\sum_{r=3}^{5}\sum_{j=0}^{N-1}(w_j^{\bar
{     (i)}},   \bar{w}_{j+Nv_r}^{\bar      {       (i)}})     $      &
$\sum_{j=0}^{N-1}[(w_j^{\bar    {(i)}},  \bar{w}_j^{\bar      {(i)}})+
\sum_{r=3}^5(w_j^{\bar  {  (i)}}, \bar{w}_{j+Nv_r}^{\bar {(i)}})] $ \\
$\bf{7_i\bar{7_i}}$ & & $\sum_{j=0}^{N-1}(u_j^{(i)}, \bar{w}_{j}^{\bar
{(i)}})$ & &  $\sum_{j=0}^{N-1}[(u_j^{(i)},  \bar{w}_{j}^{\bar{(i)}})+
\sum_{r=3}^5  (u_{j}^{(i)},  \bar{w}_{j-Nv_{r}}^{\bar{(i)}})]   $   \\
$\bf{\bar{7_i}7_i}$    &     &   $\sum_{j=0}^{N-1}(w_j^{({\bar   i})},
\bar{u}_{j}^{(i)})$      &    &  $\sum_{j=0}^{N-1}[(w_j^{({\bar  i})},
\bar{u}_{j}^{(i)})+      \sum_{r=3}^5       (w_{j}^{({\bar       i})},
\bar{u}_{j-Nv_{r}}^{(i)})]        $    \\    $\bf{7_i7_l}$     &     &
&$\sum_{j=0}^{N-1}(u_j^{(i)},   \bar{u}_{j-\frac{N}{2}v_m}^{(l)})  $ &
$\sum_{j=0}^{N-1}(u_j^{(i)},  \bar{u}_{j-\frac{N}{2}v_m}^{(l)})  $  \\
$\bf{7_l7_i}$            &           &   &$\sum_{j=0}^{N-1}(u_j^{(l)},
\bar{u}_{j-\frac{N}{2}v_m}^{(i)})  $   &  $\sum_{j=0}^{N-1}(u_j^{(l)},
\bar{u}_{j-\frac{N}{2}v_m}^{(i)}) $ \\ $\bf{\bar{7}_i\bar{7_l}}$ & & &
$\sum_{j=0}^{N-1}(w_j^{\bar{(i)}},
\bar{w}_{j-\frac{N}{2}v_m}^{\bar{(l)}})            $                 &
$\sum_{j=0}^{N-1}(w_j^{\bar                                    {(i)}},
\bar{w}_{j-\frac{N}{2}v_m}^{\bar{(l)}}) $ \\ $\bf{\bar{7}_l\bar{7_i}}$
&          &          &             $\sum_{j=0}^{N-1}(w_j^{\bar{(l)}},
\bar{w}_{j-\frac{N}{2}v_m}^{\bar{(i)}})             $                &
$\sum_{j=0}^{N-1}(w_j^{\bar                                    {(l)}},
\bar{w}_{j-\frac{N}{2}v_m}^{\bar{(i)}}) $ \\ $\bf{7_i\bar{7_l}}$ & & &
$\sum_{j=0}^{N-1}(u_j^{(i)},     \bar{w}_{j+\frac{N}{2}(-v_i         +
v_l)}^{\bar{(l)}})$          &            $\sum_{j=0}^{N-1}(u_j^{(i)},
\bar{w}_{j+\frac{N}{2}v_m}^{\bar{(l)}})$  \\ $\bf{\bar{7_l}7_i}$ & & &
$\sum_{j=0}^{N-1}(w_j^{({\bar  l})},  \bar{u}_{j+\frac{N}{2}(-v_i    +
v_l)}^{(i)})$        &    $\sum_{j=0}^{N-1}(w_j^{({\bar          l})},
\bar{u}_{j+\frac{N}{2}v_m}^{(i)})$  \\ $\bf{\bar{7_i}7_l}$  & &      &
$\sum_{j=0}^{N-1}(w_j^{({\bar i})},     \bar{u}_{j+\frac{N}{2}(-v_i  +
v_l)}^{(l)})$          &         $\sum_{j=0}^{N-1}(w_j^{({\bar   i})},
\bar{u}_{j+\frac{N}{2}v_m}^{(l)})$  \\    $\bf{7_l\bar{7_i}}$ &    & &
$\sum_{j=0}^{N-1}(u_j^{(l)},       \bar{w}_{j+\frac{N}{2}(-v_i       +
v_l)}^{\bar{(i)}})$            &          $\sum_{j=0}^{N-1}(u_j^{(l)},
\bar{w}_{j+\frac{N}{2}v_m}^{\bar{(i)}})$    \\   $\bf{37_i}$ &     & &
$\sum_{j=0}^{N-1}(n_j,        \bar{u}_{j-\frac{N}{2}v_i}^{(i)})      $
&$\sum_{j=0}^{N-1}(n_j,       \bar{u}_{j-\frac{N}{2}v_i}^{(i)})   $ \\
$\bf{7_i3}$               &                     &                    &
$\sum_{j=0}^{N-1}(u_j^{i},\bar{n}_{j-\frac{N}{2}v_i})                $
&$\sum_{j=0}^{N-1}(u_j^{(i)},   \bar{n}_{j-\frac{N}{2}v_i})   $     \\
$\bf{\bar{3}\bar{7}_i}$      &  &    &          $\sum_{j=0}^{N-1}(m_j,
\bar{w}_{j-\frac{N}{2}v_i}^{\bar{(i)}})    $&   $\sum_{j=0}^{N-1}(m_j,
\bar{w}_{j-\frac{N}{2}v_i}^{\bar{(i)}}) $ \\ $\bf{\bar{7}_i\bar{3}}$ &
&  & $\sum_{j=0}^{N-1}(w_j^{({\bar i)}}, \bar{m}_{j-\frac{N}{2}v_i})$&
$\sum_{j=0}^{N-1}(w_j^{({\bar i)}},  \bar{m}_{j-\frac{N}{2}v_i})  $ \\
$\bf{3{\bar      7}_i}$    &       &      &     $\sum_{j=0}^{N-1}(n_j,
\bar{w}_{j+\frac{N}{2}(-v_l     +        v_m)}^{\bar{(i)}})     $    &
$\sum_{j=0}^{N-1}(n_j,    \bar{w}_{j+\frac{N}{2}v_i}^{\bar{(i)}})  $\\
$\bf{{\bar      7}_i3}$  &  &    &  $\sum_{j=0}^{N-1}(w_j^{(\bar  i)},
\bar{n}_{j+\frac{N}{2}(-v_l  + v_m)}) $ & $\sum_{j=0}^{N-1}(w_j^{(\bar
i)},     \bar{m}_{j+\frac{N}{2}v_i})$\\     $\bf{\bar{3}7_i}$  &  &  &
$\sum_{j=0}^{N-1}(m_j, \bar{u}_{j+\frac{N}{2}(-v_l+v_m)}^{(i)})  $   &
$\sum_{j=0}^{N-1}(m_j,    \bar{u}_{j+\frac{N}{2}v_i}^{(i)})    $    \\
$\bf{7_i\bar{3}}$   &      &          &   $\sum_{j=0}^{N-1}(u_j^{(i)},
\bar{m}_{j+\frac{N}{2}(-v_l+v_m)})$   &   $\sum_{j=0}^{N-1}(u_j^{(i)},
\bar{m}_{j+\frac{N}{2}v_i}) $\\ \hline \hline
\end{tabular}}
\end{center}
\begin{center}
\bf{Table 2: Spectrum in the $\mathbf{37}$ \bf configuration}
\end{center}
\vspace{0.1cm}
and their contribution  towards the  $SU(n_{j})$ and $SU(u_{j}^{(i)})$
anomalies in the world-volumes of D3 and D$7_i$ branes respectively:
\begin{equation} \label{37ga1}
A_{n_j}=        \frac{2i}{N}\sum_{k=0}^{N-1}e^{-2\pi    i       kj/N}
\{ [\prod_{r=3}^{5} 2 \sin    ( \pi  k    v_r )  ]  Tr\gamma_{k,3}  +
\sum_{r=3}^{5} [ 2 \sin (\pi k v_r ) ] Tr \gamma_{k, 7_r} \}
\end{equation} 
and
\begin{equation} \label{37ga2}
A_{u_j^{(i)}}=   \frac{2i}{N}\sum_{k=0}^{N-1}e^{-2\pi i  kj/N}      \{
[ \prod_{r=3}^{5} 2 \sin ( \pi k  v_r ) ] Tr\gamma_{k,7_i} + \sum_{l,m
\not= i} [  2  \sin (\pi  k v_m )  ] Tr  \gamma_{k, 7_l} +  2\sin (\pi
kv_i)Tr \gamma_{k, 3}\}.
\end{equation} 
By using (\ref{C5})-(\ref{C8}),  the  ${\mathbf RR}$  twisted  tadpole
cancellation conditions for   a system in  the presence  of D3  and D7
branes reads:
\begin{eqnarray} \label{37tcc}
{\mathbf   [  \prod_{r=3}^{5}2\sin  (\pi    kv_r) ]  Tr  \gamma_{k,3}+
\sum_{i=3}^{5}    [   2\sin   (   \pi    kv_i)  ]   Tr\gamma_{k,7_i}=0
}\end{eqnarray}
which automatically guarantees the absence of the D3-brane gauge group
$SU(n_j)$  anomalies.   To  verify  that (\ref{37tcc})  guarantees the
absence of the D$7_i$-brane  $SU(u_j^{(i)})$ gauge anomalies,  we have
to realize that to calculate  the contribution  to the anomalies  from
the chiral matter in a particular D$7_i$  brane (for simplicity, let's
assume that we only have one type of D$7_i$ brane), we should not only
consider the contribution  from the  chiral  fermions arising  in  the
${\mathbf 7_i7_i}$ but  also  all the  chiral matter  arising from all
possible ${\mathbf 37_i}$ sectors, since  the D$7_i$ brane embeds  all
the D3 branes present  at the fixed  points  in the $l$-th and  $m$-th
$(l, m \not=  i)$ complex planes (which  are  nine for the  particular
case of the ${\mathbf Z}_3$ orbifold).  It  is easy to check that this
is the case.  So again tadpole  cancellation ensures non-abelian gauge
anomaly cancellation.

\subsection{D3-D9 brane orbifold}

We now consider a system with a number $n$  of D3-branes, a number $u$
of D$9$-branes, a number $m$ of  D$\bar 3$-branes and  a number $w$ of
D${\bar 9}$ branes. The spectrum is shown in the following table: \\
\vspace{0.1cm}
{\footnotesize
\begin{center}
\begin{tabular}{|c|c|c|c|c|c|}
\hline \hline Sector & Gauge bosons & Tachyons& Massless scalar fields
& $ Fermion_{-}    $& $ Fermion_{+} $    \\\hline \hline $\bf{33}$   &
${\bigotimes_{j=0}^{N-1}}           U(n_{j})$           &            &
$\sum_{r=3}^5\sum_{j=0}^{N-1}(n_j,       \bar{n}_{j+Nv_r})       $   &
$\sum_{j=0}^{N-1}[(n_j,            \bar{n}_j)+       \sum_{r=3}^5(n_j,
\bar{n}_{j+Nv_r})]     $  &    c.     c.\\ $\bf{\bar{3}\bar{3}}$     &
${\bigotimes_{j=0}^{N-1}}                 U(m_{j})$          &       &
$\sum_{r=3}^{5}\sum_{j=0}^{N-1}(m_j,   \bar{m}_{j+N  v_r})   $       &
$\sum_{j=0}^{N-1}[(m_j,          \bar{m}_j)+         \sum_{r=3}^5(m_j,
\bar{m}_{j+Nv_r})]   $     &  c.  c.       \\   $\bf{3\bar{3}}$  &   &
$\sum_{j=0}^{N-1}(n_j,    \bar{m}_{j})$    & & $\sum_{j=0}^{N-1}[(n_j,
\bar{m}_{j})+ \sum_{r=3}^5(n_j,  \bar{m}_{j-Nv_r})]   $ & c.   c.   \\
$\bf{\bar{3}3}$ &  &    $\sum_{j=0}^{N-1}(m_j,   \bar{n}_{j})$ &     &
$\sum_{j=0}^{N-1}[(m_j,          \bar{n}_{j})+       \sum_{r=3}^5(m_j,
\bar{n}_{j-Nv_r})]   $    &     c.      c.    \\       $\bf{99}$     &
${\bigotimes_{j=0}^{N-1}}          U(u_{j})$           &             &
$\sum_{r=3}^5\sum_{j=0}^{N-1}(u_j,       \bar{u}_{j+Nv_r})     $     &
$\sum_{j=0}^{N-1}[(u_j,            \bar{u}_j)+       \sum_{r=3}^5(u_j,
\bar{u}_{j+Nv_r})]   $   &    c.   c.   \\   $\bf{\bar{9}\bar{9}}$   &
${\bigotimes_{j=0}^{N-1}}             U(w_{j})$          &           &
$\sum_{r=3}^{5}\sum_{j=0}^{N-1}(w_j,  \bar{w}_{j+N   v_r})       $   &
$\sum_{j=0}^{N-1}[(w_j,         \bar{w}_j)+          \sum_{r=3}^5(w_j,
\bar{w}_{j+Nv_r})]   $ &     c.    c.    \\   $\bf{9\bar{9}}$  &     &
$\sum_{j=0}^{N-1}(u_j,  \bar{w}_{j})$ &   &    $\sum_{j=0}^{N-1}[(u_j,
\bar{w}_{j})+ \sum_{r=3}^5(u_j, \bar{w}_{j-Nv_r})]  $  &  c.   c.   \\
$\bf{\bar{9}9}$     & & $\sum_{j=0}^{N-1}(w_j,    \bar{u}_{j})$   &  &
$\sum_{j=0}^{N-1}[(w_j,       \bar{u}_{j})+          \sum_{r=3}^5(w_j,
\bar{u}_{j-Nv_r})]   $   &    c.  c.    \\     $\bf{39}$  &   &   &  &
$\sum_{j=0}^{N-1}(n_j, \bar{u}_{j}) $  & - \\ $\bf{93}$  &  & & & -  &
$\sum_{j=0}^{N-1}(u_j, \bar{n}_{j}) $ \\ $\bf{\bar{3}\bar{9}}$ & & & &
$\sum_{j=0}^{N-1}(m_j, \bar{w}_{j}) $ & - \\ $\bf{\bar{9}\bar{3}}$ & &
& & - & $\sum_{j=0}^{N-1}(w_j, \bar{m}_{j}) $  \\ $\bf{3{\bar 9}}$ & &
& & - & $\sum_{j=0}^{N-1}(n_j, \bar{w}_{j})  $ \\ $\bf{{\bar 9}3}$ & &
& & $\sum_{j=0}^{N-1}(w_j, \bar{n}_{j}) $ & - \\ $\bf{\bar{3}9}$ & & &
& - & $\sum_{j=0}^{N-1}(m_j, \bar{u}_{j}^{(i)}) $ \\ $\bf{\bar{9}3}$ &
& & & $\sum_{j=0}^{N-1}(u_j, \bar{m}_{j}^{(i)}) $ & - \\ \hline \hline
\end{tabular}
\end{center}}
\begin{center}
\bf{Table 3: Spectrum in the $\mathbf{93}$ \bf configuration}
\end{center}
\vspace{0.1cm}
Note   that  this  type  of brane   system  has   tachyons  and breaks
supersymmetry.  The   mixed  $\mathbf{39}$ sectors   obey NN  boundary
conditions in  the $j  =  2$  complex   plane but mixed  DN   boundary
conditions  in   the remaining complex planes  $j   = 3,  4,  5$.  The
$\mathbf{NS}$ ground state is massive, therefore there are no massless
bosons present in these sectors.   The $\mathbf{R}$ sectors consist of
only one degenerate component  $|s_2>$.  Taking the fermion number  to
be $F =   \frac{1}{2}+\sum_i s_i$, the two  possible   choices for the
$\mathsf{GSO}$ projections are the following:
\begin{equation} 
\label{gso3} \sum_{a}s_{a} = \frac{1}{2}
\qquad (\,\textrm{mod} \, 2\,)
\end{equation}
or
\begin{equation} 
\label{gso4} \sum_{a}s_{a} = -\frac{1}{2}
\qquad (\,\textrm{mod} \, 2\,).
\end{equation}
The contributions from the fermions to the $SU(n_{j})$ and $SU(u_{j})$
anomalies are:
\begin{equation} \label{39ga1}
A_{n_j}=   \frac{1}{N}\sum_{k=0}^{N-1}e^{-2\pi       i    kj/N}\   \{i
[ \prod_{r=3}^{5}  2  \sin  ( \pi k  v_r    ) ] Tr\gamma_{k,3}   +  Tr
\gamma_{k, 9}\}
\end{equation} 
and
\begin{equation} \label{39ga2}
A_{u_j}=       \frac{1}{N}\sum_{k=0}^{N-1}e^{-2\pi  i   kj/N}\     \{i
[ \prod_{r=3}^{5} 2  \sin   ( \pi k   v_r   ) ] Tr\gamma_{k,9}   +  Tr
\gamma_{k, 3}\}
\end{equation} 
respectively. Using (\ref{C1})\,(\ref{C5}) and (\ref{C9}), the tadpole
condition can be written as:
\begin{eqnarray} \label{39tcc} 
{\mathbf [\prod_{r=3}^{5}2\sin (   \pi kv_r) ]   Tr \gamma_{k,3} -  Tr
\gamma_{k,9}=0}.
\end{eqnarray} 
Clearly, the cancellation  of tadpoles does  not guarantee the absence
of $SU(n_j)$ or $SU(u_j)$ anomalies.

\subsection{D3-D5 brane orbifold}

We now  consider a system  with  a number  $n$ of  D3-branes, a number
$u^{(i)}$ of   D$5_i$-branes, a number $m$  of  D$\bar 3$-branes and a
number $w^{\bar{(i)}}$ of D${\bar 5}_{i}$  branes. The spectrum reads:
\\
\vspace{0.1cm}
\begin{center}\footnotesize{
\begin{tabular}{|c|c|c|c|c|}
\hline  \hline Sector &   Gauge  bosons &   Tachyonic scalar  fields &
Massless scalar fields & Fermion (s = - 1/2) \\\hline \hline $\bf{33}$
&      ${\bigotimes_{j=0}^{N-1}}         U(n_{j})$             &     &
$\sum_{r=3}^5\sum_{j=0}^{N-1}(n_j,   \bar{n}_{j+Nv_r})         $     &
$\sum_{j=0}^{N-1}[(n_j,          \bar{n}_j)+         \sum_{r=3}^5(n_j,
\bar{n}_{j+Nv_r})]         $        \\    $\bf{\bar{3}\bar{3}}$      &
${\bigotimes_{j=0}^{N-1}}              U(m_{j})$         &           &
$\sum_{r=3}^{5}\sum_{j=0}^{N-1}(m_j,        \bar{m}_{j+N   v_r}) $   &
$\sum_{j=0}^{N-1}[(m_j,         \bar{m}_j)+          \sum_{r=3}^5(m_j,
\bar{m}_{j+Nv_r})]   $ \\ $\bf{3\bar{3}}$   & & $\sum_{j=0}^{N-1}(n_j,
\bar{m}_{j})$      &         & $\sum_{j=0}^{N-1}[(n_j,   \bar{m}_{j})+
\sum_{r=3}^5(n_j, \bar{m}_{j-Nv_r})]   $    \\ $\bf{\bar{3}3}$  &    &
$\sum_{j=0}^{N-1}(m_j,    \bar{n}_{j})$  &  &  $\sum_{j=0}^{N-1}[(m_j,
\bar{n}_{j})+ \sum_{r=3}^5(m_j, \bar{n}_{j-Nv_r})] $ \\  $\bf{5_i5_i}$
&            ${\bigotimes_{j=0}^{N-1}}         U(u_{j}^{(i)})$       &
&$\sum_{r=3}^{5}\sum_{j=0}^{N-1}(u_j^{(i)},  \bar{u}_{j+Nv_r}^{(i)}) $
&          $\sum_{j=0}^{N-1}[(u_j^{(i)},             \bar{u}_j^{(i)})+
\sum_{r=3}^5(u_j^{(i)},          \bar{u}_{j+Nv_r}^{(i)})]          $\\
$\bf{\bar{5_i}\bar{5_i}}$       &            ${\bigotimes_{j=0}^{N-1}}
U(w_{j}^{\bar{(i)}})$   &   &$\sum_{r=3}^{5}\sum_{j=0}^{N-1}(w_j^{\bar
{      (i)}},    \bar{w}_{j+Nv_r}^{\bar       {  (i)}})       $      &
$\sum_{j=0}^{N-1}[(w_j^{\bar   {(i)}},     \bar{w}_j^{\bar    {(i)}})+
\sum_{r=3}^5(w_j^{\bar {  (i)}},  \bar{w}_{j+Nv_r}^{\bar {(i)}})] $ \\
$\bf{5_i\bar{5_i}}$ & & $\sum_{j=0}^{N-1}(u_j^{(i)}, \bar{w}_{j}^{\bar
{(i)}})$   & & $\sum_{j=0}^{N-1}[(u_j^{(i)}, \bar{w}_{j}^{\bar{(i)}})+
\sum_{r=3}^5     (u_{j}^{(i)},  \bar{w}_{j-Nv_{r}}^{\bar{(i)}})]  $ \\
$\bf{\bar{5_i}5_i}$      & &   $\sum_{j=0}^{N-1}(w_j^{({\bar     i})},
\bar{u}_{j}^{(i)})$ &    &      $\sum_{j=0}^{N-1}[(w_j^{({\bar   i})},
\bar{u}_{j}^{(i)})+       \sum_{r=3}^5         (w_{j}^{({\bar    i})},
\bar{u}_{j-Nv_{r}}^{(i)})]      $     \\  $\bf{5_i5_l}$        &     &
&$\sum_{j=0}^{N-1}(u_j^{(i)},  \bar{u}_{j-\frac{N}{2}v_m}^{(l)}) $   &
$\sum_{j=0}^{N-1}(u_j^{(i)}, \bar{u}_{j-\frac{N}{2}v_m}^{(l)})  $   \\
$\bf{5_l5_i}$     &                &     &$\sum_{j=0}^{N-1}(u_j^{(l)},
\bar{u}_{j-\frac{N}{2}v_m}^{(i)})  $  &   $\sum_{j=0}^{N-1}(u_j^{(l)},
\bar{u}_{j-\frac{N}{2}v_m}^{(i)}) $ \\ $\bf{\bar{5}_i\bar{5_l}}$ & & &
$\sum_{j=0}^{N-1}(w_j^{\bar{(i)}},
\bar{w}_{j-\frac{N}{2}v_m}^{\bar{(l)}})              $               &
$\sum_{j=0}^{N-1}(w_j^{\bar                                    {(i)}},
\bar{w}_{j-\frac{N}{2}v_m}^{\bar{(l)}}) $ \\ $\bf{\bar{5}_l\bar{5_i}}$
&           &            &          $\sum_{j=0}^{N-1}(w_j^{\bar{(l)}},
\bar{w}_{j-\frac{N}{2}v_m}^{\bar{(i)}})               $              &
$\sum_{j=0}^{N-1}(w_j^{\bar                                    {(l)}},
\bar{w}_{j-\frac{N}{2}v_m}^{\bar{(i)}}) $ \\ $\bf{5_i\bar{5_l}}$ & & &
$\sum_{j=0}^{N-1}(u_j^{(i)},          \bar{w}_{j+\frac{N}{2}(-v_i    +
v_l)}^{\bar{(l)}})$           &           $\sum_{j=0}^{N-1}(u_j^{(i)},
\bar{w}_{j+\frac{N}{2}v_m}^{\bar{(l)}})$  \\ $\bf{\bar{5_l}5_i}$ & & &
$\sum_{j=0}^{N-1}(w_j^{({\bar  l})},   \bar{u}_{j+\frac{N}{2}(-v_i   +
v_l)}^{(i)})$      &    $\sum_{j=0}^{N-1}(w_j^{({\bar            l})},
\bar{u}_{j+\frac{N}{2}v_m}^{(i)})$   \\ $\bf{\bar{5_i}5_l}$     & &  &
$\sum_{j=0}^{N-1}(w_j^{({\bar  i})},    \bar{u}_{j+\frac{N}{2}(-v_i  +
v_l)}^{(l)})$    &          $\sum_{j=0}^{N-1}(w_j^{({\bar        i})},
\bar{u}_{j+\frac{N}{2}v_m}^{(l)})$  \\   $\bf{5_l\bar{5_i}}$   &  &  &
$\sum_{j=0}^{N-1}(u_j^{(l)},       \bar{w}_{j+\frac{N}{2}(-v_i       +
v_l)}^{\bar{(i)}})$            &          $\sum_{j=0}^{N-1}(u_j^{(l)},
\bar{w}_{j+\frac{N}{2}v_m}^{\bar{(i)}})$    \\   $\bf{35_i}$   &     &
$\sum_{j=0}^{N-1}(n_j,    \bar{u}_{j+\frac{N}{2}v_i}^{(i)})    $     &
&$\sum_{j=0}^{N-1}[(n_j,\bar{u}_{j+\frac{N}{2}v_i}^{(i)})+
(n_j,\bar{u}_{j-\frac{N}{2}v_i}^{(i)})]   $   \\   $\bf{5_i3}$     & &
$\sum_{j=0}^{N-1}(u_j^{i},\bar{n}_{j+\frac{N}{2}v_i})        $       &
&$\sum_{j=0}^{N-1}[(u_j^{(i)}, \bar{n}_{j+\frac{N}{2}v_i})+(u_j^{(i)},
\bar{n}_{j-\frac{N}{2}v_i})]  $   \\ $\bf{\bar{3}\bar{5}_i}$   &     &
$\sum_{j=0}^{N-1}(m_j,   \bar{w}_{j+\frac{N}{2}v_i}^{\bar{(i)}}) $&  &
$\sum_{j=0}^{N-1}[(m_j,        \bar{w}_{j+\frac{N}{2}v_i}^{(i)})+(m_j,
\bar{w}_{j-\frac{N}{2}v_i}^{(i)})] $   \\ $\bf{\bar{5}_i\bar{3}}$ &  &
$\sum_{j=0}^{N-1}(w_j^{({\bar i)}}, \bar{m}_{j+\frac{N}{2}v_i})$&    &
$\sum_{j=0}^{N-1}[(w_j^{(i)},  \bar{m}_{j+\frac{N}{2}v_i})+(w_j^{(i)},
\bar{m}_{j-\frac{N}{2}v_i})]     $  \\    $\bf{3{\bar   5}_i}$       &
&$\sum_{j=0}^{N-1}(n_j,\bar{w}_{j-\frac{N}{2}v_i}^{\bar{(i)}}) $ &   &
$\sum_{j=0}^{N-1}[(n_j,\bar{w}_{j+
\frac{N}{2}(v_l-v_m)}^{\bar{(i)}})+(n_j,\bar{w}_{j-
\frac{N}{2}(v_l-v_m)}^{\bar{(i)}})]   $   \\    $\bf{{\bar   5}_i3}$ &
&$\sum_{j=0}^{N-1}(w_j^{({\bar i)}},  \bar{n}_{j-\frac{N}{2}v_i})$   &
&$\sum_{j=0}^{N-1}[(w_j^{({\bar                                  i)}},
\bar{n}_{j+\frac{N}{2}(v_l-v_m)})+(w_j^{({\bar                   i)}},
\bar{n}_{j-\frac{N}{2}(v_l-v_m)})]    $  \\   $\bf{\bar{3}5_i}$ &    &
$\sum_{j=0}^{N-1}(m_j,   \bar{u}_{j-\frac{N}{2}v_i}^{(i)})   $       &
&$\sum_{j=0}^{N-1}[(m_j, \bar{u}_{j+\frac{N}{2}(v_l-v_m)}^{(i)})+(m_j,
\bar{u}_{j-\frac{N}{2}(v_l-v_m)}^{(i)})]  $ \\  $\bf{5_i\bar{3}}$    &
&$\sum_{j=0}^{N-1}(u_j^{i},\bar{m}_{j-\frac{N}{2}v_i})$              &
&$\sum_{j=0}^{N-1}[(u_j^{i},\bar{m}_{j+\frac{N}{2}(v_l-v_m)})+
(u_j^{i},\bar{m}_{j+\frac{N}{2}(v_l-v_m)})] $ \\ \hline \hline
\end{tabular}}
\end{center}
\begin{center}
\bf{Table 4: Spectrum in the $\mathbf{35}$ \bf configuration}
\end{center}
\vspace{0.1cm}
The contribution from     the fermionic  states   towards  the   cubic
$SU(n_{j})$ and $SU(u^{(i)}_{j})$ anomalies are
\begin{eqnarray} \label{35ga}
&     &    A_{n_j}=   \frac{2i}{N}\sum_{k=0}^{N-1}e^{-2\pi   i   kj/N}
[ \prod_{r=3}^{5}  2 \sin  ( \pi k  v_r  )  ]  Tr\gamma_{k,3} \\  &  &
A_{u_j^{(i)}}=        \frac{2i}{N}\sum_{k=0}^{N-1}e^{-2\pi           i
kj/N}\{[\prod_{r=3}^{5} 2 \sin  ( \pi k v_r  )  ] Tr\gamma_{k,5_i  } +
\!\!\!\!  \sum_{l,m \neq i} [ 2 \sin( \pi k v_m)] Tr\gamma_{k, 5_l}\}
\end{eqnarray}
respectively and the tadpole cancellation conditions:
\begin{eqnarray} \label{35tcc} {\mathbf Tr\gamma_{k,3}-\sum_{i=3}^{5}
\frac{Tr\gamma_{k,5_i}}{(2\sin \pi kv_i)}=0} \end{eqnarray}
where we  have used (\ref{C2})\,(\ref{C3})\,(\ref{C5}) and (\ref{C13})
for its computation.  Again,  in this system, cancellation of tadpoles
does    not  guarantee  absence  of gauge     anomalies.   It  too  is
non-supersymmetric.

\subsection{D9-D7 brane orbifold}

We now consider  a  system with a  number  $n$ of D9-branes,  a number
$u^{(i)}$ of  D$7_i$-branes, a  number $m$ of   D$\bar 9$-branes and a
number  $w^{\bar{(i)}}$    of    D${\bar  7}_{i}$    branes.       The
(non-supersymmetric) spectrum reads: \\
\vspace{0.1cm}
\begin{center}\footnotesize{
\begin{tabular}{|c|c|c|c|c|}
\hline \hline  Sector  &  Gauge bosons   & Tachyonic  scalar fields  &
Massless scalar fields & Fermion (s = - 1/2) \\\hline \hline $\bf{99}$
&    ${\bigotimes_{j=0}^{N-1}}        U(n_{j})$              &       &
$\sum_{r=3}^5\sum_{j=0}^{N-1}(n_j,     \bar{n}_{j+Nv_r})       $     &
$\sum_{j=0}^{N-1}[(n_j,            \bar{n}_j)+       \sum_{r=3}^5(n_j,
\bar{n}_{j+Nv_r})]           $     \\      $\bf{\bar{9}\bar{9}}$     &
${\bigotimes_{j=0}^{N-1}}        U(m_{j})$             &             &
$\sum_{r=3}^{5}\sum_{j=0}^{N-1}(m_j, \bar{m}_{j+N    v_r})        $  &
$\sum_{j=0}^{N-1}[(m_j,         \bar{m}_j)+          \sum_{r=3}^5(m_j,
\bar{m}_{j+Nv_r})]  $  \\ $\bf{9\bar{9}}$  & &  $\sum_{j=0}^{N-1}(n_j,
\bar{m}_{j})$     &     &     $\sum_{j=0}^{N-1}[(n_j,    \bar{m}_{j})+
\sum_{r=3}^5(n_j,   \bar{m}_{j-Nv_r})] $   \\  $\bf{\bar{9}9}$  &    &
$\sum_{j=0}^{N-1}(m_j, \bar{n}_{j})$ &    &    $\sum_{j=0}^{N-1}[(m_j,
\bar{n}_{j})+ \sum_{r=3}^5(m_j,  \bar{n}_{j-Nv_r})] $ \\ $\bf{7_i7_i}$
&       ${\bigotimes_{j=0}^{N-1}}            U(u_{j}^{(i)})$         &
&$\sum_{r=3}^{5}\sum_{j=0}^{N-1}(u_j^{(i)}, \bar{u}_{j+Nv_r}^{(i)})  $
&             $\sum_{j=0}^{N-1}[(u_j^{(i)},          \bar{u}_j^{(i)})+
\sum_{r=3}^5(u_j^{(i)},           \bar{u}_{j+Nv_r}^{(i)})]         $\\
$\bf{\bar{7_i}\bar{7_i}}$            &       ${\bigotimes_{j=0}^{N-1}}
U(w_{j}^{\bar{(i)}})$  &    &$\sum_{r=3}^{5}\sum_{j=0}^{N-1}(w_j^{\bar
{       (i)}},      \bar{w}_{j+Nv_r}^{\bar   {      (i)}})     $     &
$\sum_{j=0}^{N-1}[(w_j^{\bar   {(i)}},       \bar{w}_j^{\bar  {(i)}})+
\sum_{r=3}^5(w_j^{\bar { (i)}},  \bar{w}_{j+Nv_r}^{\bar {(i)}})]  $ \\
$\bf{7_i\bar{7_i}}$ & & $\sum_{j=0}^{N-1}(u_j^{(i)}, \bar{w}_{j}^{\bar
{(i)}})$  &  & $\sum_{j=0}^{N-1}[(u_j^{(i)}, \bar{w}_{j}^{\bar{(i)}})+
\sum_{r=3}^5 (u_{j}^{(i)}, \bar{w}_{j-Nv_{r}}^{\bar{(i)}})]   $     \\
$\bf{\bar{7_i}7_i}$  &  &    $\sum_{j=0}^{N-1}(w_j^{({\bar       i})},
\bar{u}_{j}^{(i)})$     &  &  $\sum_{j=0}^{N-1}[(w_j^{({\bar     i})},
\bar{u}_{j}^{(i)})+        \sum_{r=3}^5       (w_{j}^{({\bar     i})},
\bar{u}_{j-Nv_{r}}^{(i)})]     $    \\         $\bf{7_i7_l}$   &     &
&$\sum_{j=0}^{N-1}(u_j^{(i)},  \bar{u}_{j-\frac{N}{2}v_m}^{(l)})  $  &
$\sum_{j=0}^{N-1}(u_j^{(i)},   \bar{u}_{j-\frac{N}{2}v_m}^{(l)}) $  \\
$\bf{7_l7_i}$          &       &         &$\sum_{j=0}^{N-1}(u_j^{(l)},
\bar{u}_{j-\frac{N}{2}v_m}^{(i)})   $  &  $\sum_{j=0}^{N-1}(u_j^{(l)},
\bar{u}_{j-\frac{N}{2}v_m}^{(i)}) $ \\ $\bf{\bar{7}_i\bar{7_l}}$ & & &
$\sum_{j=0}^{N-1}(w_j^{\bar{(i)}},
\bar{w}_{j-\frac{N}{2}v_m}^{\bar{(l)}})                  $           &
$\sum_{j=0}^{N-1}(w_j^{\bar                                    {(i)}},
\bar{w}_{j-\frac{N}{2}v_m}^{\bar{(l)}}) $ \\ $\bf{\bar{7}_l\bar{7_i}}$
&       &          &                $\sum_{j=0}^{N-1}(w_j^{\bar{(l)}},
\bar{w}_{j-\frac{N}{2}v_m}^{\bar{(i)}})             $                &
$\sum_{j=0}^{N-1}(w_j^{\bar                                    {(l)}},
\bar{w}_{j-\frac{N}{2}v_m}^{\bar{(i)}}) $ \\ $\bf{7_i\bar{7_l}}$ & & &
$\sum_{j=0}^{N-1}(u_j^{(i)},        \bar{w}_{j+\frac{N}{2}(-v_i      +
v_l)}^{\bar{(l)}})$            &          $\sum_{j=0}^{N-1}(u_j^{(i)},
\bar{w}_{j+\frac{N}{2}v_m}^{\bar{(l)}})$  \\ $\bf{\bar{7_l}7_i}$ & & &
$\sum_{j=0}^{N-1}(w_j^{({\bar    l})}, \bar{u}_{j+\frac{N}{2}(-v_i   +
v_l)}^{(i)})$     &         $\sum_{j=0}^{N-1}(w_j^{({\bar        l})},
\bar{u}_{j+\frac{N}{2}v_m}^{(i)})$   \\  $\bf{\bar{7_i}7_l}$  &  &   &
$\sum_{j=0}^{N-1}(w_j^{({\bar  i})},  \bar{u}_{j+\frac{N}{2}(-v_i    +
v_l)}^{(l)})$        &        $\sum_{j=0}^{N-1}(w_j^{({\bar      i})},
\bar{u}_{j+\frac{N}{2}v_m}^{(l)})$ \\  $\bf{7_l\bar{7_i}}$  &  &     &
$\sum_{j=0}^{N-1}(u_j^{(l)},        \bar{w}_{j+\frac{N}{2}(-v_i      +
v_l)}^{\bar{(i)}})$           &           $\sum_{j=0}^{N-1}(u_j^{(l)},
\bar{w}_{j+\frac{N}{2}v_m}^{\bar{(i)}})$   \\   $\bf{97_i}$     &    &
$\sum_{j=0}^{N-1}(n_j,    \bar{u}_{j+\frac{N}{2}v_i}^{(i)}) $    &   &
$\sum_{j=0}^{N-1}[(n_j,\bar{u}_{j+\frac{N}{2}v_i}^{(i)})+
(n_j,\bar{u}_{j-\frac{N}{2}v_i}^{(i)})] $   \\  $\bf{7_i9}$   &      &
$\sum_{j=0}^{N-1}(u_j^{i},\bar{n}_{j+\frac{N}{2}v_i})    $   &       &
$\sum_{j=0}^{N-1}[(u_j^{(i)},  \bar{n}_{j+\frac{N}{2}v_i})+(u_j^{(i)},
\bar{n}_{j-\frac{N}{2}v_i})]    $   \\   $\bf{\bar{9}\bar{7}_i}$ &   &
$\sum_{j=0}^{N-1}(m_j,   \bar{w}_{j+\frac{N}{2}v_i}^{\bar{(i)}})  $& &
$\sum_{j=0}^{N-1}[(m_j,        \bar{w}_{j+\frac{N}{2}v_i}^{(i)})+(m_j,
\bar{w}_{j-\frac{N}{2}v_i}^{(i)})]  $  \\  $\bf{\bar{7}_i\bar{9}}$ & &
$\sum_{j=0}^{N-1}(w_j^{({\bar   i)}},  \bar{m}_{j+\frac{N}{2}v_i})$& &
$\sum_{j=0}^{N-1}[(w_j^{(i)},  \bar{m}_{j+\frac{N}{2}v_i})+(w_j^{(i)},
\bar{m}_{j-\frac{N}{2}v_i})]      $     \\$\bf{9{\bar    7}_i}$      &
&$\sum_{j=0}^{N-1}(n_j,\bar{w}_{j-\frac{N}{2}v_i}^{\bar{(i)}}) $  &  &
$\sum_{j=0}^{N-1}[(n_j,\bar{w}_{j+
\frac{N}{2}(v_l-v_m)}^{\bar{(i)}})+(n_j,\bar{w}_{j-
\frac{N}{2}(v_l-v_m)}^{\bar{(i)}})]    $   \\  $\bf{{\bar   7}_i9}$  &
&$\sum_{j=0}^{N-1}(w_j^{({\bar  i)}},  \bar{n}_{j-\frac{N}{2}v_i})$  &
&$\sum_{j=0}^{N-1}[(w_j^{({\bar                                  i)}},
\bar{n}_{j+\frac{N}{2}(v_l-v_m)})         +(w_j^{({\bar          i)}},
\bar{n}_{j-\frac{N}{2}(v_l-v_m)})]  $  \\  $\bf{\bar{9}7_i}$    &    &
$\sum_{j=0}^{N-1}(m_j,   \bar{u}_{j-\frac{N}{2}v_i}^{(i)})   $       &
&$\sum_{j=0}^{N-1}[(m_j,  \bar{u}_{j+\frac{N}{2}(v_l-v_m)}^{(i)})    +
(m_j, \bar{u}_{j-\frac{N}{2}(v_l-v_m)}^{(i)})] $ \\ $ \bf{7_i\bar{9}}$
&   & $  \sum_{j=0}^{N-1}(u_j^{i},\bar{m}_{j-\frac{N}{2}v_i})$  & &  $
\sum_{j=0}^{N-1}[(u_j^{i},\bar{m}_{j+\frac{N}{2}(v_l-v_m)})
+(u_j^{i},\bar{m}_{j-\frac{N}{2}(v_l-v_m)})] $ \\ \hline \hline
\end{tabular}}
\end{center}
\begin{center}
\bf{Table 5: Spectrum in the $\mathbf{97}$ \bf configuration}
\end{center}
\vspace{0.1cm}
From the above spectra, we can easily compute the contributions to the
cubic $SU(n_{j})$ and $SU(u^{(i)}_{j})$ anomalies:
\begin{eqnarray} \label{97ga}
&   &        A_{n_j}= \frac{2i}{N}\sum_{k=0}^{N-1}e^{-2\pi   i   kj/N}
[ \prod_{r=3}^{5}  2 \sin  ( \pi  k  v_r  ) ]  Tr\gamma_{k,9}  \\  & &
A_{u_j^{(i)}}=          \frac{2i}{N}\sum_{k=0}^{N-1}e^{-2\pi         i
kj/N}\{[\prod_{r=3}^{5} 2 \sin  ( \pi  k v_r  ) ] Tr\gamma_{k,7_i  } +
\!\!\!\!  \sum_{l,m \neq i} [ 2 \sin( \pi k v_m)] Tr\gamma_{k, 7_l}\}.
\end{eqnarray}
The twisted ${\mathbf RR}$ tadpole cancellation condition:
\begin{eqnarray} \label{97tcc} 
{\mathbf   Tr  \gamma_{k,9}-\sum_{i=3}^{5}   [   2\sin (   \pi  kv_i)]
Tr\gamma_{k,7_i}=0}
\end{eqnarray}
does not guarantee the absence  of either $SU(n_j)$ or $SU(u_j^{(i)})$
anomalies.

\subsection{D5-D7 brane orbifold}

We now consider a system  with a number  $n^{(i)}$ of D$5_i$-branes, a
number $u^{(i)}$ of D$7_i$-branes, a number $m^{\bar{(i)}}$ of D${\bar
5}_{i}$   branes and   a  number  $w^{\bar{(i)}}$   of D${\bar 7}_{i}$
branes. The spectrum reads: \\
\vspace{0.1cm}
\begin{center}\tiny{
\hspace{-10.5mm}\begin{tabular}{|c|c|c|c|c|c|} \hline \hline Sector  &
Gauge bosons & Tachyons  & Massless scalar fields  & $Fermion_{-} $ \\
\hline      \hline    $\bf{5_i5_i}$     &    ${\bigotimes_{j=0}^{N-1}}
U(n_{j}^{(i)})$     &      &$\sum_{r=3}^{5}\sum_{j=0}^{N-1}(n_j^{(i)},
\bar{n}_{j+Nv_r}^{(i)})        $     &   $\sum_{j=0}^{N-1}[(n_j^{(i)},
\bar{n}_j^{(i)})+ \sum_{r=3}^5(n_j^{(i)}, \bar{n}_{j+Nv_r}^{(i)})] $\\
$\bf{\bar{5_i}\bar{5_i}}$        &           ${\bigotimes_{j=0}^{N-1}}
U(m_{j}^{\bar{(i)}})$   &   &$\sum_{r=3}^{5}\sum_{j=0}^{N-1}(m_j^{\bar
{       (i)}},     \bar{m}_{j+Nv_r}^{\bar       {   (i)}})     $     &
$\sum_{j=0}^{N-1}[(m_j^{\bar   {(i)}},  \bar{m}_j^{\bar       {(i)}})+
\sum_{r=3}^5(m_j^{\bar  { (i)}}, \bar{m}_{j+Nv_r}^{\bar  {(i)}})] $ \\
$\bf{5_i\bar{5_i}}$ & & $\sum_{j=0}^{N-1}(n_j^{(i)}, \bar{m}_{j}^{\bar
{(i)}})$ &  &  $\sum_{j=0}^{N-1}[(n_j^{(i)}, \bar{m}_{j}^{\bar{(i)}})+
\sum_{r=3}^5   (n_{j}^{(i)},  \bar{m}_{j-Nv_{r}}^{\bar{(i)}})] $    \\
$\bf{\bar{5_i}5_i}$   &    &      $\sum_{j=0}^{N-1}(m_j^{({\bar  i})},
\bar{n}_{j}^{(i)})$    &     &  $\sum_{j=0}^{N-1}[(m_j^{({\bar   i})},
\bar{n}_{j}^{(i)})+     \sum_{r=3}^5        (m_{j}^{({\bar       i})},
\bar{n}_{j-Nv_{r}}^{(i)})]        $    \\     $\bf{5_i5_l}$       &  &
&$\sum_{j=0}^{N-1}(n_j^{(i)}, \bar{n}_{j-\frac{N}{2}v_m}^{(l)})   $  &
$\sum_{j=0}^{N-1}(n_j^{(i)},  \bar{n}_{j-\frac{N}{2}v_m}^{(l)})  $  \\
$\bf{5_l5_i}$         &        &         &$\sum_{j=0}^{N-1}(n_j^{(l)},
\bar{n}_{j-\frac{N}{2}v_m}^{(i)})   $  &  $\sum_{j=0}^{N-1}(n_j^{(l)},
\bar{n}_{j-\frac{N}{2}v_m}^{(i)}) $ \\ $\bf{\bar{5}_i\bar{5_l}}$ & & &
$\sum_{j=0}^{N-1}(m_j^{\bar{(i)}},
\bar{m}_{j-\frac{N}{2}v_m}^{\bar{(l)}})             $                &
$\sum_{j=0}^{N-1}(m_j^{\bar                                    {(i)}},
\bar{m}_{j-\frac{N}{2}v_m}^{\bar{(l)}}) $ \\ $\bf{\bar{5}_l\bar{5_i}}$
&            &         &            $\sum_{j=0}^{N-1}(m_j^{\bar{(l)}},
\bar{m}_{j-\frac{N}{2}v_m}^{\bar{(i)}})                $             &
$\sum_{j=0}^{N-1}(m_j^{\bar                                    {(l)}},
\bar{m}_{j-\frac{N}{2}v_m}^{\bar{(i)}}) $ \\ $\bf{5_i\bar{5_l}}$ & & &
$\sum_{j=0}^{N-1}(n_j^{(i)}       \bar{m}_{j+\frac{N}{2}(-v_i        +
v_l)}^{\bar{(l)}})$            &          $\sum_{j=0}^{N-1}(n_j^{(i)},
\bar{m}_{j+\frac{N}{2}v_m}^{\bar{(l)}})$  \\ $\bf{\bar{5_l}5_i}$ & & &
$\sum_{j=0}^{N-1}(m_j^{({\bar  l})},    \bar{n}_{j+\frac{N}{2}(-v_i  +
v_l)}^{(i)})$         &     $\sum_{j=0}^{N-1}(m_j^{({\bar        l})},
\bar{n}_{j+\frac{N}{2}v_m}^{(i)})$   \\   $\bf{\bar{5_i}5_l}$   &  & &
$\sum_{j=0}^{N-1}(m_j^{({\bar   i})}, \bar{n}_{j+\frac{N}{2}(-v_i    +
v_l)}^{(l)})$        &        $\sum_{j=0}^{N-1}(m_j^{({\bar      i})},
\bar{n}_{j+\frac{N}{2}v_m}^{(l)})$  \\     $\bf{5_l\bar{5_i}}$ &  &  &
$\sum_{j=0}^{N-1}(n_j^{(l)},        \bar{m}_{j+\frac{N}{2}(-v_i      +
v_l)}^{\bar{(i)}})$            &          $\sum_{j=0}^{N-1}(n_j^{(l)},
\bar{m}_{j+\frac{N}{2}v_m}^{\bar{(i)}})$    \\   $\bf{7_i7_i}$       &
${\bigotimes_{j=0}^{N-1}}                U(u_{j}^{(i)})$             &
&$\sum_{r=3}^{5}\sum_{j=0}^{N-1}(u_j^{(i)}, \bar{u}_{j+Nv_r}^{(i)})  $
&            $\sum_{j=0}^{N-1}[(u_j^{(i)},           \bar{u}_j^{(i)})+
\sum_{r=3}^5(u_j^{(i)},       \bar{u}_{j+Nv_r}^{(i)})]         $    \\
$\bf{\bar{7_i}\bar{7_i}}$       &            ${\bigotimes_{j=0}^{N-1}}
U(w_{j}^{\bar{(i)}})$  &    &$\sum_{r=3}^{5}\sum_{j=0}^{N-1}(w_j^{\bar
{     (i)}},      \bar{w}_{j+Nv_r}^{\bar     {      (i)}})      $    &
$\sum_{j=0}^{N-1}[(w_j^{\bar    {(i)}},   \bar{w}_j^{\bar     {(i)}})+
\sum_{r=3}^5(w_j^{\bar { (i)}},  \bar{w}_{j+Nv_r}^{\bar  {(i)}})] $ \\
$\bf{7_i\bar{7_i}}$ & & $\sum_{j=0}^{N-1}(u_j^{(i)}, \bar{w}_{j}^{\bar
{(i)}})$ &  &  $\sum_{j=0}^{N-1}[(u_j^{(i)}, \bar{w}_{j}^{\bar{(i)}})+
\sum_{r=3}^5      (u_{j}^{(i)}, \bar{w}_{j-Nv_{r}}^{\bar{(i)}})] $  \\
$\bf{\bar{7_i}7_i}$     &  &  $\sum_{j=0}^{N-1}(w_j^{({\bar      i})},
\bar{u}_{j}^{(i)})$ &      &      $\sum_{j=0}^{N-1}[(w_j^{({\bar i})},
\bar{u}_{j}^{(i)})+        \sum_{r=3}^5    (w_{j}^{({\bar        i})},
\bar{u}_{j-Nv_{r}}^{(i)})]       $  \\       $\bf{7_i7_l}$    &      &
&$\sum_{j=0}^{N-1}(u_j^{(i)}, \bar{u}_{j-\frac{N}{2}v_m}^{(l)})  $   &
$\sum_{j=0}^{N-1}(u_j^{(i)},  \bar{u}_{j-\frac{N}{2}v_m}^{(l)}) $   \\
$\bf{7_l7_i}$         &        &         &$\sum_{j=0}^{N-1}(u_j^{(l)},
\bar{u}_{j-\frac{N}{2}v_m}^{(i)})  $  &   $\sum_{j=0}^{N-1}(u_j^{(l)},
\bar{u}_{j-\frac{N}{2}v_m}^{(i)}) $ \\ $\bf{\bar{7}_i\bar{7_l}}$ & & &
$\sum_{j=0}^{N-1}(w_j^{\bar{(i)}},
\bar{w}_{j-\frac{N}{2}v_m}^{\bar{(l)}})              $               &
$\sum_{j=0}^{N-1}(w_j^{\bar                                    {(i)}},
\bar{w}_{j-\frac{N}{2}v_m}^{\bar{(l)}}) $ \\ $\bf{\bar{7}_l\bar{7_i}}$
&          &             &          $\sum_{j=0}^{N-1}(w_j^{\bar{(l)}},
\bar{w}_{j-\frac{N}{2}v_m}^{\bar{(i)}})              $               &
$\sum_{j=0}^{N-1}(w_j^{\bar                                    {(l)}},
\bar{w}_{j-\frac{N}{2}v_m}^{\bar{(i)}}) $ \\ $\bf{7_i\bar{7_l}}$ & & &
$\sum_{j=0}^{N-1}(u_j^{(i)},       \bar{w}_{j+\frac{N}{2}(-v_i       +
v_l)}^{\bar{(l)}})$             &         $\sum_{j=0}^{N-1}(u_j^{(i)},
\bar{w}_{j+\frac{N}{2}v_m}^{\bar{(l)}})$  \\ $\bf{\bar{7_l}7_i}$ & & &
$\sum_{j=0}^{N-1}(w_j^{({\bar    l})},  \bar{u}_{j+\frac{N}{2}(-v_i  +
v_l)}^{(i)})$         &      $\sum_{j=0}^{N-1}(w_j^{({\bar       l})},
\bar{u}_{j+\frac{N}{2}v_m}^{(i)})$   \\    $\bf{\bar{7_i}7_l}$  & &  &
$\sum_{j=0}^{N-1}(w_j^{({\bar     i})},  \bar{u}_{j+\frac{N}{2}(-v_i +
v_l)}^{(l)})$           &    $\sum_{j=0}^{N-1}(w_j^{({\bar       i})},
\bar{u}_{j+\frac{N}{2}v_m}^{(l)})$   \\   $\bf{7_l\bar{7_i}}$    & & &
$\sum_{j=0}^{N-1}(u_j^{(l)},        \bar{w}_{j+\frac{N}{2}(-v_i      +
v_l)}^{\bar{(i)}})$            &          $\sum_{j=0}^{N-1}(u_j^{(l)},
\bar{w}_{j+\frac{N}{2}v_m}^{\bar{(i)}})$  \\  $\bf{5_i7_i}$  &   &   &
&$\sum_{j=0}^{N-1}(n_j^{(i)}, \bar{u}^{(i)}_{j})$ \\ $\bf{7_i5_i}$ & &
&    &       -     \\$\bf{\bar{5_i}\bar{7_i}}$     &    &       &    &
$\sum_{j=0}^{N-1}(m_j^{\bar{(i)}},    \bar{w}^{\bar{(i)}}_{j})$     \\
$\bf{\bar{7_i}\bar{5_i}}$ & & & &-\\ $\bf{5_i\bar{7_i}}$ &  & & & - \\
$\bf{\bar{7_i}5_i}$ &       &    & &$\sum_{j=0}^{N-1}(w_j^{\bar{(i)}},
\bar{n}^{(i)}_{j})$ \\   $\bf{\bar{5_i}7_i}$   &    &   &   & -     \\
$\bf{7_i\bar{5_i}}$    &    &    &     &  $\sum_{j=0}^{N-1}(u_j^{(i)},
\bar{m}^{\bar{(i)}}_{j})$      \\    $\bf{5_i7_l}$           &       &
$\sum_{j=0}^{N-1}(n_j^{(i)}, \bar{u}^{(l)}_{j+\frac{N}{2}v_m} )  $&  &
$\sum_{j=0}^{N-1}[(n_j^{(i)},    \bar{u}^{(l)}_{j+\frac{N}{2}v_m}   )+
(n_j^{(i)}, \bar{u}^{(l)}_{j-\frac{N}{2}v_m})]$   \\ $\bf{7_l5_i}$ & &
$\sum_{j=0}^{N-1}(u_j^{(l)}, \bar{n}^{(i)}_{j+\frac{N}{2}v_m} ) $    &
&$\sum_{j=0}^{N-1}[(u_j^{(l)},        \bar{n}^{(i)}_{j+\frac{N}{2}v_m}
)+(u_j^{(l)},    \bar{n}^{(i)}_{j-\frac{N}{2}v_m     })]   $        \\
$\bf{\bar{5_i}\bar{7_l}}$    & &    $\sum_{j=0}^{N-1}(m_j^{\bar{(i)}},
\bar{w}^{\bar{(l)}}_{j+\frac{N}{2}v_m}  )  $               &         &
$\sum_{j=0}^{N-1}[(m_j^{\bar{(i)}},
\bar{w}^{\bar{(l)}}_{j+\frac{N}{2}v_m}             )+(m_j^{\bar{(i)}},
\bar{w}^{\bar{(l)}}_{j-\frac{N}{2}v_m            })]       $        \\
$\bf{\bar{7_l}\bar{5_i}}$  &   &    $\sum_{j=0}^{N-1}(w_j^{\bar{(l)}},
\bar{m}^{\bar{(i)}}_{j+\frac{N}{2}v_m}        )        $       &     &
$\sum_{j=0}^{N-1}[(w_j^{\bar{(l)}},
\bar{m}^{\bar{(i)}}_{j+\frac{N}{2}v_m}             )+(w_j^{\bar{(l)}},
\bar{m}^{\bar{(i)}}_{j-\frac{N}{2}v_m})] $ \\    $\bf{5_i\bar{7_l}}$ &
&$\sum_{j=0}^{N-1}(n_j^{(i)}, \bar{w}^{\bar{(l)}}_{j-\frac{N}{2}v_m} )
$                      &                &$\sum_{j=0}^{N-1}[(n_j^{(i)},
\bar{w}^{\bar{(l)}}_{j+\frac{N}{2}(v_i-v_l)}             )+(n_j^{(i)},
\bar{w}^{\bar{(l)}}_{j-\frac{N}{2}(v_i-v_l)}       )]       $       \\
$\bf{\bar{7_l}5_i}$        &       &$\sum_{j=0}^{N-1}(w_j^{\bar{(l)}},
\bar{n}^{(i)}_{j-\frac{N}{2}v_m}           )               $         &
&$\sum_{j=0}^{N-1}[(w_j^{\bar{(l)}},
\bar{n}^{(i)}_{j+\frac{N}{2}(v_i-v_l)}             )+(w_j^{\bar{(l)}},
\bar{n}^{(i)}_{j-\frac{N}{2}(v_i-v_l)} )]   $ \\ $\bf{\bar{5_i}7_l}$ &
&$\sum_{j=0}^{N-1}(m_j^{\bar{(i)}}, \bar{u}^{(l)}_{j-\frac{N}{2}v_m} )
$            &           &         $\sum_{j=0}^{N-1}[(m_j^{\bar{(i)}},
\bar{u}^{(l)}_{j+\frac{N}{2}(v_i-v_l)}             )+(m_j^{\bar{(i)}},
\bar{u}^{(l)}_{j-\frac{N}{4}(v_i-v_l)}  )] $ \\   $\bf{7_l\bar{5_i}}$&
&$\sum_{j=0}^{N-1}(u_j^{(l)}, \bar{m}^{\bar{(i)}}_{j-\frac{N}{2}v_m} )
$                   &                   &$\sum_{j=0}^{N-1}[(u_j^{(l)},
\bar{m}^{\bar{(i)}}_{j+\frac{N}{2}(v_i-v_l)}             )+(u_j^{(l)},
\bar{m}^{\bar{(i)}}_{j-\frac{N}{2}(v_i-v_l)} )] $ \\ \hline \hline
\end{tabular}}
\end{center}
\begin{center}
\bf{Table 6: Spectrum in the $\mathbf{57}$ \bf configuration}
\end{center}
\vspace{0.1cm}
The  contribution   to the  $SU(n_{j})$   and  $SU(u^{(i)}_{j})$ cubic
anomalies is:
\begin{eqnarray}\label{57ga1}
A(n_j^{(i)})=                      \frac{1}{N}\sum_{k=0}^{N-1}e^{-2\pi
ikj/N}\{i[\prod_{r=3}^{5}2\sin   (\pi  kv_r)]  Tr \gamma_{k,   5_i}+ i
\!\!\!  \sum_{l,  m  \not=  i}[2\sin(\pi kv_m)]Tr\gamma_{k,  5_l}+  Tr
\gamma_{k, 7_i}\}
\end{eqnarray}
\begin{eqnarray}\label{57ga2}
A(u_j^{(i)})=                      \frac{1}{N}\sum_{k=0}^{N-1}e^{-2\pi
ikj/N}\{i[\prod_{r=3}^{5}2\sin (\pi  kv_r)]    Tr \gamma_{k,  7_i}+  i
\!\!\!  \sum_{l,   m \not=  i}[2\sin(\pi  kv_m)]Tr\gamma_{k, 7_l}-  Tr
\gamma_{k,5_i}\}
\end{eqnarray}
respectively.   Using (\ref{C2})\,(\ref{C3})\,(\ref{C6})\, (\ref{C11})
and (\ref{C12}), we obtain  the  following expression for  the tadpole
cancellation conditions:
\begin{equation} \label{57tcc}
{\mathbf[ \prod_{r=3}^{5}  2   \sin (  \pi  kv_r )    ] \sum_{i=3}^{5}
\frac{Tr\gamma_{k,5_i}}{2\sin ( \pi kv_i)}  - \sum_{i=3}^{5} [ 2\sin (
\pi kv_i) ] Tr\gamma_{k,7_i}=0},
\end{equation}
which  does  not     guarantee  the absence  of    $SU(n_j^{(i)})$  or
$SU(_j^{(i)})$ gauge anomalies.

\subsection{The most general D-brane orbifold}

In the most general case, tadpole cancellation conditions read:
\begin{equation}
[    \prod_{j=3}^{5} 2   \sin (\pi   kv_j)   ] \{ Tr\gamma_{k,3} -
\sum_{i=3}^{5}  \frac{Tr\gamma_{k,5_i}}  { 2  \sin (\pi kv_i)} \}- 
\{ Tr\gamma_{k,9} - \sum_{i=3}^{5} [ 2 \sin ( \pi kv_i)] Tr
\gamma_{k,7_i} \} =0.
\end{equation}

\section{Conclusions}

For   many  years   the   attempts   to   construct phenomenologically
semi-realistic   models from    String  Theory   mostly  focussed   on
supersymmetric models \cite{Al1}  - \cite{Leigh}  \cite{Al4},  despite
the  fact that  no  supersymmetric particles have  ever been observed.
The reason for this is that the  weakly coupled Heterotic String has a
string scale close to  the Planck scale  and supersymmetry is the only
known  method of evading the hierarchy  problem.  The discovery of the
D-brane  world,  in  which    gauge theories   may  inhabit a    lower
dimensionality,   with    gravitational     interaction     in     the
(higher-dimensional) bulk,  has    led to the   construction  of  new,
non-supersymmetric models  with intermediate \cite{Al2} - \cite{Huan1}
or even TeV \cite{Blum1}  - \cite{Cvetic}  string scales.  The  former
arises,  for example, when D-branes wrap   the coordinate planes of an
orbifold  (or    orientifold) compactified   space, such  as   we have
considered,   and hidden-sector anti-D-branes transmit   supersymmetry
breaking to the (observable sector) D-branes.   The latter arises when
intersecting D-branes wrap  a toroidally compactified space transverse
to an orbifold or orientifold.  A particularly attractive scenario for
model building using the first technique is the ``bottom-up'' approach
\cite{Al2}  - \cite{Huan1}   in   which  some  approximation  to   the
(supersymmetric) Standard Model is constructed using D$3$-branes at an
orbifold fixed point.   The cancellation of the twisted  $\mathbf{RR}$
charge at this  point requires the introduction  of other D-branes and
to preserve supersymmetry these  should be D$7$-branes.  However it is
of interest  to  consider, as  we have  done,  alternatives which  are
non-supersymmetric,  especially since  the  string scale  is no longer
tied  to the Planck   scale.  One immediate   objection might  be that
non-supersymmetric string theories  often  but not always, as  we have
shown,    possess (scalar)  tachyons.  However  these   too have  been
rehabilitated in recent    years.   If they  are   electroweak $SU(2)$
doublets, they may be interpreted  as Higgs bosons \cite{Marchesano1}.
Also  singlet tachyons imply the existence  of a scalar potential with
possibly  interesting   cosmological  consequences   \cite{Burgess}  -
\cite{Lerda}.  At  a more technical level,  it is widely believed that
twisted tadpole  cancellation implies the  cancellation of non-abelian
anomalies  in the emergent gauge field  theory,  but hitherto this has
only been verified in supersymmetric theories.

In  this paper, we have studied  the construction of Type IIB orbifold
models in  four  dimensions in   the  presence of  different  types of
parallel branes and antibranes.  We computed the open string spectrum,
its contribution to the gauge anomalies and the twisted ${\mathbf RR}$
tadpole  cancellation conditions in  each  case.  We obtained  several
supersymmetric and  non-supersymmetric configurations of D-branes. For
the  supersymmetric systems (${\mathbf  37}$  and ${\mathbf  95}$), we
verified that tadpole cancellation  conditions guarantees the  absence
of          all     non-abelian       gauge    anomalies    \cite{Al2}
\cite{Bada}\cite{Blum}\cite{Leigh2}.     For   the  non-supersymmetric
systems, the   presence of tachyonic  excitations is  a common feature
(although for some particular cases  tachyons get projected out of the
spectrum)  and  the  cancellation of  the  twisted tadpoles  does  not
guarantee the cancellation  of the non-abelian  gauge anomalies.  As a
result,  additional  constraints coming from  the  cancellation of the
non-abelian anomalies   should  be  imposed  in  order   to  obtain  a
consistent theory.  The bottom up construction of standard-like models
using non-supersymmetric configurations of  D-branes is being  studied
elsewhere \cite{Yo2}.

\subsection*{Acknowledgments}

M. E. Angulo would like to thank the Centre for Theoretical Physics at
the Univesity of Sussex for the  financial support and hospitality and
Mar   Bastero, Malcolm Fairbairn,    Pedro Castelo Ferreira, Alejandro
Ibarra, S.  G.  Nibbelink, Liliana Velasco and  Ivonne Zavala for very
useful discussions.  H.X.  Yang is  supported by Pao's scholarship for
Chinese studying overseas.  This research   is partially supported  by
PPARC.

\section*{Appendix A} \renewcommand{\theequation}{A.\arabic{equation}}
\setcounter{equation}{0}

\subsection{Theta $ \vartheta $ functions}

In this appendix  we write the properties  of the theta  $ \vartheta $
functions used in the computation of the partition functions.  

The theta $\vartheta$ function with characteristics a and b is given by:
\begin{equation} \label{eq:A1}
\vartheta     \left[   \begin{array}{c}      a  \\     b   \end{array}
\right](t)=\sum_{n}q^{\frac{1}{2}(n+a)^2}e^{2i\pi(n+a)b}
\end{equation} 
where the  variable  q is defined as  $q=e^{-2  \pi t}$.  The Dedekind
$\eta$ function is:
\begin{equation} \label{eq:A2}
\eta=q^{\frac{1}{24}} \prod^{\infty}_{n=1}(1-q^n). \end{equation}
The modular transformation properties are given by:
\begin{eqnarray} \label{eq:A3}
\vartheta \left[ \begin{array}{c} a \\ b  \end{array} \right](t) & = &
e^{2 i \pi a b }t^{-\frac{1}{2}}  \vartheta \left[ \begin{array}{c} -b
\\ a
\end{array} \right](1/t) \\
\eta(t) & = & t^{-\frac{1}{2}}\eta (1/t).
\end{eqnarray}
A fundamental   expression for the  ratio  of these two   functions in
product form is given by:
\begin{equation} \label{eq:A5} 
\frac{\vartheta \left[
\begin{array}{c} a \\ b \end{array} \right] (t)}{\eta (t)} = [e^{2i\pi
ab}q^{\frac{1}{2}a^2        -           \frac{1}{24}}]           \cdot
\prod^{\infty}_{n=1}(1+q^{n+a-\frac{1}{2}}e^{2\pi
ib})(1+q^{n-a-\frac{1}{2}}e^{-2\pi ib}).
\end{equation}
The    $\vartheta$  functions   satisfy   several  $abstruse$  Riemann
identities, $e.g.$,
\begin{equation} \label{eq:A6} 
\sum_{a,b}\eta _{a,b}\vartheta \left[
\begin{array}{c} a \\ b \end{array}\right] \prod_{r=3}^{5}\vartheta
\left[ \begin{array}{c} a \\ b+v_r
\end{array} \right] = 0 \end{equation} 
\begin{equation}\label{eq:A7} \sum_{a,b}\eta
_{ab}\vartheta  \left[  \begin{array}{c} a   \\ b  \end{array}\right ]
\vartheta \left[
\begin{array}{c} a \\ b+v_5 \end{array}\right] \prod_{r=3}^{4}
\vartheta \left[ \begin{array}{c} a+  \frac{1}{2} \\ b+v_r \end{array}
\right] = 0 \end{equation}
where $\eta_{ab}=(-1)^{2(a+b+2ab)}$ and $ v_3 + v_4 + v_5 = 0$.  It is
helpful to analyze the various  limits that appear in the  calculation
of the partition functions and the tadpole cancellation conditions:
\begin{equation} \label{eq:A8} \lim_{b \rightarrow 0} \frac{-2\sin \pi
b}{\vartheta \left[ \begin{array}{c} \frac{1}{2} \\ \frac{1}{2}+b
\end{array} \right]}=\frac{1}{\eta^3} \end{equation} 
\begin{equation} \label{eq:A9} \lim_{t
\rightarrow  0}    \frac{\vartheta   \left[  \begin{array}{c}    0  \\
\frac{1}{2} \end{array} \right](t)}{\eta^3 (t)}=2t \end{equation}
\begin{equation} \label{eq:A10} 
\lim_{t \rightarrow 0} \frac{\vartheta \left[
\begin{array}{c} 0 \\ \frac{1}{2}+kv_i \end{array}
\right](t)}{\vartheta       \left[  \begin{array}{c}  \frac{1}{2}   \\
\frac{1}{2}+kv_i
\end{array} \right](t)}= \left \{ \begin{array}{r@{\quad \textrm{if} 
\quad}l} (-1)^{[kv_i] + 1} & v_j >0 \\ (-1)^{[-kv_i]} & v_j <0
\end{array} \right. 
\end{equation} 
where $[kv_i]$  denotes the integer  part  of $kv_i$.  It can  also be
verified for all  supersymmetric orbifold groups ${\bf Z}_{N}$  listed
in \cite{Al2} that
\begin{eqnarray} \label{eq:A11} (-1)^{[k | v_3 |] +[k | v_4 |] + 
 [k    |      v_5   |]  } \prod_{r=3}^{5}       (2\sin    \pi kv_r)  =
-\prod_{r=3}^{5}|2\sin \pi kv_r| \end{eqnarray}
if the components  of the twist vector satisfy  the constraint $v_3  +
v_4 + v_5 =0$.

\section*{Appendix B} 
\renewcommand{\theequation}{B.\arabic{equation}}\setcounter{equation}{0}

\subsection{Partition functions}

The   other partition functions relevant   for  the computation of the
tadpole cancellation conditions are:
\begin{eqnarray}\label{33pf}
Z_{33}(\theta           ^k)           =          iV_4(8\pi    ^2\alpha
't)^{-2}(Tr\gamma_{k,3})(Tr\gamma_{k,3}^{-1})\cdot        \!\!\!\!\!\!
\sum_{a, b = 0, 1/2} \!\!\!\!\!\!  \eta_{a b}\frac{\vartheta \left [
\begin{array}{c} a \\ b \end{array} \right ] (t)}{\eta ^3 (t)} \cdot
\prod _{r=3}^{5}\frac{(-2sin\pi kv_r)\vartheta \left [
\begin{array}{c} a \\ b+kv_r \end{array} \right ] (t)}{\vartheta \left
[ \begin{array}{c} 1/2 \\ 1/2+kv_r
\end{array} \right ] (t)} \end{eqnarray} 
where we have considered that the  D$3$-brane world-volume embedds the
full  non-compact space-time.  All  compact complex dimensions obey DD
boundary conditions.  Like always,  the non-compact dimensions obey NN
boundary conditions.
\begin{eqnarray}\label{7i7ipf}
Z_{7_i7_i}(\theta       ^k)     =     &i&V_4(8\pi             ^2\alpha
't)^{-2}(Tr\gamma_{k,7_i})(Tr\gamma_{k,7_i}^{-1})                \cdot
\frac{(2\sin\pi kv_i)^2}{\prod_{j=3}^{5}(2\sin \pi  kv_j)}\nonumber \\
&\cdot& \!\!\!\!\!\!  \sum_{a, b  =   0, 1/2} \!\!\!\!\!\!    \eta_{a
b}\frac{\vartheta \left [
\begin{array}{c} a \\ b \end{array} \right ] (t)}{\eta ^3 (t)}
\cdot \prod _{r=3}^{5}\frac{(-2sin\pi kv_r)\vartheta \left [
\begin{array}{c} a \\ b+kv_r \end{array} \right ] (t)}{\vartheta \left
[ \begin{array}{c} 1/2 \\ 1/2+kv_r \end{array} \right ] (t)}.
\end{eqnarray}
By   D$7_i$ we denote   a D$7$-brane transverse   to the $z_i$ complex
plane.  Therefore,   in  the  $\mathbf{7_i7_i}$ system  there   are NN
boundary conditions in the $l$-th and $m$-th complex directions and DD
boundary conditions in the $i$-th complex plane.
\begin{eqnarray} \label{37pf}
Z_{37_i}(\theta       ^k)        =          &i&V_4(8\pi       ^2\alpha
't)^{-2}(Tr\gamma_{k,3})(Tr\gamma_{k,7_i}^{-1}) \nonumber \\ & & \cdot
\!\!\!\!\!\!    \sum_{a,    b   =    0,   1/2}\!\!\!\!\!\!    \eta_{a
b}\frac{\vartheta \left [
\begin{array}{c} a \\ b \end{array} \right ] (t)}{\eta ^3 (t)} \cdot
\frac{(-2sin\pi kv_i)\vartheta \left [ \begin{array}{c} a \\ b+kv_i
\end{array} \right ] (t)}{\vartheta \left [ \begin{array}{c} 1/2 \\
1/2+kv_i
\end{array} \right ] (t)}\cdot \prod_{l, m}
\frac{\vartheta \left [ \begin{array}{c}  1/2-a \\  b+kv_l \end{array}
\right ] (t)}{\vartheta \left [
\begin{array}{c} 0 \\ 1/2+kv_l \end{array} \right ] (t)}. 
\end{eqnarray} 
The $\mathbf{37_i}$ system obeys  DD boundary conditions in the $i$-th
complex plane and   mixed DN  boundary conditions  in  the  other  two
complex directions.
\begin{eqnarray}\label{7i7lpf}
Z_{7_i7_l}(\theta          ^k)       =      &i&V_4(8\pi       ^2\alpha
't)^{-2}(Tr\gamma_{k,7_i})(Tr\gamma_{k,7_j}^{-1})     \cdot  (2\sin\pi
kv_i)(2\sin\pi kv_l)   \cdot      \prod_{r=3}^{5}(2\sin\pi  kv_r)^{-1}
\nonumber \\ & & \cdot \!\!\!\!\!\!  \sum_{a, b = 0, 1/2} \!\!\!\!\!\!
\eta_{a b}\frac{\vartheta \left [ \begin{array}{c} a \\ b
\end{array} \right ] (t)}{\eta ^3 (t)} \cdot \frac{-\vartheta 
\left  [ \begin{array}{c} a     \\    b+kv_m \end{array}   \right    ]
(t)}{\vartheta \left [
\begin{array}{c} 1/2 \\ 1/2+kv_m \end{array} \right ] (t)}\cdot 
\frac{\vartheta \left [
\begin{array}{c} 1/2-a \\ b+kv_i \end{array} \right ] (t)}{\vartheta 
\left [
\begin{array}{c} 0 \\ 1/2+kv_i \end{array} \right ] (t)}\cdot 
\frac{\vartheta \left [
\begin{array}{c} 1/2-a \\ b+kv_l \end{array} \right ] (t)}{\vartheta 
\left [
\begin{array}{c} 0 \\ 1/2+kv_l \end{array} \right ] (t)}
\end{eqnarray} 
where $i\not=l \not= m \not= i$.  This brane system obeys mixed ND(ND)
boundary  conditions in the    $i$-th  and $l$-th  complex  directions
respectively and NN boundary conditions in the m-th complex plane.
\begin{eqnarray}\label{39pf}
Z_{39}(\theta      ^k)      =      &i&V_4(8\pi                ^2\alpha
't)^{-2}(Tr\gamma_{k,3})(Tr\gamma_{k,9}^{-1})    \cdot    \!\!\!\!\!\!
\sum_{a, b = 0, 1/2} \!\!\!\!\!\!  \eta_{a b}\frac{\vartheta \left [
\begin{array}{c} a \\ b \end{array} \right ] (t)}{\eta ^3 (t)} \cdot
\prod_{r=3}^{5} \frac{\vartheta  \left  [  \begin{array}{c}  1/2-a  \\
b+kv_r \end{array} \right  ] (t)}{\vartheta \left [ \begin{array}{c} 0
\\ 1/2+kv_r
\end{array} \right ] (t)}. \end{eqnarray} 
This system obeys DN boundary conditions in all complex directions.
\begin{eqnarray}\label{97pf}
Z_{97_i}(\theta         ^k)          =       &i&V_4(8\pi      ^2\alpha
 't)^{-2}(Tr\gamma_{k,9})(Tr\gamma_{k,7_i}^{-1})                 \cdot
 \prod_{r=3}^{5}(2\sin\pi   kv_r)^{-2}     \nonumber    \\ &\cdot    &
 \!\!\!\!\!\!   \sum_{a,   b    =    0,   1/2}\!\!\!\!\!\!    \eta_{a
 b}\frac{\vartheta \left [
\begin{array}{c} a \\ b \end{array} \right ] (t)}{\eta ^3 (t)}
\cdot \frac{\vartheta \left [ \begin{array}{c} 1/2-a \\ b+kv_i
\end{array} \right ] (t)}{\vartheta \left [ \begin{array}{c} 0 \\
1/2+kv_i \end{array} \right  ] (t)} \cdot \prod_{l, m}\frac{(-2\sin\pi
kv_l)\vartheta \left [ \begin{array}{c} a \\ b+kv_l
\end{array} \right ] (t)}{\vartheta \left [ \begin{array}{c} 1/2 \\
1/2+kv_l
\end{array} \right ] (t)} \end{eqnarray}
where  the $\mathbf{97_i}$ system obeys  NN boundary conditions in the
$l$-th and $m$-th complex planes  and mixed ND boundary conditions  in
the $i$-th complex plane perpendicular to the D$7_i$ brane.
\begin{eqnarray}\label{7i5ipf}
Z_{7_i5_i}(\theta            ^k)             =    iV_4(8\pi   ^2\alpha
't)^{-2}(Tr\gamma_{k,7_i})(Tr\gamma_{k,5_i}^{-1})   \cdot \!\!\!\!\!\!
\sum_{a, b = 0, 1/2} \!\!\!\!\!\!  \eta_{a b}\frac{\vartheta \left [
\begin{array}{c} a \\ b \end{array} \right ] (t)}{\eta ^3 (t)} \cdot
\prod_{r=3}^{5}  \frac{\vartheta  \left [  \begin{array}{c}  1/2-a  \\
b+kv_r \end{array} \right ] (t)}{\vartheta  \left [ \begin{array}{c} 0
\\ 1/2+kv_r
\end{array} \right ] (t)} \end{eqnarray} 
This system obeys mixed DN(ND) boundary conditions  in all the complex
planes.
\begin{eqnarray}\label{7i5lpf} & &
Z_{7_i5_l}(\theta          ^k)         =      iV_4(8\pi       ^2\alpha
't)^{-2}(Tr\gamma_{k,7_i})(Tr\gamma_{k,5_l}^{-1})(2\sin\pi  kv_l)^{-2}
\nonumber \\ & & \cdot \!\!\!\!\!\!   \sum_{a, b = 0, 1/2}\!\!\!\!\!\!
\eta_{a b}\frac{\vartheta \left [ \begin{array}{c} a \\ b \end{array}
\right ] (t)}{\eta ^3 (t)} \cdot \frac{(-2\sin\pi kv_i)\vartheta \left
[  \begin{array}{c} a  \\  b+kv_i \end{array} \right ]  (t)}{\vartheta
\left   [ \begin{array}{c} 1/2 \\  1/2+kv_i  \end{array} \right ] (t)}
\cdot \frac{(-2\sin\pi kv_l)\vartheta \left   [ \begin{array}{c} a  \\
b+kv_l
\end{array} \right ] (t)}{\vartheta \left [ \begin{array}{c} 1/2\\
1/2+kv_l
\end{array} \right ] (t)}\cdot \frac{\vartheta \left
[ \begin{array}{c} 1/2-a \\ b+kv_m
\end{array} \right ] (t)}{\vartheta \left [ \begin{array}{c} 0 
\\ 1/2+kv_m \end{array} \right ] (t)}. \end{eqnarray}
The $\mathbf{7_i5_l}$   system  obeys DD boundary    conditions in the
$i$-th complex plane, NN  boundary conditions in the  $l$-th direction
and   mixed   boundary  conditions  in   the    $m$-th  complex  plane
perpendicular to both the $7_i$ and $5_l$ branes.
\begin{eqnarray}\label{35ipf}
Z_{35_i}(\theta           ^k)   =      &i&V_4(8\pi            ^2\alpha
't)^{-2}(Tr\gamma_{k,3})(Tr\gamma_{k,5_i}^{-1}) \nonumber \\ & & \cdot
\!\!\!\!\!\!     \sum_{a,  b   =   0,    1/2}  \!\!\!\!\!\!   \eta_{a
b}\frac{\vartheta \left [
\begin{array}{c} a \\ b \end{array} \right ] (t)}{\eta ^3 (t)} \cdot
\prod_{r=3}^{5} \frac{(-2\sin\pi kv_r)\vartheta \left [
\begin{array}{c} a \\ b+kv_r \end{array} \right ] (t)}{\vartheta \left [
\begin{array}{c} 1/2 \\ 1/2+kv_r \end{array} \right ] (t)}
 \cdot \frac{\vartheta \left [ \begin{array}{c} 1/2-a \\ b+kv_i
\end{array} \right ] (t)}{\vartheta \left [ \begin{array}{c} 0 \\
1/2+kv_i
\end{array} \right ] (t)}
\end{eqnarray} 
where  the system obeys   mixed DN boundary  conditions in  the $i$-th
complex plane and  DD boundary  conditions  in the other two   complex
directions.

\section*{Appendix C}
\renewcommand{\theequation}{C.\arabic{equation}}\setcounter{equation}{0}

\subsection{Asymptotic behaviours}

We chose $v_1>0$,  $v_2>0$ but $v_3<0$ in order  to guarantee $\pm v_1
\pm v_2  \pm  v_3  =0$.  Using  expressions    (A.9) and  (A.10),  the
asymptotic       behaviour  of    the    $\mathbf{RR}$      ${\mathcal
Z}_{pq}(\theta^k)$ amplitudes is given by:
\begin{eqnarray} \label{C1}
{\mathcal    Z}_{99}^{(RR)}   \cong -iV_4(8\pi  ^2  \alpha')^{-2}\cdot
\frac{2}{t}\cdot \frac{1}{\prod_{r=3}^{5}(2\sin \pi kv_r)} \cdot  {(Tr
\gamma_{k,9})(Tr \gamma_{k, 9}^{-1})} (-1)^{[kv_1]+[kv_2]+[-kv_3]+1}
\end{eqnarray} 
\begin{eqnarray} \label{C2}
{\mathcal Z}_{5_i5_i}^{(RR)} \cong  -iV_4(8\pi  ^2 \alpha')^{-2} \cdot
\frac{2}{t} \cdot  \frac{\prod_{r=3}^{5}(2\sin   \pi kv_r)}{(2\sin \pi
kv_i)^2}\cdot   (Tr    \gamma_{k,5_i})(Tr       \gamma_{k,  5_i}^{-1})
(-1)^{[kv_1]+[kv_2]+[-kv_3]+1}
\end{eqnarray} 
\begin{eqnarray}\label{C3} 
{\mathcal  Z}_{5_i5_l}^{(RR)} \cong  -iV_4(8\pi ^2 \alpha')^{-2} \cdot
\frac{2}{t}  \cdot \frac{\prod_{r=3}^{5}(2\sin  \pi kv_r)}{(2\sin  \pi
kv_i)(2\sin   \pi  kv_l)}   \cdot (Tr   \gamma_{k,5_i})(Tr  \gamma_{k,
5_l}^{-1}) (-1)^{[kv_1]+[kv_2]+[-kv_3]+1} \end{eqnarray}
\begin{eqnarray}\label{C4}
{\mathcal  Z}_{95_i}^{(RR)}  \cong -iV_4(8\pi ^2  \alpha')^{-2}  \cdot
\frac{2}{t}\cdot  \frac{1}{(2\sin \pi kv_i)}\cdot (Tr \gamma_{k,9})(Tr
\gamma_{k, 5_i}^{-1}) (-1)^{[kv_1]+[kv_2]+[-kv_3]+1} \end{eqnarray}
\begin{eqnarray}\label{C5}
{\mathcal   Z}_{33}^{(RR)}  \cong -iV_4(8\pi  ^2  \alpha')^{-2}  \cdot
\frac{2}{t}\cdot      \prod_{r=3}^{5}(2\sin    \pi   kv_r)\cdot    (Tr
\gamma_{k,3})(Tr \gamma_{k, 3}^{-1}) (-1)^{[kv_1]+[kv_2]+[-kv_3]+1}
\end{eqnarray}
\begin{eqnarray}\label{C6}
{\mathcal  Z}_{7_i7_i}^{(RR)} \cong  -iV_4(8\pi ^2 \alpha')^{-2} \cdot
\frac{2}{t}\cdot \frac{(2\sin  \pi kv_i)^2} {\prod_{r=3}^{5}(2\sin \pi
kv_r)}\cdot     (Tr    \gamma_{k,7_i})(Tr  \gamma_{k,       7_i}^{-1})
(-1)^{[kv_1]+[kv_2]+[-kv_3]+1} \end{eqnarray}
\begin{eqnarray} \label{C7}
{\mathcal  Z}_{37_i}^{(RR)}  \cong  -iV_4(8\pi ^2 \alpha')^{-2}  \cdot
\frac{2}{t}\cdot  (2\sin  \pi  kv_i)    \cdot (Tr  \gamma_{k,   3})(Tr
\gamma_{k, 7_i}^{-1}) (-1)^{[kv_1]+[kv_2]+[-kv_3]+1}
\end{eqnarray} 
\begin{eqnarray}\label{C8} 
{\mathcal Z}_{7_i7_l}^{(RR)} \cong  -iV_4(8\pi  ^2 \alpha')^{-2} \cdot
\frac{2}{t}\cdot  \frac{(2\sin     \pi   kv_i)(2\sin   \pi      kv_l)}
{\prod_{r=3}^{5}(2\sin   \pi   kv_r)}    \cdot(Tr  \gamma_{k, 7_i})(Tr
\gamma_{k, 7_l}^{-1}) (-1)^{[kv_1]+[kv_2]+[-kv_3]+1} \end{eqnarray}
\begin{eqnarray} \label{C9}
{\mathcal   Z}_{39}^{(RR)} \cong  -iV_4(8\pi  ^2  \alpha')^{-2}  \cdot
\frac{2}{t}\cdot    (Tr    \gamma_{k,  3})(Tr   \gamma_{k,    9}^{-1})
(-1)^{[kv_1]+[kv_2]+[-kv_3]+1}
\end{eqnarray} 
\begin{eqnarray} \label{C10}
{\mathcal Z}_{97_i}^{(RR)} \cong   -iV_4(8\pi  ^2 \alpha')^{-2}  \cdot
\frac{2}{t}\cdot     \frac{1}{\prod_{l, m}(2\sin  \pi  kv_l)}\cdot (Tr
\gamma_{k, 9})(Tr \gamma_{k, 7_i}^{-1}) (-1)^{[kv_1]+[kv_2]+[-kv_3]+1}
\end{eqnarray} 
\begin{eqnarray} \label{C11}
{\mathcal  Z}_{7_i5_i}^{(RR)}  \cong -iV_4(8\pi ^2 \alpha')^{-2} \cdot
\frac{2}{t}\cdot    (Tr \gamma_{k,   7_i})(Tr   \gamma_{k,  5_i}^{-1})
(-1)^{[kv_1]+[kv_2]+[-kv_3]}
\end{eqnarray} 
\begin{eqnarray} \label{C12}
{\mathcal Z}_{7_i5_l}^{(RR)} \cong  -iV_4(8\pi  ^2 \alpha')^{-2} \cdot
\frac{2}{t}\cdot \frac{(2\sin \pi   kv_i)(2\sin \pi kv_l)}{(2\sin  \pi
kv_m)}\cdot     (Tr     \gamma_{k, 7_i})(Tr    \gamma_{k,   5_l}^{-1})
(-1)^{[kv_1]+[kv_2]+[-kv_3]}
\end{eqnarray} 
\begin{eqnarray} \label{C13}
{\mathcal   Z}_{35_i}^{(RR)} \cong  -iV_4(8\pi  ^2 \alpha')^{-2} \cdot
\frac{2}{t}\cdot  \frac{\prod_{r=3}^{5}(2\sin \pi  kv_r)} {(2\sin  \pi
kv_r)}\cdot   (Tr      \gamma_{k,  3})(Tr     \gamma_{k,    5_i}^{-1})
(-1)^{[kv_1]+[kv_2]+[-kv_3]} \end{eqnarray}

\end{document}